\documentclass[sigconf]{acmart}
\usepackage[vlined,boxed]{algorithm2e}
\usepackage{amsmath}
\usepackage{amsfonts}
\usepackage{epsfig}
\usepackage{graphics}
\usepackage{amssymb}
\usepackage{color}
\usepackage{comment}
\usepackage{paralist}
\usepackage{mathtools}
\usepackage{multirow}
\usepackage{subfigure}
\usepackage{eepic}
\usepackage{stmaryrd}
\usepackage{textcomp}
\usepackage{balance}
\usepackage[inline]{enumitem}
\usepackage{ulem}

\begin{document}

\newcommand{\mi}[1]{\mathit{#1}}
\newcommand{\ins}[1]{\mathbf{#1}}
\newcommand{\adom}[1]{\mathsf{dom}(#1)}
\renewcommand{\paragraph}[1]{\textbf{#1}}
\newcommand{\ra}{\rightarrow}
\newcommand{\fr}[1]{\mathsf{fr}(#1)}
\newcommand{\dep}{{\mathcal T}}
\newcommand{\sch}[1]{\mathsf{sch}(#1)}
\newcommand{\ar}[1]{\mathsf{ar}(#1)}
\newcommand{\body}[1]{\mathsf{body}(#1)}
\newcommand{\head}[1]{\mathsf{head}(#1)}
\newcommand{\guard}[1]{\mathsf{guard}(#1)}
\newcommand{\class}[1]{\mathbb{#1}}
\newcommand{\pos}[2]{\mathsf{pos}(#1,#2)}
\newcommand{\app}[2]{\langle #1,#2 \rangle}
\newcommand{\crel}[1]{\prec_{#1}}

\newcommand{\ccrel}[1]{\prec_{#1}^+}

\newcommand{\tcrel}[1]{\prec_{#1}^{\star}}
\newcommand{\rctaa}{\class{CT}_{\forall \forall}^{\mathsf{res}}}
\newcommand{\rctaapr}{\mathsf{CT}_{\forall \forall}^{\mathsf{res}}}
\newcommand{\rctae}{\class{CT}_{\forall \exists}^{\mathsf{res}}}
\newcommand{\rctaepr}{\mathsf{CT}_{\forall \exists}^{\mathsf{res}}}
\newcommand{\base}[1]{\mathsf{base}(#1)}
\newcommand{\eqt}[1]{\mathsf{eqtype}(#1)}
\newcommand{\result}[1]{\mathsf{result}(#1)}
\newcommand{\chase}[2]{\mathsf{ochase}(#1,#2)}
\newcommand{\pred}[1]{\mathsf{pr}(#1)}
\newcommand{\origin}[1]{\mathsf{org}(#1)}
\newcommand{\eq}[1]{\mathsf{eq}(#1)}
\newcommand{\et}[1]{\mathsf{et}(#1)}
\newcommand{\can}[1]{\mathsf{can}(#1)}
\newcommand{\etypes}[1]{\mathsf{etypes}(#1)}
\newcommand{\etypest}[2]{\mathsf{etypes}_{#1}(#2)}
\newcommand{\eqtt}[2]{\mathsf{et}_{#1}(#2)}

\newcommand{\dept}[1]{\mathsf{depth}(#1)}

\newcommand{\ccc}{{\mathfrak c}}


\def\qed{\hfill{\qedboxempty}      
  \ifdim\lastskip<\medskipamount \removelastskip\penalty55\medskip\fi}

\def\qedboxempty{\vbox{\hrule\hbox{\vrule\kern3pt
                 \vbox{\kern3pt\kern3pt}\kern3pt\vrule}\hrule}}

\def\qedfull{\hfill{\qedboxfull}   
  \ifdim\lastskip<\medskipamount \removelastskip\penalty55\medskip\fi}

\def\qedboxfull{\vrule height 4pt width 4pt depth 0pt}

\newcommand{\markfull}{\qedboxfull}
\newcommand{\markempty}{\qed} 

\newcommand{\OMIT}[1]{}

\newtheorem{claim}[theorem]{Claim}
\newtheorem{fact}[theorem]{Fact}
\newtheorem{observation}{Observation}
\newtheorem{remark}{Remark}
\newtheorem{apptheorem}{Theorem}[section]
\newtheorem{appcorollary}[apptheorem]{Corollary}
\newtheorem{appproposition}[apptheorem]{Proposition}
\newtheorem{applemma}[apptheorem]{Lemma}
\newtheorem{appclaim}[apptheorem]{Claim}
\newtheorem{appfact}[apptheorem]{Fact}

\fancyhead{}

\title{All-Instances Restricted Chase Termination}

\author{Tomasz Gogacz}
\affiliation{%
	\institution{Institute of Informatics}
	\city{University of Warsaw}
}
\email{t.gogacz@mimuw.edu.pl}

\author{Jerzy Marcinkowski}
\affiliation{%
	\institution{Institute of Computer Science}
	\city{University of Wroclaw}
}
\email{jma@cs.uni.wroc.pl}

\author{Andreas Pieris}
\affiliation{%
	\institution{School of Informatics}
	\city{University of Edinburgh}
}
\email{apieris@inf.ed.ac.uk}

\begin{abstract}
The chase procedure is a fundamental algorithmic tool in database theory with a variety of applications.
A key problem concerning the chase procedure is all-instances termination: for a given set of tuple-generating dependencies (TGDs), is it the case that the chase terminates for every input database? In view of the fact that this problem is undecidable, it is natural to ask whether known well-behaved classes of TGDs ensure decidability.
We consider here the main paradigms that led to robust TGD-based formalisms, that is, guardedness and stickiness.
%
Although all-instances termination is well-understood for the oblivious version of the chase, the more subtle case of the restricted (a.k.a.~the standard) chase is rather unexplored. We show that all-instances restricted chase termination for guarded and sticky single-head TGDs is decidable.
\end{abstract}

\maketitle

\section{Introduction}\label{sec:introduction}

The \emph{chase procedure} (or simply chase) is a fundamental algorithmic tool that has been applied to several database problems such as computing data exchange solutions~\cite{FKMP05}, and query answering and containment under
constraints~\cite{CaGK13,AhSU79},
to name a few.
The chase takes as input a database $D$ and a set $\dep$
of constraints -- which, for this work, are tuple-generating dependencies (TGDs) of the form $\forall \bar x \forall \bar y \left(\phi(\bar x,\bar y) \ra \exists \bar z\, \psi(\bar x,\bar z)\right)$ with $\phi$ and $\psi$ being conjunctions of atoms -- and, if it terminates, its result is a finite instance $D_\dep$ that is a {\em universal model} of $D$ and $\dep$, i.e., a model that can be homomorphically embedded into every other model of $D$ and $\dep$. This is the reason for the ubiquity of the chase as discussed in~\cite{DeNR08}.
Indeed, many central database problems, which involve reasoning with TGDs, can be solved by simply exhibiting a universal model. And this is not only in theory. Despite the fact that the instance constructed by the chase can be very large, efficient implementations of the chase procedure have been successfully applied during the last few years in many different contexts~\cite{BKMMPST17,KrMR19,NPMHWB15,UKJDC18}.

\medskip

\noindent
\paragraph{The Chase In a Nutshell.}
%
%
%
%
Roughly speaking, the chase adds new tuples to the database $D$ (possibly involving null values that act as witnesses for the existentially quantified variables), as dictated by the TGDs of $\dep$, and it keeps doing this until all the TGDs of $\dep$ are satisfied.
%
%
%
%
%
%
There are, in principle, two different ways for formalizing this simple idea, which lead to different versions of the chase procedure.
The first one, which gives rise to the {\em oblivious chase}, is as follows: for each pair $\bar t,\bar u$ of tuples of terms from the instance $I$ constructed so far, trigger a TGD $\forall \bar x \forall \bar y \left(\phi(\bar x,\bar y) \ra \exists \bar z\, \psi(\bar x,\bar z)\right)$ if $\phi(\bar t,\bar u) \subseteq I$, and add to $I$ the set of atoms $\psi(\bar t,\bar v)$, where $\bar v$ is a tuple of new terms not occurring in $I$.
The second way, which leads to the {\em restricted} (a.k.a.~{\em standard}) {\em chase}, is a refinement of the above with the additional condition that, for a pair $\bar t,\bar u$ of tuples of terms, a TGD $\forall \bar x \forall \bar y \left(\phi(\bar x,\bar y) \ra \exists \bar z\, \psi(\bar x,\bar z)\right)$ is triggered not only if $\phi(\bar t,\bar u) \subseteq I$, but also if there is no tuple $\bar v$ of terms from $I$ such that $\psi(\bar t,\bar v) \subseteq I$, i.e., if the TGD is not already satisfied.
Thus, the key difference between the oblivious and restricted versions of the chase is that the former triggers a TGD whenever the left-hand side of the implication is satisfied, while the latter triggers a TGD only if it is violated.

It should be clear that the restricted chase, in general, builds much smaller instances than the oblivious one. Actually, it is very easy to devise an example where, according to the restricted chase, none of the TGDs should be triggered, while the oblivious chase builds an infinite instance. Consider, e.g., the database $D = \{R(a,b)\}$ and the TGD $\forall x \forall y (R(x,y) \ra \exists z\, R(x,z))$. The restricted chase will detect that the database already satisfies the TGD, while the oblivious chase will build the infinite instance $\{R(a,b),R(a,\nu_1),R(a,\nu_2),\ldots\}$, where $\nu_1,\nu_2,\ldots$ are (labeled) nulls.
Consequently, the restricted chase has a clear advantage over the oblivious chase when it comes to the size of the result. But, of course, this advantage does not come for free: at each step, the restricted chase has to check that there is no way to satisfy the right-hand side of the TGD at hand, and this is costly. However, as it has been recently observed, the benefit from producing much smaller instances can justify the effort of checking whether a TGD is already satisfied; see, e.g.,~\cite{BKMMPST17,KrMR19}.

\subsection{The Challenge of Non-termination}

As said above, there are nowadays efficient implementations of the restricted chase that allows us to solve central database problems by adopting a materialization-based approach~\cite{BKMMPST17,KrMR19,NPMHWB15,UKJDC18}. But, of course, for this to be feasible in practice we need a guarantee that the restricted chase terminates, which is not always the case.
This fact motivated a long line of research on identifying fragments of TGDs that ensure the termination of the restricted chase, for every input database.
A prime example is the class of \emph{weakly-acyclic} TGDs~\cite{FKMP05}, which is the standard language for data exchange purposes. A similar formalism, called \emph{constraints with stratified-witness}, has been proposed in~\cite{DeTa03}.
%
%
%
Many other sufficient conditions for the termination of the restricted chase can be found in the literature; see, e.g.,~\cite{DeNR08,DeTa03,GHKK+13,GrST11,Marn09,MeSL09}
-- this list is by no means exhaustive, and we refer the reader to~\cite{GrMS12} for a comprehensive survey.

With so much effort spent on identifying sufficient conditions for the termination of the restricted chase, the question that comes up is whether a sufficient condition that is also \emph{necessary} exists. In other words, given a set $\dep$ of TGDs, is it possible to decide whether, for every database $D$, the restricted chase on $D$ and $\dep$ terminates?
This has been addressed in~\cite{GoMa14}, where it is shown that the answer is negative, even for the oblivious chase.


The undecidability proof in~\cite{GoMa14} constructs a sophisticated set of TGDs that goes beyond existing well-behaved classes of TGDs that enjoy certain syntactic properties, which in turn ensure favorable model-theoretic properties.
Such well-behaved classes of TGDs have been proposed in the context of ontological reasoning.
The two main paradigms that led to robust TGD-based formalisms, without forcing the restricted chase to terminate, are {\em guardedness}~\cite{BLMS11,CaGK13,CaGL12} and {\em stickiness}~\cite{CaGP12}.
A TGD is guarded if the left-hand side of the implication, known as the body of the TGD, has an atom that contains (or ``guards'') all the universally quantified variables.
%
On the other hand, sticky sets of TGDs are inherently unguarded, and their main goal is to express joins among relations that cannot be expressed via guarded TGDs (details are given in Section~\ref{sec:preliminaries}).
%

The fact that the set of TGDs given in the undecidability proof of~\cite{GoMa14} is far from being guarded or sticky brings us to the following question: is the restricted chase termination problem, as described above, decidable for guarded or sticky TGDs?
This question is rather well-understood for the oblivious chase. In the case of guarded TGDs, the problem is 2EXPTIME-complete, and becomes PSPACE-complete for linear (one body-atom) TGDs~\cite{CaGP15}. The sticky case has been recently addressed in~\cite{CaPi19}, where it is shown that the problem is PSPACE-complete.
However, despite its clear advantage over the oblivious chase, we know very little about the restricted chase. It has been shown, independently of our work, that the problem is decidable for single-head (one atom in the head) linear TGDs~\cite{LMTU19}. However, nothing so far was known about guarded or sticky TGDs.

\subsection{Research Challenges}

We concentrate on guarded and sticky TGDs (in fact, single-head TGDs), and study the restricted chase termination problem. More precisely, we study the following: given a set $\dep$ of single-head guarded or sticky TGDs, is it the case that for {\em every} database $D$, {\em every} restricted chase derivation of $D$ w.r.t.~$\dep$ is finite? It might be the case that some derivations are finite and some others are not, depending on the order that TGDs are triggered, which is not the case for the oblivious chase.
The reason for this non-deterministic behavior is the fact that the restricted chase applies a TGD only if it is necessary (recall the restricted vs. oblivious chase discussion above). On the other hand, the oblivious chase applies TGDs whenever the body is satisfied,
which ensures a deterministic behavior.
%
%
Our ultimate goal is to show that the problem in question is decidable.
Towards this direction, one has to overcome a couple of non-trivial technical issues, which were not so difficult in the case of the oblivious chase.

\medskip

\noindent \paragraph{Dealing with Fairness.} The fairness condition is crucial in the definition of the chase in order to ensure that the result is indeed a model of the input database and set of TGDs. It states that each TGD that is violated at some point of the execution of the chase eventually will be satisfied.
One of the main difficulties underlying our problem is to ensure fairness. In other words, focussing on the complement of our problem, it is not enough to simply check whether there exists a database that leads to an infinite derivation w.r.t.~the set of TGDs, but we have to ensure that it is also fair.

As shown in~\cite{CaGP15}, for the oblivious chase, the existence of a (possibly unfair) infinite chase derivation implies the existence of a fair one, which in turn implies that we can completely neglect the fairness condition.
The question that comes up is whether we can establish the same for the restricted chase, which will crucially simplify our task. Actually, this question has been already posed by Jan Van den Bussche some years ago in a different context~\cite{Jan}.
Showing such a result for the restricted chase is significantly more difficult than showing it for the oblivious chase.
Note that the recent work~\cite{LMTU19}, which considers the restricted chase, establishes such a result, but only for single-head linear TGDs. Generalizing this to single-head guarded or sticky TGDs, or ideally to arbitrary single-head TGDs, is a non-trivial task. As we shall see, here is the place where we need the TGDs to be single-head.

\medskip

\noindent \paragraph{Existence of a Critical Database.} It would be extremely useful to have a special database $D^*$ in place, let us call it {\em critical}, of a very simple form, that ensures the following: given a set $\dep$ of TGDs, if there is a database that leads to an infinite chase derivation w.r.t.~$\dep$, then already $D^*$ does. With such a critical database in place, one can focus on the complement of our problem, and check whether $D^*$ leads to an infinite chase derivation w.r.t.~the given TGDs.

For the oblivious chase such a critical database exists: it simply collects all the atoms of the form $R(c,\ldots,c)$, where $R$ is a relation that occurs in the given set of TGDs~\cite{Marn09}, and $c$ an arbitrary constant.
All the known decidability results about the oblivious chase heavily rely on the critical database $D^*$~\cite{CaGP15,CaPi19}.
It is an easy exercise, however, to show that $D^*$, as defined above, does not serve as a critical database in the case of the restricted chase.
This brings us to the other technical challenge that we need to overcome, that is, the lack of an obvious database that can serve as a critical database.
Let us say that~\cite{LMTU19}, which considers the restricted chase, follows the critical database approach. However, it is easy to see that for single-head linear TGDs (the main concern of~\cite{LMTU19}) such a critical database is simply a database consisting of a single atom. This is far from being true for single-head guarded or sticky TGDs.

\subsection{Summary of Contributions}

Our main results (Theorem~\ref{the:guarded-tgds} and Theorem~\ref{the:sticky-tgds}) state that, for a set $\dep$ of single-head guarded or sticky TGDs, checking whether, for every database $D$, every restricted chase derivation of $D$ w.r.t.~$\dep$ is finite, is decidable in elementary time. To show these results, we had to establish a series of auxiliary results, related to the technical challenges discussed above.
Our main contributions follow:

In Section~\ref{sec:fairness-lemma}, we establish the {\em Fairness Theorem}, which essentially states that, for single-head (not necessarily guarded or sticky) TGDs, we can neglect the fairness condition. This overcomes the first challenge raised in the previous subsection. Let us stress that this result does {\em not} hold once we go beyond single-head TGDs,
which means that our decision to focus on single-head TGDs is not for simplicity, but it might be crucial for the validity of our main results. This has been also observed, independently of our work, in~\cite{LMTU19}.

In Section~\ref{sec:guardedness}, we focus on guarded TGDs.
We first characterize the existence of an infinite (possibly unfair) restricted chase derivation of a database $D$ w.r.t.~a set $\dep$ of single-head (not necessarily guarded) TGDs via the existence of an infinite subset $S$, called {\em chaseable}, of the instance $C_{D,\dep}$ constructed by applying a variant of the oblivious chase on $D$ using $\dep$. Such a chaseable set $S$ enjoys certain properties that allow us to convert it into an infinite restricted chase derivation of $D$ w.r.t.$\dep$.
%
We then show that, for a set $\dep$ of single-head guarded TGDs, the problem of deciding whether there is a database $D$ such that an infinite chaseable subset of $C_{D,\dep}$ exists can be reduced to the satisfiability problem of Monadic Second-Order Logic (MSOL) over infinite trees of bounded degree.
The correctness of this reduction relies on another key result of independent interest: if there is a database that leads to a (possibly unfair) infinite chase derivation w.r.t.~$\dep$, then there is an {\em acyclic} one with the same property.

Finally, in Section~\ref{sec:stickiness}, we concentrate on sticky TGDs. Given a set $\dep$ of sticky TGDs, we reduce the problem of deciding whether there exists a database $D$ such that an infinite (possibly unfair) restricted chase derivation of $D$ w.r.t.~$\dep$ exists to the emptiness problem of deterministic B\"{u}chi automata.
This reduction relies on another key result of independent interest: there exists a database $D$ such that an infinite (possibly unfair) restricted chase derivation of $D$ w.r.t.~$\dep$ exists iff a so-called {\em finitary caterpillar} for $\dep$ exists. The latter is essentially an infinite ``path-like'' restricted chase derivation of some database w.r.t.~$\dep$, and is precisely the existence of such an object that we check via a deterministic B\"{u}chi automaton.

\section{Preliminaries}\label{sec:preliminaries}

We consider the disjoint countably infinite sets $\ins{C}$, $\ins{N}$, and $\ins{V}$ of {\em constants}, {\em (labeled) nulls}, and {\em variables} (used in dependencies), respectively. We refer to constants, nulls and variables as {\em terms}. For an integer $n > 0$, we may write $[n]$ for the set $\{1,\ldots,n\}$.

\medskip

\noindent
\paragraph{Relational Databases.} A {\em schema} $\ins{S}$ is a finite set of relation symbols (or predicates) with associated arity. We write $R/n$ to denote that $R$ has arity $n > 0$; we may also write $\ar{R}$ for $n$. A position of $\ins{S}$ is a pair $(R,i)$, where $R/n \in \ins{S}$ and $i \in [n]$, that essentially identifies the $i$-th argument of $R$.
An {\em atom} over $\ins{S}$ is an expression of the form $R(\bar t)$, where $R/n \in \ins{S}$ and $\bar t$ is an $n$-tuple of terms. A {\em fact} is an atom whose arguments consist only of constants.
We write $R(\bar t)[i]$ for the term of $R(\bar t)$ at position $(R,i)$, i.e., the $i$-th element of $\bar t$. For brevity, we may refer to the position $(R,i)$ in $R(\bar t)$ simply as the $i$-th position of $R(\bar t)$ and write $(R(\bar t),i)$.
Moreover, for a variable $x$ in $\bar t$, let $\pos{R(\bar t)}{x} = \{(R,i) : R(\bar t)[i] = x\}$, i.e., is the set of positions at which $x$ occurs according to $R(\bar t)$.
%
%
%
%
%
%
An {\em instance} over $\ins{S}$ is a (possibly infinite) set of atoms over $\ins{S}$ that contain constants and nulls, while a {\em database} over $\ins{S}$ is a finite set of facts over $\ins{S}$. The {\em active domain} of an instance $I$, denoted $\adom{I}$, is the set of all terms in $I$.

\medskip

\noindent
\paragraph{Substitutions and Homomorphisms.}
A {\em substitution} from a set of terms $T$ to a set of terms $T'$ is a function $h : T \ra T'$ defined as follows: $\emptyset$ is a substitution, and if $h$ is a substitution, then $h \cup \{t \mapsto t'\}$, where $t \in T$ and $t' \in T'$, is a substitution. The restriction of $h$ to $S \subseteq T$ is denoted $h_{|S}$.
%
A {\em homomorphism} from a set of atoms $A$ to a set of atoms $B$ is a substitution $h$ from the terms of $A$ to the terms of $B$ such that (i) $t \in \ins{C}$ implies $h(t) = t$, and (ii) $R(t_1,\ldots,t_n) \in A$ implies $h(R(t_1,\ldots,t_n)) =  R(h(t_1),\ldots,h(t_n)) \in B$.
%

\medskip

\noindent
\paragraph{Single-Head Tuple-Generating Dependencies.} A {\em single-head tuple-generating dependency} $\sigma$ is a constant-free first-order sentence
$\forall \bar x \forall \bar y \left(\phi(\bar x,\bar y) \ra \exists \bar z\, R(\bar x,\bar z)\right)$,
where $\bar x, \bar y, \bar z$ are tuples of variables of $\ins{V}$, $\phi(\bar x,\bar y)$ is a conjunction of atoms, and $R(\bar x,\bar z)$ is a single atom.
For brevity, we write $\sigma$ as $\phi(\bar x,\bar y) \ra \exists \bar z\, R(\bar x,\bar z)$, and use comma instead of $\wedge$ for joining atoms. We refer to $\phi(\bar x,\bar y)$ and $R(\bar x,\bar z)$ as the {\em body} and {\em head} of $\sigma$, denoted $\body{\sigma}$ and $\head{\sigma}$, respectively.
%
Henceforth, we simply say tuple-generating dependency (TGD) instead of single-head TGD.
The {\em frontier} of the TGD $\sigma$, denoted $\fr{\sigma}$, is the set of variables $\bar x$, i.e., the variables that appear both in the body and the head of $\sigma$. The schema of a set $\dep$ of TGDs, denoted $\sch{\dep}$, is the set of predicates in $\dep$, and we write $\ar{\dep}$ for the maximum arity over all those predicates.
An instance $I$ satisfies a TGD $\sigma$ as the one above, written $I \models \sigma$, if the following holds: whenever there exists a homomorphism $h$ such that $h(\phi(\bar x, \bar y)) \subseteq I$, then there exists $h' \supseteq h_{|\bar x}$ such that $h'(R(\bar x,\bar z)) \in I$. Note that, by abuse of notation, we sometimes treat a tuple of variables as a set of variables,
and a conjunction of atoms as a set of atoms. The instance $I$ satisfies a set $\dep$ of TGDs, written $I \models \dep$, if $I \models \sigma$ for each $\sigma \in \dep$.

\medskip

\noindent
\paragraph{Guardedness.} A TGD $\sigma$ is {\em guarded} if there exists an atom $\alpha$ in its body that contains all the variables occurring in $\body{\sigma}$~\cite{CaGK13}. The atom $\alpha$ is the {\em guard} of $\sigma$. In case there are more than one atoms that can serve as the guard of $\sigma$, then we fix the left-most such atom in $\body{\sigma}$ as the guard. We write $\guard{\sigma}$ for the guard of $\sigma$. The class of guarded TGDs, denoted $\class{G}$, is defined as the family of all possible finite sets of guarded single-head TGDs.

\medskip

\noindent
\paragraph{Stickiness.}
The goal of stickiness is to capture joins that are not expressible via guarded TGDs~\cite{CaGP12}. The key property is that variables occurring more than once in the body of a TGD should be inductively propagated (or ``stick'') to the head-atom as follows

\centerline{\includegraphics{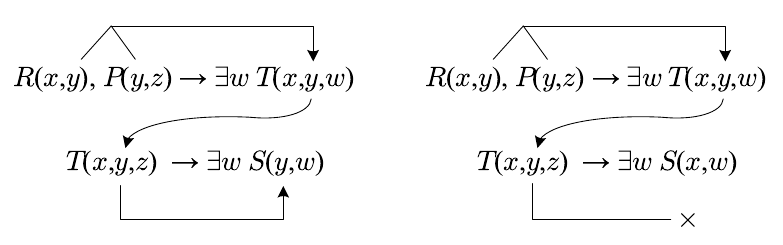}}

\noindent where the first set of TGDs is sticky, while the second is not.
The formal definition is based on an inductive procedure that marks the variables that may violate the above property. Roughly, the base step marks a body-variable that does occur in the head.
Then, the marking is inductively propagated from head to body as follows

\centerline{\includegraphics{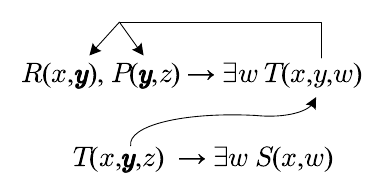}}

\noindent Stickiness requires every marked variable to appear only once in the body of a TGD. The formal definition follows.

Consider a set $\dep$ of single-head TGDs; we assume, w.l.o.g., that the TGDs in $\dep$ do not share variables. Let $\sigma \in \dep$ and $x$ a variable in $\body{\sigma}$. We inductively define when {\em $x$ is marked in $\dep$}:
\begin{enumerate}
\item if $x$ does not occur in $\head{\sigma}$, then $x$ is marked in $\dep$, and
\item assuming that $\head{\sigma} = R(\bar t)$ and $x \in \bar t$, if there is $\sigma' \in \dep$ with $R(\bar t')$ in its body, and each variable in $R(\bar t')$ at a position of $\pos{R(\bar t)}{x}$ is marked in $\dep$, then $x$ is marked in $\dep$.
\end{enumerate}
The set $\dep$ is {\em sticky} if there is no TGD with two occurrences of a variable that is marked in $\dep$. Let $\class{S}$ be the corresponding class.

\section{The Chase Procedure}\label{sec:chase-procedure}

The chase procedure accepts as input a database $D$ and a set $\dep$ of TGDs, and constructs an instance that contains $D$ and satisfies $\dep$. Central notions in this context are the notion of trigger, and the notion of trigger application (see, e.g.,~\cite{GrOn18}).

\begin{definition}\label{j-trigger}
A {\em trigger} for a set $\dep$ of TGDs on an instance $I$ is a pair $(\sigma,h)$, where $\sigma \in \dep$ and $h$ is a homomorphism from $\body{\sigma}$ to $I$. We call $(\sigma,h)$ {\em active} if there is no extension $h'$ of $h_{|\fr{\sigma}}$ such that $h'(\head{\sigma}) \in I$.
We denote by $\result{\sigma,h}$ the atom $v(\head{\sigma})$, where $v$ is a mapping from the variables of $\head{\sigma}$ to $\ins{N}$ defined as
\begin{eqnarray*}
v(x)\
=\ \left\{
\begin{array}{ll}
h(x) & \text{if } x \in \fr{\sigma},\\
&\\
c_{\sigma,h}^{x} & \text{otherwise.}
\end{array} \right.
\end{eqnarray*}
An {\em application} of $(\sigma,h)$ to $I$ returns the instance
\[
J\ =\ I \cup \{\result{\sigma,h}\},
\]
and such an application is denoted as $I \app{\sigma}{h} J$. \hfill\markfull
\end{definition}

In the definition of $\result{\sigma,h}$, each existentially quantified variable $x$ occurring in $\head{\sigma}$ is mapped by $v$ to a ``fresh'' null value of $\ins{N}$ whose name is uniquely determined by the trigger $(\sigma,h)$ and $x$ itself. Thus, given a trigger $(\sigma,h)$, we can unambiguously write down the atom $\result{\sigma,h}$.
In our analysis, it would be useful to be able to refer to the terms in $\result{\sigma,h}$ that have been propagated (not invented) during the application of $(\sigma,h)$. Formally, the {\em frontier} of $\result{\sigma,h}$, denoted $\fr{\result{\sigma,h}}$, are the terms of $\result{\sigma,h}$ that occur at the positions of $\bigcup_{x \in \fr{\sigma}} \pos{\head{\sigma}}{x}$.

\subsection{The Real Oblivious Chase}

Although this work is about the termination of the restricted chase, we use a variant of the oblivious chase, which we introduce below, as an auxiliary tool.
The oblivious chase of a database $D$ w.r.t.~a set $\dep$ of TGDs is essentially the $\subseteq$-minimal instance $I_{D,\dep}$ that contains $D$ and is closed under trigger applications, i.e., for every trigger $(\sigma,h)$ for $\dep$ on $I_{D,\dep}$, $\result{\sigma,h} \in I_{D,\dep}$.
It is well-known that it can be realized by starting from the database $D$, and applying (active or non-active) triggers, which have not been applied before, for the given set $\dep$ of TGDs on the instance constructed so far, and keep doing this until a fixpoint is reached.
It is also well-known that $I_{D,\dep}$ is unique since it does not depend on the order in which we apply the triggers; for more details see, e.g.,~\cite{CaPi19,GrOn18}.

Our intention is to use the (unique) oblivious chase of $D$ w.r.t.~$\dep$ as a predefined instance in which all the restricted chase derivations live (the formal definition of the restricted chase is given below). Thus, our task will be essentially to search in this instance for an infinite restricted chase derivation of $D$ w.r.t.~$\dep$.
To this end, we need the {\em parent relation} over the oblivious chase, which essentially gives us the atoms that were involved in the trigger application that produced a certain atom.
However, as the following simple example shows, this relation is, in general, not unique:

\begin{example}\label{prawdziwy}
Consider the set $\dep$ of TGDs consisting of:
\begin{eqnarray*}
\sigma_1\ :\ P(x,y)\ \ra\ R(x,y) &\quad& \sigma_3\ :\ R(x,y)\ \ra\ S(x)\\
\sigma_2\ :\ P(x,y)\ \ra\ S(x) &\quad& \sigma_4\ :\ S(x)\ \ra\ \exists y \, R(x,y).
\end{eqnarray*}
The oblivious chase of $D = \{P(a,b)\}$ w.r.t.~$\dep$ is the instance
\[
\{P(a,b), R(a,b), S(a), R(a,c)\},
\]
where $c$ is a null.
However, its atoms could have been produced in different ways: by applying $\sigma_1,\sigma_2,\sigma_4$, or by applying $\sigma_1,\sigma_3,\sigma_4$. In the first case, the parent of $S(a)$ is $P(a,b)$, while, in the second case, is the atom $R(a,b)$. Thus, although the oblivious chase is unique, its ambiguous which atom is the parent of $S(a)$. \hfill\markfull
\end{example}

As the above example illustrates, if we want to know in an unambiguous way who are the parents of a certain atom by simply inspecting the oblivious chase, we need to rely on a more refined structure. This is the purpose of the so-called real oblivious chase.

\begin{definition}\label{realoblivious}
The {\em real oblivious chase} of a database $D$ w.r.t.~a set $\dep$ of TGDs is the smallest labeled directed graph $\chase{D}{\dep} = \langle V, \crel{p}, \lambda, \tau \rangle$, where $\lambda$ and $\tau$ assign atoms over $\sch{\dep}$ and TGD-mapping pairs (including the empty pair $\bot$) to nodes, such that:
\begin{itemize}
\item For each atom $\alpha\in D$, there is a node $v\in V$ with $\lambda(v)=\alpha$, $\tau(v)=\bot$, and, for each $u \crel{p} w$, $w \neq v$.

\item For each TGD $\sigma \in \dep$, with $\body{\sigma} = \{\gamma_1,\ldots,\gamma_m\}$, for each mapping $h$ from the variables in $\body{\sigma}$ to $\ins{C} \cup \ins{N}$, and for each $(v_1,\ldots v_m) \in V^m$,
    if $h(\gamma_1)=\lambda(v_1),\ldots, h(\gamma_m)=\lambda(v_m)$, then there exists $v \in V$ such that $v_1 \crel{p} v, \ldots, v_m \crel{p} v$, $\lambda(v) = \result{\sigma,h}$, and $\tau(v) = (\sigma, h)$.
\end{itemize}
The elements of $\{\lambda(v) : v \in V\}$ are the atoms of $\chase{D}{\dep}$, and the relation $\crel{p}$ is the {\em parent relation} of $\chase{D}{\dep}$. \hfill\markfull
\end{definition}

Here is a simple example that illustrates the real oblivious chase:

\begin{example}\label{exa:real-oblivious}
Let $D$ and $\dep$ be the database and the set of TGDs from Example~\ref{prawdziwy}. Then, $\chase{D}{\dep}$ is the following directed graph (for clarity, the homomorphisms are omitted)

\smallskip

\centerline{\includegraphics{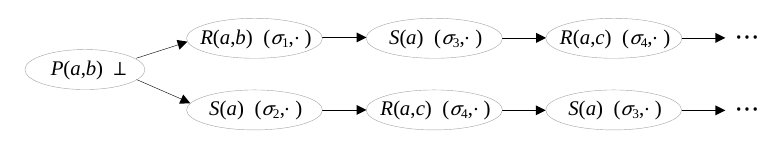}}

\smallskip

\noindent where $c$ is the null determined by the trigger $(\sigma_4,\{x \mapsto a\})$. \hfill\markfull
\end{example}



Strictly speaking, $\crel{p}$ is a relation over the node set of the real oblivious chase of $D$ w.r.t.~$\dep$. However, for notational convenience, from now on we will usually identify $\chase{D}{\dep}$ with its atoms, which clearly form a {\em multiset}, and we will see $\crel{p}$ as a relation over this multiset of atoms.
Let us also clarify that, although the real oblivious chase may generate several copies of the same atom, it will never produce an atom that is not generated by the oblivious chase, i.e., the oblivious chase coincides with the set consisting of the atoms of the real oblivious chase.
%
The advantage of the real oblivious chase is that it provides a unique multiset instance where all the different restricted chase derivations live, and at the same time we can unambiguously refer to the parents of a certain atom.

\medskip

\textsc{Remark.} The name ``real oblivious'' reflects the fact that an atom is generated and added to the instance under construction even if its already present. On the other hand, the oblivious chase, since it builds a set (not a multiset) of atoms, it implicitly checks, before applying a trigger $(\sigma,h)$, whether the atom $\result{\sigma,h}$ is already present. This somehow tells us that what we normally call oblivious chase is not completely oblivious, unlike the real oblivious one, which generates an atom no matter if it has been generated before.


\medskip

\noindent
\paragraph{Stop Relation.} Before we proceed further, let us introduce one more basic relation, in addition to the parent relation, which will be heavily used throughout the paper. This is the ``stop'' relation $\crel{s}$ over $\chase{D}{\dep}$. Intuitively, $\alpha \crel{s} \beta$ means that in the presence of $\alpha$ the atom $\beta$ is superfluous in the sense that the trigger $(\sigma,h)$ for $\dep$ on an instance that contains $\alpha$, with $\beta = \result{\sigma,h}$, is not active due to the presence of $\alpha$.
Formally, given two vertices $v,u$ of $\chase{D}{\dep}$ such that $\tau(u) = (\sigma,h)$, we say that $\lambda(v)$ {\em stops} $\lambda(u)$, denoted $\lambda(v) \crel{s} \lambda(u)$, if there exists a homomorphism $h'$ such that (i) $h'(\lambda(u)) = \lambda(v)$, and (ii) $h'(h(x)) = h(x)$ for every $x \in \fr{\sigma}$.
Notice that two copies of the same atom in the real oblivious chase always stop each other. It is also easy to verify that the following holds, which relates the notion of active trigger with $\crel{s}$:

\begin{fact}\label{fact:stop-relation}
Let $I \subseteq \chase{D}{\dep}$, and $(\sigma,h)$ a trigger for $\dep$ on $I$. Then, $(\sigma,h)$ is active iff there is no $\alpha \in I$ such that $\alpha \crel{s} \result{\sigma,h}$.
\end{fact}

\subsection{The Restricted Chase}

We now come to the main object of our study, that is, the restricted (a.k.a.~standard) chase. Similarly to the oblivious chase, the main idea of the restricted chase is, starting from a database $D$, to apply triggers for the given set $\dep$ of TGDs on the instance constructed so far, and keep doing this until a fixpoint is reached. However, unlike the oblivious chase, it only applies active triggers.
This is formalized as follows. Consider a database $D$ and a set $\dep$ of TGDs. We distinguish the two cases where the chase is terminating or not:

\begin{itemize}
\item A finite sequence $(I_i)_{0 \leq i \leq n}$ of instances, with $D = I_0$ and $n \geq 0$, is a {\em restricted chase derivation} of $D$ w.r.t.~$\dep$ if: for each $0 \leq i < n$, there is an active trigger $(\sigma,h)$ for $\dep$ on $I_i$ with $I_i \app{\sigma}{h} I_{i+1}$, and there is no active trigger $(\sigma,h)$ for $\dep$ on $I_n$.

\item An infinite sequence $(I_i)_{i \geq 0}$ of instances, with $D = I_0$, is a {\em restricted chase derivation} of $D$ w.r.t.~$\dep$ if, for each $i \geq 0$, there exists an active trigger $(\sigma,h)$ for $\dep$ on $I_i$ such that $I_i \app{\sigma}{h} I_{i+1}$. Moreover, $(I_i)_{i \geq 0}$ is called {\em fair} if, for each $i \geq 0$, and every active trigger $(\sigma,h)$ for $\dep$ on $I_i$, there exists $j > i$ such that $(\sigma,h)$ is a non-active trigger for $\dep$ on $I_j$. Notice that in a fair derivation all the active triggers will eventually be deactivated, which is not true for unfair derivations.
\end{itemize}

A restricted chase derivation is called {\em valid} if it is finite, or infinite and fair. Infinite but unfair restricted chase derivations are not valid since they do not serve the main purpose of the chase procedure, i.e., build an instance that satisfies the given set of TGDs.

\medskip

\noindent
\paragraph{\underline{Restricted Chase Termination Problem}}

\smallskip

\noindent It is well-known that 
even for simple databases and sets of TGDs, we may have infinite chase derivations. The key question is, given a set $\dep$ of TGDs, can we check whether, for every database $D$, every valid chase derivation of $D$ w.r.t.~$\dep$ is finite? Before formalizing this problem, let us recall a central class of TGDs:
\[
\begin{array}{rcl}
\rctaa &=& \left\{\dep\ :
\begin{array}{l}
\text{ for {\em every} database } D,\\
\text{ {\em every} valid restricted chase derivation}\\
\text{ of } D \text{ w.r.t.~} \dep \text{ is finite}.
\end{array} \right\}
\end{array}
\]
The superscript $\mathsf{res}$ in $\rctaa$ indicates that we concentrate on restricted chase derivations. The main problem tackled in this work is defined as follows, where $\class{C}$ is a class of TGDs:

\medskip

\begin{center}
\fbox{\begin{tabular}{ll}
{\small PROBLEM} : & $\rctaapr(\class{C})$
\\
{\small INPUT} : & A set $\dep \in \class{C}$ of TGDs.
\\
{\small QUESTION} : &  Is it the case that $\dep \in \rctaa$?
\end{tabular}}
\end{center}

\medskip

The above decision problem is, in general, undecidable. In fact, assuming that $\class{TGD}$ is the class of arbitrary (single-head) TGDs:

\begin{theorem}[\cite{GoMa14}]\label{the:undecidable}
$\rctaapr(\class{TGD})$ is undecidable, even if we focus on binary and ternary predicates.
\end{theorem}

But what about $\rctaapr(\class{G})$ and $\rctaapr(\class{S})$? These are non-trivial problems, and showing that are decidable is our main contribution. 
\section{The Fairness Theorem}\label{sec:fairness-lemma}

As one might expect, to establish the decidability of the problem $\rctaapr(\class{C})$, for $\class{C} \in \{\class{G},\class{S}\}$, we focus on its complement and show that, for a set $\dep \in \class{C}$ of TGDs, we can decide whether there is a database $D$ such that there exists a fair infinite chase derivation of $D$ w.r.t.~$\dep$. However, as observed in~\cite{CaGP15}, where the same problem for the simpler case of the oblivious chase is studied, one of the main difficulties is to ensure fairness.
For the oblivious chase, the existence of an (unfair) infinite chase derivation of $D$ w.r.t.~$\dep$ implies the existence of a fair one~\cite{CaGP15}. Does the same hold for the restricted chase? This is a non-trivial question that is affirmatively answered by the following result dubbed {\em Fairness Theorem}:

\begin{theorem}[\textbf{Fairness}]\label{the:fairness-lemma}
Consider a database $D$ and a set $\dep$ of single-head TGDs. If there exists an infinite restricted chase derivation of $D$ w.r.t.~$\dep$, then there exists a fair one.
\end{theorem}

Note that, to our surprise, the above theorem does {\em not} hold for multi-head TGDs, i.e., TGDs where the head is an arbitrary conjunction of atoms; a counterexample can be found in the appendix.\footnote{This has been also observed, independently of our work, in the recent paper~\cite{LMTU19} that concentrates on single-head linear TGDs.} This reveals the subtlety of the restricted chase, and explains that our decision to focus on single-head TGDs is not just for simplicity, but it is crucial for our results. The decidability status of $\rctaapr(\class{G})$ and $\rctaepr(\class{S})$ for multi-head TGDs are challenging open problems.
%

We now proceed to show the Fairness Theorem. By hypothesis, there exists an infinite restricted chase derivation $(I_i)_{i \geq 0}$ of $D$ w.r.t.~$\dep$. By exploiting $(I_i)_{i \geq 0}$, we are going to construct an infinite sequence $s_{D,\dep} = ((I_{i}^{j})_{i \geq 0})_{j \geq 0}$ of chase derivations of $D$ w.r.t.~$\dep$ such that $(I_{i}^{i})_{i \geq 0}$ is fair.
In other words, $s_{D,\dep}$ can be seen as an infinite matrix $M$, where the $j$-th row is the chase derivation $(I_{i}^{j})_{i \geq 0}$, while the diagonal is a fair chase derivation of $D$ w.r.t.~$\dep$.


\medskip

\noindent
\paragraph{The Diagonal Property.} We start by first exposing a crucial property that $s_{D,\dep}$ should enjoy:

\begin{definition}\label{def:diagonal-property}
A sequence $((J_{i}^{j})_{i \geq 0})_{j \geq 0}$ of infinite restricted chase derivations of $D$ w.r.t.~$\dep$ enjoys the {\em diagonal property} if, for each $i,j,k \geq 0$, $i \leq j$ and $i \leq k$ implies that $J_{i}^{j} = J_{i}^{k}$. \hfill\markfull
\end{definition}

In other words, by saying that the sequence $s_{D,\dep}$ enjoys the diagonal property, we simply mean that on the $i$-th column of the matrix $M$, all instances below the diagonal element $I_{i}^{i}$ coincide with $I_{i}^{i}$ (hence the name diagonal property). This allows us to show that the diagonal gives rise to an infinite chase derivation of $D$ w.r.t.~$\dep$:

\begin{lemma}\label{lem:diagonal-property}
Consider a sequence $((J_{i}^{j})_{i \geq 0})_{j \geq 0}$ of infinite restricted chase derivations of $D$ w.r.t.~$\dep$ that enjoys the diagonal property. Then, $(J_{i}^{i})_{i \geq 0}$ is a restricted chase derivation of $D$ w.r.t.~$\dep$
\end{lemma}


Of course, the diagonal property alone does not guarantee that the chase derivation $(I_{i}^{i})_{i \geq 0}$ is fair. Thus, our main task is to construct $s_{D,\dep} = ((I_{i}^{j})_{i \geq 0})_{j \geq 0}$ in such a way that (i) it enjoys the diagonal property, and (ii) $(I_{i}^{i})_{i \geq 0}$ is a fair chase derivation.


\medskip

\noindent
\paragraph{The Construction of $s_{D,\dep}$.} The high-level idea is as follows. The sequence $(I_{i}^{0})_{i \geq 0}$ is defined as $(I_i)_{i \geq 0}$, which exists by hypothesis. Now, our intention is to obtain $(I_{i}^{n+1})_{i \geq 0}$ from $(I_{i}^{n})_{i \geq 0}$. To this end, we carefully choose a large enough index $\ell > 0$ and (i) we define $(I_{i}^{n+1})_{0 \leq i \leq \ell}$ as $(I_{i}^{n})_{0 \leq i \leq \ell}$, i.e., by simply copying the first $\ell+1$ instances of $(I_{i}^{n})_{i \geq 0}$, (ii) we obtain $I_{\ell+1}^{n+1}$ from $I_{\ell}^{n+1} = I_{\ell}^{n}$ by deactivating one of the early active triggers due to which $(I_{i}^{n})_{i \geq 0}$ is not fair, and (iii) we obtain $(I_{i}^{n+1})_{i \geq \ell+2}$ by mimicking $(I_{i}^{n})_{i \geq \ell+1}$.
The formal construction of $s_{D,\dep}$ follows.

As said above, $(I_{i}^{0})_{i \geq 0}$ is defined as $(I_i)_{i \geq 0}$. Assume now that $(I_{i}^{n})_{i \geq 0}$ has been defined for some $n \geq 0$. We are going to define $(I_{i}^{n+1})_{i \geq 0}$. Let $m \geq 0$ be the smallest index such that there exists an active trigger $(\sigma,h)$ for $\dep$ on $I_{m}^{n}$ that remains active forever in $(I_{i}^{n})_{i \geq 0}$.
(Notice that if such an $m \geq 0$ does not exist, then $(I_{i}^{n})_{i \geq 0}$ is fair and we are done.) Assume that $I_{i+1}^{n}$ is obtained from $I_{i}^{n}$ via the trigger $(\sigma_i,h_i)$.
Let $A = \{i \geq 0 : \result{\sigma,h} \crel{s} \result{\sigma_i,h_i}\}$. By exploiting the properties of $\crel{s}$, it is not difficult to show that:

\begin{lemma}\label{lem:A-finite}
The set $A$ is finite.
\end{lemma}

%

Let $\ell$ be an integer greater than all the elements of $\{n,m\} \cup A$, which exists by Lemma~\ref{lem:A-finite}. We define:
\begin{eqnarray*}
I_{i}^{n+1}\
=\ \left\{
\begin{array}{ll}
I_{i}^{n} & 0 \leq i \leq \ell\\
&\\
I_{i-1}^{n} \cup \{\result{\sigma,h}\} & i > \ell
\end{array} \right.
\end{eqnarray*}

We can show the following; the proof is in the appendix:

\begin{lemma}\label{lem:correctness}
$(I_{i}^{n+1})_{i \geq 0}$ is a restricted chase derivation of $D$ w.r.t.~$\dep$.
\end{lemma}

\noindent
\paragraph{Finalizing the Proof.} Lemma~\ref{lem:correctness} implies that indeed $s_{D,\dep} = ((I_{i}^{j})_{i \geq 0})_{j \geq 0}$ is an infinite sequence of chase derivations of $D$ w.r.t.~$\dep$.
The fact that in the definition of $(I_{i}^{n+1})_{i \geq 0}$ above we choose the integer $\ell$ to be greater than $n$ ensures that $s_{D,\dep}$ enjoys the diagonal property. Therefore, by Lemma~\ref{lem:diagonal-property}, we conclude that $(I_{i}^{i})_{i \geq 0}$ is an infinite chase derivation of $D$ w.r.t.~$\dep$. Moreover, since there are only finitely many active triggers for $\dep$ on an instance $I_{i}^{j}$ since $I_{i}^{j}$ is finite, it follows from the construction of $s_{D,\dep}$ that $(I_{i}^{i})_{i \geq 0}$ is fair. Hence, $(I_{i}^{i})_{i \geq 0}$ is a fair infinite chase derivation of $D$ w.r.t.~$\dep$.

%

%
%
%
%
%
%
\section{Chase Termination \& Guardedness}\label{sec:guardedness}

We now concentrate on guarded TGDs, and show that:

\begin{theorem}\label{the:guarded-tgds}
$\rctaapr(\class{G})$ is decidable in elementary time.
\end{theorem}

By Theorem~\ref{the:fairness-lemma}, to establish the above result it suffices to show that, for a set $\dep \in \class{G}$ of TGDs, we can decide in elementary time whether there is a database $D$ such that there exists an infinite (possibly unfair) restricted chase derivation of $D$ w.r.t.~$\dep$.
To this end, we first characterize the existence of an infinite (possibly unfair) restricted chase derivation of $D$ w.r.t.~$\dep$ via the existence of an infinite subset of $\chase{D}{\dep}$, called chaseable, that enjoys certain properties.
We then show that the problem of deciding whether there is a database $D$ such that an infinite chaseable subset of $\chase{D}{\dep}$ exists can be reduced to the satisfiability problem of Monadic Second-Order Logic (MSOL) over infinite trees of bounded degree, which in turn implies that $\rctaapr(\class{G})$ is decidable.
%
At first glance, such a reduction looks unfeasible since the above statement talks about arbitrary databases $D$, and thus $\chase{D}{\dep}$ can be structurally very complex, i.e., not close to a tree.
Nevertheless, we can show that it suffices to concentrate on acyclic databases $D$, which in turn implies (due to the fact that we consider single-head guarded TGDs) that $\chase{D}{\dep}$ is acyclic.

\subsection{Non-Termination via Chaseable Sets}

We proceed to introduce the notion of chaseable set for a database $D$ and a set $\dep$ of TGDs. The key idea is to isolate certain properties of an infinite subset of $\chase{D}{\dep}$ that allow us to convert it into an infinite restricted chase derivation of $D$ w.r.t.~$\dep$.
To this end, we need the ``before'' relation $\crel{b}$ over $\chase{D}{\dep}$. Intuitively, $\alpha \crel{b} \beta$ means that, if the atoms $\alpha$ and $\beta$ have been generated by some restricted chase derivation $\delta$, then necessarily $\alpha$ has been generated before $\beta$; otherwise, $\delta$ is not a restricted chase derivation.
Given a sequence of instances $I_0,I_1,\ldots$, where each $I_i$ is a subset of $\chase{D}{\dep}$, there are essentially three reasons why it is not, or it cannot be converted (by merging some of the initial instances) into a restricted chase derivation of $D$ w.r.t.~$\dep$: there are atoms $\alpha \in I_i \setminus I_{i-1}$ and $\beta \in I_j \setminus I_{j-1}$ such that:
\begin{enumerate}
\item $\alpha \in D$, $\beta \not\in D$ and $j < i$, i.e., $\alpha$ is generated after $\beta$.

\item $\alpha \crel{p} \beta$ but $j < i$, i.e., the parent of $\beta$ is generated after $\beta$.

\item $\alpha \crel{s} \beta$ but $i < j$, i.e., $\beta$ is generated after $\alpha$, while $\alpha$ stops $\beta$.
\end{enumerate}
The goal of the relation $\crel{b}$ is to ensure that none of the above holds. Having the parent relation $\crel{p}$, and the stop relation $\crel{s}$ (together with Fact~\ref{fact:stop-relation}), it should be clear that the before relation $\crel{b}$ is
\[
\{\langle \alpha,\beta \rangle : \alpha \in D \text{ and } \beta\in \chase{D}{\dep} \setminus D\}\,\, \cup\ \crel{p}\ \cup\ \crel{s}^{-1},
\]
where $\crel{s}^{-1}$ refers to the inverse relation of $\crel{s}$. We write $\ccrel{b}$ for the transitive closure of $\crel{b}$.
The notion of chaseable set follows.



\begin{definition}\label{j-chaseable-def}
Consider a database $D$, and a set $\dep$ of TGDs. A set $A \subseteq \chase{D}{\dep}$ is called {\em chaseable} if the following hold:
\begin{enumerate}
\item For each  $\alpha\in A$, the set $\{\beta \in A : \beta \ccrel{b} \alpha\}$ is finite.

\item For each $\alpha \in A$ and $\beta \in \chase{D}{\dep}$, $\beta \crel{p} \alpha$ implies $\beta \in A$.

\item $\{\langle \alpha,\beta \rangle : \alpha,\beta \in A \text{ and } \alpha \crel{b} \beta\}$ is a directed acyclic graph, i.e., there are no cycles in the relation $\crel{b}$ over $A$. \hfill\markfull
\end{enumerate}
\end{definition}

The first condition states that, for each $\alpha \in A$, only finitely many atoms of $A$ should come before $\alpha$. The second condition says that the parent of an atom $\alpha \in A$ should be in $A$. Finally, the third condition states that, for every pair of distinct atoms $\alpha,\beta \in A$, either $\alpha$ should come before $\beta$, or $\beta$ should come before $\alpha$.
It is not difficult to show that indeed the existence of an infinite chaseable set characterizes the existence of an infinite restricted chase derivation.

\begin{theorem}\label{j-chaseable-lemma}
Consider a database $D$ and a set $\dep$ of TGDs. The following are equivalent:
\begin{enumerate}
\item There exists an infinite restricted chase derivation of $D$ w.r.t.~$\dep$.

\item There exists an infinite set $A \subseteq \chase{D}{\dep}$ that is chaseable.
\end{enumerate}
\end{theorem}

Let us clarify that Theorem~\ref{j-chaseable-lemma} holds for arbitrary, not necessarily guarded TGDs. The importance of guardedness is revealed in the next section, where we show that we can focus on acyclic databases.

\subsection{The Treeification Theorem}

We first need to recall the standard notion of acyclicity for instances. Intuitively, an instance $I$ is acyclic if its atoms can be rearranged in a tree $T$ in such a way that, for each term $t \in \adom{I}$, the set of atoms that mention $t$ induces a connected subtree of $T$.

\begin{definition}\label{def:acyclic-dbs}
A {\em join tree} of an instance $I$ is a pair $(T,\lambda)$, where $T = (V,E)$ is a tree, and $\lambda$ is the labeling function $V \ra I$, such that:
\begin{enumerate}
\item For each $\alpha \in I$, there exists $v \in V$ with $\lambda(v) = \alpha$.

\item For each term $t \in \adom{I}$, the set $\{v \in V : t \text{ occurs in } \lambda(v)\}$ induces a connected subtree of $T$.
\end{enumerate}
We say that $I$ is {\em acyclic} if it admits a joint tree. \hfill\markfull
\end{definition}

We then show the following result dubbed {\em Treeification Theorem}:


\begin{theorem}[\textbf{Treeification}]\label{the:treefication}
Let $\dep \in \class{G}$. If there exists a database $D$ such that there is an infinite restricted chase derivation of $D$ w.r.t.~$\dep$, then there is an acyclic database with the same property.
\end{theorem}

This is a rather involved result and its proof can be found in the appendix. In what follows, we give the high-level idea underlying the construction of the desired acyclic database.
By hypothesis, there exists an infinite restricted chase derivation $(I_i)_{i \geq 0}$ of some database $D$ w.r.t.~$\dep$.
From $\chase{D}{\dep} = (V, \crel{p}, \lambda, \tau)$ we can naturally obtain the {\em guard-parent} (resp., {\em side-parent}) relation $\crel{\mathit{gp}}$ (resp., $\crel{\mathit{sp}}$) over $V$ as the subrelation of $\crel{p}$ by keeping only the pairs of nodes $(v,u)$ where $v$ corresponds to the guard atom (resp., to a side atom, i.e., an atom other than the guard) of the TGD in $\tau(u)$.
Let $\crel{\mathit{gp}}^+$ be the transitive closure of $\crel{\mi{gp}}$.
Observe that, due to guardedness, $\chase{D}{\dep}$ can be seen as a forest w.r.t.~$\crel{\mi{gp}}$, where the nodes of $V$ labeled with atoms of $D$ are the roots of the trees, and all the other nodes are the non-root nodes.
As with $\crel{p}$, for convenience, we will usually see $\crel{\mathit{gp}}$ and $\crel{\mathit{sp}}$ as relations over the multiset consisting of the atoms of $\chase{D}{\dep}$.


Let $\mathcal{I} = \bigcup_{i \geq 0} I_i$. For an atom $\beta \in \mathcal{I}$, we define $\mathcal{I}_\beta$ as the set $\{\alpha \in \mathcal{I} : \beta \ccrel{\mi{gp}} \alpha\}$.
Since $D$ is finite, while $\mathcal{I}$ is infinite, we can conclude that there exists an atom $\alpha^\infty \in D$ such that the set  $\mathcal{I}_{\alpha^\infty}$ is infinite.
At this point, one may think that the desired acyclic database consists of the atom $\alpha^\infty$ together with the atoms of $D$ that can serve as its side atoms, i.e., the database
\[
\{\alpha^\infty\}\ \cup\ \{R(t_1,\ldots,t_n) \in D : t_1,\ldots,t_n \text{ occur in } \alpha^\infty\}.
\]
Unfortunately, as shown below, this is not the case:

\begin{example}\label{exa:acyclic-db-counterexample}
Assume that $\dep$ consists of the TGDs
\begin{eqnarray*}
\sigma_1 &:& S(x,y)\ \ra\ T(x)\\
\sigma_2 &:& R(x,y),T(y)\ \ra\ P(x,y)\\
\sigma_3 &:& P(x,y)\ \ra\ \exists z \, P(y,z).
\end{eqnarray*}
It is clear that there exists an infinite restricted chase derivation of $\{R(a,b),S(b,c)\}$ w.r.t.~$\dep$: first apply $\sigma_1$ and obtain $T(b)$, then apply $\sigma_2$ and obtain $P(a,b)$, and then apply $\sigma_3$ infinitely many times. Observe that the key atom $\alpha^\infty$ is $R(a,b)$. However, there is no infinite restricted chase derivation of $\{R(a,b)\}$ w.r.t~$\dep$. In fact, there are no active triggers for $\dep$ on $\{R(a,b)\}$ \hfill\markfull
\end{example}

As it can be seen from the above example, the reason why $\alpha^\infty$, together with its potential side atoms from $D$, do not give rise to an infinite restricted chase derivation is the need of what we call here remote side-parents.
In particular, referring to Example~\ref{exa:acyclic-db-counterexample}, we have an infinite restricted chase derivation of $\{R(a,b),S(b,c)\}$ w.r.t.~$\dep$ due to the atom $P(a,b)$, which has as a guard-parent the atom $\alpha^\infty = R(a,b)$, and as a side-parent the atom $T(b)$. However, $T(b)$ is not a database atom, but is obtained due to the database atom $S(b,c)$, which cannot serve as a side atom of $R(a,b)$. So, somehow, the atom $S(b,c)$ is a remote side-parent of $P(a,b)$. This situation can be formalized as follows.

\begin{definition}\label{def:remote-side-parent}
Consider two distinct atoms $\alpha,\beta \in D$, and two atoms $\alpha', \beta'\in \mathcal{I}$.
The tuple $\langle \alpha,\alpha',\beta,\beta' \rangle$ is a {\em remote-side-parent situation} if the following hold: $\alpha \ccrel{\mi{gp}} \alpha'$, $\beta \ccrel{\mi{gp}} \beta'$, and $\beta' \crel{\mi{sp}} \alpha'$. If this is the case, then we say that $\alpha$ {\em longs for} $\beta$. \hfill\markfull
\end{definition}

%
It is now not difficult to show that there exists a natural number $\ell_\infty$ such that, if $\langle \alpha^\infty,\alpha',\beta,\beta' \rangle$ is a remote-side-parent situation, then $\beta' \in I_{\ell_\infty}$.
In fact, if $\langle \alpha^\infty,\alpha',\beta,\beta' \rangle$ is a remote-side-parent situation, then, due to guardedness, all the terms occurring in $\beta'$ occur also in $\alpha^\infty$ and $\beta$. This implies that there are only finitely many pairs of atoms $(\beta,\beta')$, where $\beta \in D$ and $\beta' \in \mathcal{I}$, such that, for some $\alpha' \in \mathcal{I}$ with $\alpha^\infty \ccrel{\mi{gp}} \alpha'$, $\langle \alpha^\infty,\alpha',\beta,\beta' \rangle$ is a remote-side-parent situation. The latter implies the existence of $\ell_\infty$ claimed above, which is crucial in the construction of the desired acyclic database.
We can now give the intuition underlying this construction.

Our intention is to explicitly construct from $D$ a join tree $(T_{\mi{ac}},\lambda)$, where $T_{\mi{ac}} = (V,E)$, and the desired acyclic database $D_{\mi{ac}}$ will be the set of atoms $\{\lambda(v) : v \in V\}$.
Imagine $D$ as a directed multigraph: the atoms of $D$ are the vertices of this graph, while the edge-relation is ``longs for''. Now, $T_{\mi{ac}}$ is the set of all directed paths in this directed graph, starting from $\alpha^\infty$, of length at most $\ell_\infty$. There is a natural tree ordering on such a set of paths, and this is exactly the ordering $E$ of $T_{\mi{ac}}$. Every path is labelled with an isomorphic copy of the atom being its end-point, but in a particular way: if $x$ and $y$ are two vertices of $T_{\mi{ac}}$, with $(x,y) \in E$, which means that $x$ comes from some $\alpha \in D$ and $y$ comes from some $\beta \in D$ such that $\alpha$ longs for $\beta$, then, if $\alpha,\beta$ share a term, then $\lambda(x)$ and $\lambda(y)$ share the respective terms. Thanks to that, we are able to show that (the offspring of) $\lambda(y)$ can offer to (the offspring of) $\lambda(x)$ the same service in $D_{\mi{ac}}$ as (the offspring of) $\beta$ provides to (the offspring of) $\alpha$ in $D$.

\subsection{Deciding $\rctaapr(\class{G})$ via MSOL}\label{sec:msol-sentence}

By Theorems~\ref{the:fairness-lemma},\ref{j-chaseable-lemma} and~\ref{the:treefication}, given a set $\dep \in \class{G}$, deciding whether $\dep \not\in \rctaa$ is equivalent to the problem of checking whether there is an {\em acyclic} database $D$
such that an infinite chaseable subset of $\chase{D}{\dep}$ exists.
Our goal is to reduce the latter to the satisfiability problem of Monadic Second-Order Logic (MSOL) over infinite trees of bounded degree, which is decidable in $k$-\text{ExpTime}, where $k$ is the number of quantifier alternations.

We need to devise an MSOL sentence $\phi_\dep$ such that the following statements are equivalent:
\begin{enumerate}
\item There is an acyclic database $D$ such that an infinite chaseable subset of $\chase{D}{\dep}$ exists.

\item $\phi_\dep$ is satisfiable over $\Lambda_\dep$-labeled infinite trees of bounded degree, where $\Lambda_\dep$ is a finite alphabet that depends on $\dep$.
\end{enumerate}

\medskip
\noindent \paragraph{\underline{Abstract Join Trees}}

\smallskip

\noindent
Whenever $\dep$ consists of single-head guarded TGDs and $D$ is acyclic, then $\chase{D}{\dep}$ is also acyclic, which means that it has a join tree~\cite{BaGP16}. Thus, one may think that this join tree is a natural candidate for a tree that our MSOL formula could talk about.
But this is not going to work for the simple reason that the codomain of the labeling function $\lambda$ of such a join tree is infinite. We therefore need to invent something similar to a join tree, i.e., a structure that encodes an instance as a labeled tree, but much more parsimonious with respect to the labeling function. This is precisely the purpose of what we call abstract join trees.

We define the finite alphabet $\Lambda_\dep$ as a set of triples
\[
\Lambda_\dep\ =\ \sch{\dep}\ \times\ (\{F\} \cup \dep)\ \times\ \mathcal{EQ}_\dep
\]
that encode atoms. Here is the idea underlying this encoding:
\begin{itemize}
\item The first element of each triple is a predicate; it simply tells us the predicate of the atom in question.

\item Concerning the second element, $F$ stands for ``database fact'', and indicates that the encoded atom is an atom from the original database. If an atom does not come from the database,
    then the second element of the triple tells us which TGD of $\dep$ was used to generate it.

\item Concerning the third element, we define ${\mathcal EQ}_\dep$ as the set of all equivalence relations on $\{f,m\} \times \{1,2,\ldots \ar{\dep}\}$, where $f$ and $m$ stand for ``father'' and ``me''. The idea is that, for example, the pair $[[m,i],[m,j]]$ says that the encoded atom has the same term at its $i$-th and $j$-th position, while the pair $[[m,i],[f,j]]$ says that the term at the $i$-th position in the atom in question equals to the term at the $j$-th position of its father (with respect to the tree relation).
\end{itemize}
In what follows, for brevity, given a node $v$ that is labeled by the triple $\langle x,y,z \rangle$, we write $\pred{v}$ for the predicate $x$, $\origin{v}$ for $y$, i.e., the origin of the encoded atom, and $\eq{v}$ for the equivalence relation $z$. Recall also that, for an atom $\alpha$, we write $\alpha[i]$ for its $i$-th term. We are now ready to formally define abstract join trees.

\begin{definition}\label{j-guarded-tree}
An {\em abstract join tree } for a set $\dep \in \class{G}$ of TGDs is a (finite or infinite)
$\Lambda_\dep$-labeled rooted tree $T = \langle V, \Yleft \rangle$, of degree at most $\max\{\ar{\dep},|\dep|\}$, that satisfies the following conditions:
\begin{enumerate}
\item The set $\{x \in V : \origin{x} = F\}$ is non-empty but finite.

\item If $x \Yleft y$ and $\origin{y} = F$, then $\origin{x} = F$.

\item If $x\Yleft y$ and $\origin{y} = \sigma$, then $\pred{x}$ is the predicate of $\guard{\sigma}$ and $\pred{y}$ is the predicate of $\head{\sigma}$.

\item If $x\Yleft y$, then $[[m,i],[m,j]] \in \eq{x}$ iff $[[f,i],[f,j]] \in \eq{y}$.

\item If $x\Yleft y$ and $\origin{y} = \sigma$, for some $\sigma \in \dep$ with $\alpha = \guard{\sigma}$ and $\beta = \head{\sigma}$, then:
\begin{enumerate}
\item $\alpha[i] = \beta[j]$ implies $[[f,i],[m,j]] \in \eq{y}$,

\item $\alpha[i] = \alpha[j]$ implies $[[f,i],[f,j]]\in \eq{y}$, and

\item if $\beta[j]$ is an existentially quantified variable in $\sigma$, then $[[m,i],[m,j]] \in \eq{y}$ iff $\beta[j] = \beta[i]$. \hfill\markfull
\end{enumerate}
\end{enumerate}
\end{definition}

We now need to explain how an abstract join tree is transformed into an instance. Consider an abstract join tree $T = \langle V, \Yleft \rangle$ for a set $\dep \in \class{G}$. We define $\mathsf{Eq}_T \subseteq (V \times \{1,\ldots,\ar{\dep}\}) \times (V \times \{1,\ldots,\ar{\dep}\})$ as the smallest equivalence relation such that, for every edge $x \Yleft y$, if $[[m,i],[m,j]] \in \eq{x}$ then $[[x,i],[x,j]]\in  \mathsf{Eq}_T$, and if $[[f,i],[m,j]]\in \eq{y}$ then $[[x,i],[y,j]] \in \mathsf{Eq}_T$. The instance $\Delta(T)$ is defined as the set of atoms $\{\delta(x) : x \in V\}$, where (i) for each $x \in V$, the predicate of $\delta(x)$ is $\pred{x}$, and (ii) for each $x,y \in V$, $\delta(x)[i] = \delta(y)[j]$ iff $[[x,i],[y,j]]\in \mathsf{Eq}_T$.

For an abstract join tree $T$, we write $T_{|F}$ for the restriction of $T$ to its nodes that are labeled with a label  of the form $\langle \cdot,F,\cdot \rangle$. Then, it is not hard to see that the following equivalence holds:

\begin{lemma}\label{lem:acyclic-to-treelike}
For a set $\dep \in \class{G}$, and an acyclic database $D$, the following are equivalent:
\begin{enumerate}
\item There exists an infinite chaseable subset of $\chase{D}{\dep}$.

\item There exists an abstract join tree $T$ such that
$\Delta(T_{|F})$ and $D$ are isomorphic, and $\Delta(T)$ is an infinite and chaseable subset of $\chase{\Delta(T_{|F})}{\dep}$.
\end{enumerate}
\end{lemma}

Therefore, in order to prove Theorem \ref{the:guarded-tgds}, it is now enough to construct, for a given $\dep$, an MSOL formula $\phi_\dep$ such that, for any abstract join tree $T$, it holds that: $T\models \phi_\dep$ iff ($\star$) $\Delta(T)$ is an infinite and chaseable subset of $\chase{\Delta(T_{|F})}{\dep}$.

\medskip

\noindent \paragraph{\underline{Chaseable Abstract Join Trees}}

\smallskip

\noindent
Our MSOL formula $\phi_\dep$ (under construction) is supposed to
express some property of $\Delta(T)$, for a given  abstract join tree
$T$, namely the property ($\star$).
But, it does not see $\Delta(T)$. It can only talk about $T$. Moreover, talking about nodes, let say $x$ and $y$, of $T$, and relations between these nodes, it must actually mean the atoms $\delta(x)$ and $\delta(y)$, and relations among those atoms.
Thus, it will be convenient to have a language to talk about the nodes of
$T$ but to mean atoms of $\Delta(T)$. We now define such a language,
allowing ourselves to slightly abuse the notation and overload the symbols $\crel{p}$, $\crel{s}$ and $\crel{b}$.

First, we need a way to say that an atom $\alpha \in \Delta(T)$ can act as a side atom for some other atom $\beta \in \Delta(T)$, and also to specify which terms of $\beta$ occur in $\alpha$ and at which positions. This can be achieved via the notion of sideatom type.
A {\em sideatom type} $\pi$ (w.r.t~$\dep$) is a triple $\langle P, m, \xi \rangle$, where $P/n \in \sch{\dep}$, $m \leq \ar{\dep}$ is a natural number, called the arity of $\pi$, and $\xi : [n] \ra [m]$.
Given two atoms $\alpha$ and $\beta$, we say that $\alpha$ is a {\em $\pi$-sideatom} of $\beta$, denoted $\alpha \subseteq_\pi \beta$, if the predicate of $\alpha$ is $P$, the predicate of $\beta$ has arity $m$, and $\alpha[i] = \beta(\xi(i))$ for each $i \in [n]$.
For example, the atom $\alpha = P(a,b,c)$ is a $\pi$-sideatom of $\beta = R(a,d,c,b)$ with $\pi = \langle P, 4, \{1 \mapsto 1, 2 \mapsto 4, 3 \mapsto 3\} \rangle$.
In what follows, it would be convenient to represent a guarded body by directly using sideatom types. More precisely, for a guarded TGD $\sigma$, where $\body{\sigma} = \gamma,\gamma_1,\ldots,\gamma_m$ with $\gamma = \guard{\sigma}$, its body can be represented in the obvious way as $\gamma,\pi_1,\ldots \pi_m$, where $\pi_1,\ldots,\pi_m$ are sideatom types of arity equal to the arity of the predicate of $\gamma$.

\medskip
\noindent \paragraph{Parent Relation.} Consider an abstract join tree $T = \langle V, \Yleft \rangle$ for a set $\dep$ of guarded TGDs.
The parent relation is defined as follows:
\begin{itemize}
\item Given an edge $x \Yleft y$ in $T$, with $\origin{y} = \sigma$, for some $\sigma \in \dep$ such that $\body{\sigma} = \gamma,\pi_1,\ldots,\pi_m$, we say that a node $z \in V$ is a {\em $\pi_i$-side-parent} of $y$, denoted $z \crel{sp}^{\pi_i} y$, if $\delta(z)\subseteq_{\pi_i} \delta(x)$.

\item Given two nodes $x,y \in V$, $x$ is a {\em parent} of $y$, denoted $x \crel{p} y$, if $x \Yleft y$, or $x \crel{sp}^{\pi} y$ for some sideatom type $\pi$.
\end{itemize}

\smallskip

\noindent \paragraph{Stop Relation.} Consider two nodes $x,y \in V$, with $\origin{y} = \sigma$. We say that $x$ {\em stops} y, denoted $x \crel{s} y$, if there exists a homomorphism $h$ such that $h(\delta(y)) = \delta(x)$, and, for each term $t$ in $\delta(y)$ that occurs at a position of $\bigcup_{x \in \fr{\sigma}} \pos{\head{\sigma}}{x}$, $h(t) = t$.

\medskip
\noindent \paragraph{Before Relation.} The before relation is defined as expected:
\[
\crel{b}\ =\ \{\langle x,y \rangle : x,y \in V, \origin{x} = F \text{ and } \origin{y} \neq F\}\,\, \cup\ \crel{p}\ \cup\ \crel{s}^{-1}.
\]
We write $\crel{b}^{+}$ for the transitive closure of $\crel{b}$.

\medskip

Having the above relations in place, we can now define the notion of chaseable abstract join tree:

\begin{definition}\label{def:chaseable-guarded-tree}
Consider an abstract join tree $T = \langle V, \Yleft \rangle$ for a set $\dep \in \class{G}$. We say that $T$ is {\em chaseable} if the following hold:
\begin{enumerate}
\item For each  $x \in V$, the set $\{y \in V : y \ccrel{b} x\}$ is finite.

\item For each edge $x \Yleft y $, where $\origin{y} = \sigma$ for some $\sigma\in \dep $ with $\body{\sigma} = \gamma,\pi_1,\ldots,\pi_m$, there exists $z \in V$ such that $z \crel{sp}^{\pi_i} y$ for each $i \in [m]$.

\item $\{\langle x,y \rangle : x,y \in V \text{ and } x \crel{b} y\}$ is a directed acyclic graph, i.e., there are no cycles in the relation $\crel{b}$ over $V$. \hfill\markfull
\end{enumerate}
\end{definition}

It follows, by construction, that for a set $\dep$ of guarded TGDs, and an abstract join tree $T$ for $\dep$, $\Delta(T)$ is an infinite chaseable subset of $\chase{\Delta(T_{|F})}{\dep}$ iff there exists an infinite chaseable abstract join tree $\hat{T}$ for $\dep$ such that $\Delta(T_{|F})$ is isomorphic to $\Delta(\hat{T}_{|F})$. Then:

%


\begin{lemma}\label{lem:chaseable-guarded-tree}
For a set $\dep \in \class{G}$, the following are equivalent:
\begin{enumerate}
\item There exists an abstract join tree $T$ for $\dep$ such that $\Delta(T)$ is an infinite chaseable subset of $\chase{\Delta(T_{|F})}{\dep}$.

\item There exists an infinite chaseable abstract join tree for $\dep$.
\end{enumerate}
\end{lemma}

\smallskip

\noindent \paragraph{\underline{Chaseable Abstract Join Trees are MSOL-definable}}

\smallskip

\noindent
The last task is to show the following:

\begin{lemma}\label{lem:chaseable-guarded-trees-msol}
Consider a set $\dep \in \class{G}$. There is an MSOL sentence $\phi_\dep$ such that, for a $\Lambda_\dep$-labeled tree $T$ of degree at most $\max \{\ar{\dep},|\dep|\}$, it holds that $T \models \phi_\dep$ iff $T$ is a chaseable abstract join tree for $\dep$.
\end{lemma}

The sentence $\phi_\dep$ has to check whether a tree is an abstract join tree, and also whether the three conditions of Definition~\ref{def:chaseable-guarded-tree} are satisfied.
Since, given an abstract join tree $T = \langle V, \Yleft \rangle$, for each term $t$ in $\Delta(T)$, $\{x \in V : t \text{ occurs in } \delta(x)\}$ induces a connected subtree of $T$, it should be evident that indeed the above conditions can be checked via an MSOL sentence.
More details concerning the MSOL sentence $\phi_\dep$ can be found in the appendix.

Having Lemmas~\ref{lem:acyclic-to-treelike},~\ref{lem:chaseable-guarded-tree} and~\ref{lem:chaseable-guarded-trees-msol}, we get that $\rctaapr(\class{G})$ is decidable in elementary time, and Theorem~\ref{the:guarded-tgds} follows. 
\section{Chase Termination \& Stickiness}\label{sec:stickiness}

We now concentrate on sticky sets of TGDs, and show that:

\begin{theorem}\label{the:sticky-tgds}
$\rctaapr(\class{S})$ is decidable in elementary time.
\end{theorem}

As in the case of guarded TGDs, to establish the above result we are going to show that the complement of $\rctaapr(\class{S})$ is decidable in elementary time.
%
%
In fact, our ultimate goal is to reduce the complement of $\rctaapr(\class{S})$ to the emptiness problem of deterministic B\"{u}chi automata, which is feasible in linear time in the size of the automaton.
To this end, given a set $\dep \in \class{S}$, we characterize the existence of a database $D$ such that there is an infinite restricted chase derivation of $D$ w.r.t.~$\dep$ via the existence of a finitary caterpillar for~$\dep$. The latter is essentially an infinite ``path-like'' restricted chase derivation of some database w.r.t.~$\dep$, and, as we shall see, its existence can be checked via a deterministic B\"{u}chi automaton.

\subsection{Non-Termination via Caterpillars}

To formally introduce the notion of finitary caterpillar, we first need the notion of proto-caterpillar.

\begin{definition}\label{proto-caterpillar}
Consider a set $\dep$ of TGDs. A {\em proto-caterpillar for $\dep$} is a tuple $\diamondsuit = (L^\diamondsuit,B^\diamondsuit,T^\diamondsuit,G^\diamondsuit)$,
where:
\begin{itemize}
\item $L^\diamondsuit$ is a (possibly infinite) instance over $\sch{\dep}$, the {\em legs} of $\diamondsuit$,

\item $B^\diamondsuit = (\alpha_i^\diamondsuit)_{i \geq 0}$ is a sequence of atoms over $\sch{\dep}$ with constants and nulls (no variables), the {\em body} of $\diamondsuit$,\footnote{By abuse of notation, we may sometimes treat $B^\diamondsuit$ as the set of atoms $\{\alpha_i^\diamondsuit\}_{i \geq 0}$.}

\item $T^\diamondsuit = (\sigma_i^\diamondsuit,h_i^\diamondsuit)_{i > 0}$ is a sequence of TGD-mapping pairs where $h_i^\diamondsuit$ maps the variables in $\body{\sigma_i^\diamondsuit}$ to $\ins{C} \cup \ins{N}$, and

\item $G^\diamondsuit = (\gamma_i^\diamondsuit)_{i > 0}$ is a sequence of atoms with $\gamma_i^\diamondsuit \in \body{\sigma_i^\diamondsuit}$,
\end{itemize}
such that, for each $i \geq 0$, the following holds:
\begin{itemize}
\item[(1)]
$(\sigma_{i+1}^\diamondsuit,h_{i+1}^\diamondsuit)$ is a trigger for $\dep$ on $L^\diamondsuit \cup \{\alpha_i^\diamondsuit\}$;

\item[(2)]
$\alpha_i^\diamondsuit = h_{i+1}^\diamondsuit(\gamma_{i+1}^\diamondsuit)$;

\item[(3)]
$\alpha_{i+1}^\diamondsuit = \result{\sigma_{i+1}^\diamondsuit,h_{i+1}^\diamondsuit}$. \hfill\markfull
\end{itemize}
\end{definition}

It should not be difficult to see that a proto-caterpillar $\diamondsuit$ for $\dep$ as above encodes a ``path-like'' oblivious chase derivation (modulo repetition of triggers) of the (possibly infinite) instance $L^\diamondsuit \cup \{\alpha_0^\diamondsuit\}$ w.r.t.~$\dep$.
Indeed, each atom $\alpha_i^\diamondsuit$, for $i > 0$, of the sequence $B^\diamondsuit$ can be derived from $L^\diamondsuit \cup \{\alpha_{i-1}^\diamondsuit\}$, i.e., the previous atom on the sequence and atoms of $L^\diamondsuit$, via the trigger $(\sigma_i^\diamondsuit,h_i^\diamondsuit)$.
In other words, the infinite sequence of instances $(I_i)_{i \geq 0}$, where $I_0 = L^\diamondsuit \cup \{\alpha_0^\diamondsuit\}$ and, for $i > 0$, $I_i = I_{i-1} \cup \{\alpha_i^\diamondsuit\}$, is an oblivious chase derivation (modulo repetition of triggers) of $I_0$ w.r.t.~$\dep$.
But, even if we remove the repeated triggers, there is no guarantee that it is a restricted chase derivation for the following two reasons: an atom from $L^\diamondsuit$ may stop an atom $\alpha_i^\diamondsuit$ for $i > 0$, or an atom $\alpha_i^\diamondsuit$ may stop an atom $\alpha_j^\diamondsuit$ for $j > i$.
This brings us to the notion of caterpillar, which is essentially a proto-caterpillar with the guarantee that the above two cases are excluded.

\begin{definition}\label{caterpillar}
Consider a set $\dep$ of TGDs. A {\em caterpillar for $\dep$} is a proto-caterpillar
$ \diamondsuit = (L^\diamondsuit, (\alpha_i^\diamondsuit)_{i \geq 0}, \cdot, \cdot)$ for $\dep$ such that:
\begin{enumerate}
\item for each $\beta \in L^\diamondsuit$ and $i > 0$, $\beta \not\crel{s} \alpha_i^\diamondsuit$, and

\item for each $0 \leq i < j$, $\alpha_i^\diamondsuit \not\crel{s} \alpha_j^\diamondsuit$. \hfill\markfull
\end{enumerate}
\end{definition}

It is an easy task to verify that a caterpillar $\diamondsuit$ for $\dep$ as above encodes a ``path-like'' restricted chase derivation of the (possibly infinite) instance $L^\diamondsuit \cup \{\alpha_0^\diamondsuit\}$ w.r.t.~$\dep$.
However, it should not be forgotten that we are interested on finite databases. This brings us to the central notion of finitary caterpillar.

\begin{definition}\label{finitary-caterpillar}
Consider a set $\dep$ of TGDs. A {\em finitary caterpillar for $\dep$} is a caterpillar $(L^\diamondsuit,\cdot,\cdot,\cdot)$ for $\dep$ such that $L^\diamondsuit$ is finite. \hfill\markfull
\end{definition}



%

Our goal is to characterize the existence of a database that gives rise to an infinite restricted chase derivation w.r.t.~a sticky set $\dep$ of TGDs via the existence of a finitary caterpillar for it.

\begin{theorem}\label{the:finitary-caterpillar}
Let $\dep \in \class{S}$. The following are equivalent:
\begin{enumerate}
\item There exists a database $D$ such that there is an infinite restricted chase derivation of $D$ w.r.t.~$\dep$.

\item There exists a finitary caterpillar for $\dep$.
\end{enumerate}
\end{theorem}

The fact that $(2) \Rightarrow (1)$ follows by definition, and holds
for every set of TGDs, not necessarily sticky. The interesting direction is $(1) \Rightarrow (2)$, which relies on stickiness.
%
To this end, we are going to introduce refined variants of caterpillars, which will eventually lead to finitary caterpillars. In particular, we are going to introduce the notions of {\em (uniformly) connected caterpillar}, and {\em free caterpillar}, and establish the following chain of implications:
\begin{itemize}
\item[] there exists a database $D$ such that there is an infinite restricted chase derivation of $D$ w.r.t.~$\dep$

\item[$\Rightarrow$] there exists a free connected caterpillar for $\dep$

\item[$\Rightarrow$] there exists a free uniformly connected caterpillar for $\dep$

\item[$\Rightarrow$] there exists a finitary caterpillar for $\dep$,
\end{itemize}
which shows that indeed $(1) \Rightarrow (2)$. In the rest of the section, let $\dep$ be a sticky set of single-head TGDs. For brevity, we will usually say (proto-)caterpillar meaning (proto-)caterpillar for $\dep$.

\medskip

\noindent
\paragraph{\underline{Variants of Caterpillars}}

\smallskip


%

\noindent We first need some auxiliary terminology.
Let $\alpha \in \chase{D}{\dep}$, for some database $D$, and assume that $\alpha = \result{\sigma,h}$.
Let $\gamma$ be an atom of $\body{\sigma}$, which means that $h(\gamma) \in \chase{D}{\dep}$ with $h(\gamma) \crel{p} \alpha$.
We say that the $i$-th position of $h(\gamma)$ and the $j$-th position of $\alpha$ are related, denoted as $(h(\gamma),i) \simeq (\alpha,j)$, if $\gamma[i] = \head{\sigma}[j]$.
Moreover, the $i$-th and $j$-th positions of $\alpha$ are related, written as $(\alpha,i) \simeq (\alpha,j)$, if $\head{\sigma}[i] = \head{\sigma}[j]$.
Now, for an instance ${\mathcal I} \subseteq \chase{D}{\dep}$, let $\Pi(\mathcal I) = \{(R(\bar t),i) : R(\bar t) \in {\mathcal I} \text{ and } 1 \leq i \leq \ar{R}\}$.
%
We denote by $\simeq^{*}_{\mathcal I}$ the smallest equivalence relation that contains $(\Pi(\mathcal I))^2 \cap (\simeq)$.
Intuitively, $(\alpha,i) \simeq^{*}_{\mathcal I} (\beta,j)$, for some atoms $\alpha,\beta \in \chase{D}{\dep}$, means that the terms $\alpha[i]$ and $\beta[j]$ are provably equal via a proof that uses only atoms of $\mathcal I$. Notice that $(\alpha,i) \simeq_{\mathcal I}^{*} (\beta, j)$ implies $\alpha[i] = \beta[j]$, but the opposite implication is not always true.

We also need the notion of the ``birth atom'' of a null value. Consider a parent-closed instance ${\mathcal I} \subseteq \chase{D}{\dep}$,
and let $c \in \adom{\mathcal I}$ be a null. We write $\beta^B(c)$ (which reads ``the birth atom of $c$'') for the atom of $\mathcal I$ such that: (1) $c$ occurs in $\beta^B(c)$, and (2) for every $\alpha \in \mathcal I$ with $\alpha \crel{p} \beta^B(c)$, $c$ does not occur in $\alpha$.
It is clear that there is only one birth atom of $c$. Notice also that for an atom $\beta \in \mathcal I$ such that $\beta[j] = c$, it holds that $\beta$ is the birth atom of $c$ iff for each parent $\alpha$ of $\beta$ and each position $i$ of $\alpha$, $(\alpha,i) \not\simeq (\beta,j)$.

We finally need to introduce the notion of immortal position, which relies on the marking procedure used in the definition of stickiness; see Section~\ref{sec:preliminaries}. Let $\alpha \in \chase{D}{\dep}$, for some database $D$, with $\alpha = \result{\sigma,h}$. The $i$-th position of $\alpha$ is {\em immortal} (w.r.t.~$\dep$) if the variable at the $i$-th position of $\head{\sigma}$ is {\em not} marked in $\dep$. The name ``immortal'' reflects the fact that $\alpha[i]$ will be propagated forever, i.e., for every $\beta$ such that $\alpha \crel{p} \beta$, $\alpha[i] \in \fr{\beta}$.

We are now ready to introduce the first variant of caterpillars.

\begin{definition}\label{caterpillar-c}
A caterpillar (or proto-caterpillar) $\diamondsuit = (\cdot, B^\diamondsuit, \cdot, \cdot)$, where $B^\diamondsuit = (\alpha_i^\diamondsuit)_{i \geq 0}$, is {\em connected} if there exist an infinite sequence $(\ccc_i)_{i \geq 0}$ of terms, called the {\em relay terms} of $\diamondsuit$, an infinite sequence $(b_i)_{i > 0}$ of integers with $b_1 < b_2 < b_3 < \cdots$, called the {\em pass-on points} of $\diamondsuit$, and infinite sequences $(p_i)_{i > 0}$ and $(m_i)_{i \geq 0}$ ($p$ for ``parent'' and $m$ for ``me'') of integers from $[\ar{\dep}]$, such that, for each $k \geq 0$:
\begin{enumerate}
\item $\ccc_0$ occurs in $\alpha_0^\diamondsuit$;

\item $\alpha_{b_k}^\diamondsuit = \beta^B(\ccc_k)$;

\item $\ccc_k = \alpha_{b_k}^\diamondsuit[m_k]$ and $\left(\alpha_{b_k}^\diamondsuit, m_k\right)\simeq_{B^\diamondsuit}^{*} \left(\alpha_{b_{k+1}}^\diamondsuit,p_{k+1}\right)$;

\item $\alpha_{j}^\diamondsuit[i] = \ccc_k$, for $i > 0$, $j \geq 0$, implies $\left(\alpha_{j}^\diamondsuit,i\right)$ is not immortal. \hfill\markfull
\end{enumerate}
\end{definition}

The above definition is indeed a bit technical. Intuitively, we can imagine the sequence of terms $\ccc_0, \ccc_1, \ccc_2, \ldots$ as an infinite relay race, where the $\ccc_i$'s are mortal runners, and their birth atoms are the batton passing points. In other words, connectedness ensures the continuous propagation of a new null in the underlying ``path-like'' chase derivation.
%
%
However, the distance between two consecutive pass-on points can be arbitrarily large, i.e., there is no uniform bound. As we shall see later, having such a uniform bound is crucial for going from connected caterpillars to finitary caterpillars. This brings us to the next refined variant of caterpillars.

\begin{definition}\label{caterpillar-uc}
A caterpillar $\diamondsuit$ is {\em uniformly connected} if it is connected and, with $(b_i)_{i > 0}$ being its pass-on points, there exists an integer $d \geq 0$ such that, for each $k \geq 0$, $b_{k+1} - b_{k} < d$. \hfill\markfull
\end{definition}

Let us now introduce the last variant of caterpillars that we need, namely free caterpillars.
%
%
Recall that $(\alpha,i) \simeq^{*}_{L^\diamondsuit \cup B^\diamondsuit} (\beta,j)$, for $\alpha,\beta \in \chase{L^\diamondsuit \cup B^\diamondsuit}{\dep}$, means that $\alpha[i]$ and $\beta[j]$ are provably equal via a proof that uses only atoms of $L^\diamondsuit \cup B^\diamondsuit$. It would be very useful to ensure that also the other direction holds.

\begin{definition}\label{caterpillar-f}
A (proto-)caterpillar $\diamondsuit = (L^\diamondsuit,B^\diamondsuit,\cdot,\cdot)$ is {\em free} if, for each $(\alpha,i), (\beta,j) \in \Pi(L^\diamondsuit \cup B^\diamondsuit)$, $\alpha[i] = \beta[j]$ iff $(\alpha,i) \simeq_{L^\diamondsuit \cup B^\diamondsuit}^* (\beta, j)$. \hfill\markfull
\end{definition}

%
%

%
%
%


\subsection{Implication 1: Extracting a Free Connected Caterpillar}

We are now ready to establish the chain of implications discussed above, immediately after Theorem~\ref{the:finitary-caterpillar}. We first focus on the first implication that states the following:
if there exists a database $D$ such that there is an infinite restricted chase derivation of $D$ w.r.t.~$\dep$, then there exists a free connected caterpillar for $\dep$.
Suppose $(I_i)_{i \geq 0}$ is an infinite restricted chase derivation of some database $D$ w.r.t.~$\dep$, and let ${\mathcal I} = \bigcup_{i \geq 0} I_i$. We are going to extract from $(I_i)_{i \geq 0}$ a free connected caterpillar.
The construction proceeds in three steps:
\begin{enumerate}
\item First, we are going to construct a proto-caterpillar $\clubsuit$.
\item Then, we will convert $\clubsuit$ into a connected proto-caterpillar $\spadesuit$.
\item Finally, from $\spadesuit$ we will get a free connected caterpillar $\heartsuit$.
\end{enumerate}

\smallskip

\noindent
\paragraph{\underline{Step 1: Construct a Proto-Caterpillar}}

\smallskip

\noindent We proceed to extract from $(I_i)_{i \geq 0}$ a proto-caterpillar $\clubsuit$ that is ``almost connected''.
Given a term (constant or null) $c$ and a null $c'$, both in $\adom{\mathcal I}$, we say that $c$ is a {\em parent term} of $c'$ (w.r.t.~$\mathcal I$), denoted $c \crel{p}^\mathit{t} c'$, if $c$ occurs in $\fr{\beta^B(c')}$.
Notice that, for $c$ to be a parent term of $c'$ it is not enough to be in one of the parent atoms of the birth atom of $c'$, but it needs to be propagated, via a frontier variable, during the application of the trigger that generates the birth atom of $c'$.
Now, for each term $c$ occurring in $\mathcal I$, we inductively define the {\em rank} of $c$ (w.r.t.~$\mathcal I$) as follows:
\begin{eqnarray*}
&\mathsf{rank}(c)\
=\ \left\{
\begin{array}{ll}
0 & \text{if } c \in \adom{D},\\
&\\
1 + \max\{ \mathsf{rank}(c') : c' \crel{p}^\mathit{t} c \} & \text{otherwise.}
\end{array} \right.&
\end{eqnarray*}
For a term $c \in \adom{\mathcal I}$ with $\mathsf{rank}(c) > 0$, select a term $c' \in \adom{\mathcal I}$ such that $\mathsf{rank}(c') = \mathsf{rank}(c) - 1$ and $c' \crel{p}^\mathit{t} c$. We call $c'$ {\em the favourite parent} of $c$, and we write $c' \crel{\mathit{fp}}^\mathit{t} c$.\footnote{We assume that there exists some fixed mechanism that selects $c'$. For example, $c'$ can be the lexicographically first element of $\{c'' : \mathsf{rank}(c'') = \mathsf{rank}(c) - 1 \text{ and } c'' \crel{p}^\mathit{t} c\}$.}

It should be clear that the binary relation $\crel{\mathit{fp}}^\mathit{t}$ over $\adom{\mathcal I}$ forms an infinite forest $F$ consisting of a finite number of trees, where the roots are terms from $\adom{D}$ of rank $0$.
But what about the out-degree of each node of $F$? We can show that, for each $i \geq 0$, the set $\{c \in \adom{\mathcal I} : \mathsf{rank}(c) = i\}$ is finite. This can be shown by induction on $i \geq 0$, while the key fact is that only finitely many triggers can be formed due to which a null with rank $i+1$ is generated (since, by induction hypothesis, the set of terms with rank at most $i$ is finite). Thus, the nodes of $F$ have finite out-degree.
By applying K\"{o}nig's Lemma\footnote{K\"{o}nig's Lemma is a well-known result from graph theory: for an infinite directed rooted graph, if every node is reachable from the root, and every node has finite out-degree, then there exists an infinite directed simple path from the root.} on $F$, we get that $F$ contains an infinite simple path starting from a root node; let $\ccc_0, \ccc_1, \ccc_2, \ldots$ be such a path.

By construction, for each $i \geq 0$, $\ccc_i$ occurs in the birth atom of $\ccc_{i+1}$. Moreover, there exists a sequence of atoms $\beta^i_0, \beta^i_1, \beta^i_2, \ldots, \beta^i_{m_i}=\beta^B(\ccc_{i+1})$, where $\beta^i_0 \in D$ if $i = 0$ and $\beta^i_0 = \beta^B(\ccc_{i})$ if $i > 0$, such that $\beta^i_k\crel{p}\beta^i_{k+1}$, for each $0 \leq k < m_i$, and there are positions $j$ in $\beta^B(\ccc_{i})$ and $j'$ in $\beta^B(\ccc_{i+1})$ such that $(\beta^B(\ccc_{i}),j) \simeq^*_{P_i} (\beta^B(\ccc_{i+1}),j')$, where $P_i$ is the set of atoms $\{\beta^i_0, \beta^i_1,\ldots \beta^i_{m_i}\}$.

We are now ready to define $\clubsuit$. For brevity, let $P$ be the infinite set of atoms $\bigcup_{i \geq 0} P_i$. Let $\clubsuit = (L^\clubsuit,B^\clubsuit,T^\clubsuit,G^\clubsuit)$, where
\begin{itemize}
\item $L^\clubsuit = \{\alpha \in {\mathcal I} \setminus P : \text{ there is } \beta \in P \text{ such that } \alpha \crel{p} \beta\}$,

\item $B^\clubsuit = (\alpha_i^\clubsuit)_{i \geq 0}$ with $\alpha^\clubsuit_0, \alpha^\clubsuit_1, \alpha^\clubsuit_2,\ldots $ being the enumeration of the atoms of $P$ such that $\alpha^\clubsuit_i \crel{p} \alpha^\clubsuit_{i+1}$, for each $i \geq 0$,

\item $T^\clubsuit = (\sigma_i^\clubsuit,h_i^\clubsuit)_{i > 0}$ with $(\sigma_i^\clubsuit,h_i^\clubsuit)$, for $i > 0$, being the trigger for $\dep$ on $B^\clubsuit \cup \{\alpha_{i-1}^\clubsuit\}$ such that $\alpha_i^\clubsuit = \result{\sigma_i^\clubsuit,h_i^\clubsuit}$, and

\item $G^\clubsuit = (\gamma_i^\clubsuit)_{i > 0}$ with $\gamma^\clubsuit_i \in \body{\sigma_i^\clubsuit}$ and $\alpha_{i-1}^\clubsuit = h_i^\clubsuit(\gamma_i^\clubsuit)$.
\end{itemize}
It should be clear, from the above construction, that the sequence of triggers $(\sigma_i^\clubsuit,h_i^\clubsuit)_{i > 0}$ exists, and thus, $\clubsuit$ is well-defined. Then:

\begin{lemma}\label{lem-step-1}
$\clubsuit$ is a proto-caterpillar for $\dep$.
\end{lemma}

As said at the beginning of Step 1, the goal was to extract from $(I_i)_{i \geq 0}$ a proto-caterpillar that is ``almost connected''. It is easy to verify that the proto-caterpillar $\clubsuit$ is ``almost connected'' in the sense that all the conditions of Definition~\ref{caterpillar-c} are satisfied, with $\ccc_0,\ccc_1,\ldots$ playing the role of the relay terms, apart from (4). Indeed, there is no guarantee that $\ccc_0,\ccc_1,\ldots$
do not occur at immortal positions. Can we convert $\clubsuit$ into a connected proto-caterpillar that satisfies also condition (4)? This is the goal of the next step.

\medskip

\noindent
\paragraph{\underline{Step 2: Construct a Connected Proto-Caterpillar}}

\smallskip

\noindent It is clear that if a term $\ccc_i$, for $i \geq 0$, occurs in an immortal position in some atom $\alpha^\clubsuit_j$, then it occurs in every $\alpha^\clubsuit_k$ for $k > j$. Since $\ar{\dep}$ is finite, we can have only finitely many integers $i \geq 0$ such that the term $\ccc_i$ occurs at an immortal position. Let $i_0$ be an integer greater than all such numbers $i$, which means that $\ccc_{i_0},\ccc_{i_0 + 1},\ccc_{i_0 + 2},\ldots$ do not occur at immortal positions. Let $n$ be such that $\alpha_n^\clubsuit = \beta^B(\ccc_{i_0})$, i.e., is the birth atom of $\ccc_{i_0}$.
It should be now clear how $\clubsuit$ can be converted into a connected proto-caterpillar $\spadesuit$. For brevity, let $P$ be the set of atoms $\left\{\alpha_{k+n}^\clubsuit : k \geq 0\right\}$.
Let $\spadesuit = (L^\spadesuit,B^\spadesuit,T^\spadesuit,G^\spadesuit)$, where
\begin{itemize}
\item $L^\spadesuit = \{\alpha \in {\mathcal I} \setminus P : \text{ there is } \beta \in P \text{ such that } \alpha \crel{p} \beta\}$,

\item $B^\spadesuit = (\alpha_i^\spadesuit)_{i \geq 0}$ with $\alpha^\spadesuit_i = \alpha^\clubsuit_{i+n}$ for each $i \geq 0$,

\item $T^\spadesuit = (\sigma_i^\spadesuit,h_i^\spadesuit)_{i > 0}$ with $(\sigma_i^\spadesuit,h_i^\spadesuit) = (\sigma_{i+n}^\clubsuit,h_{i+n}^\clubsuit)$ for $i > 0$, and

\item $G^\spadesuit = (\gamma_i^\spadesuit)_{i > 0}$ with $\gamma_i^\spadesuit = \gamma^\clubsuit_{i+n}$ for each $i > 0$.
\end{itemize}
Since, by Lemma~\ref{lem-step-1}, $\clubsuit$ is a proto-caterpillar for $\dep$, we can conclude that $\spadesuit$ is also a proto-caterpillar. It also follows by construction that $\spadesuit$ is connected with $\ccc_{i_0},\ccc_{i_0 + 1},\ccc_{i_0 + 2},\ldots$ being its relay terms. Then:

\begin{lemma}\label{lem-step-2}
$\spadesuit$ is a connected proto-caterpillar for $\dep$.
\end{lemma}

Observe that there is no guarantee that $\spadesuit$ is a caterpillar since the two conditions in Definition~\ref{caterpillar} might be violated. Moreover, there is no guarantee that $\spadesuit$ is free, or, equivalently, that, for each $(\alpha,i), (\beta,j) \in \Pi(L^\spadesuit \cup B^\spadesuit)$, $\alpha[i] = \beta[j]$ implies $(\alpha,i) \simeq_{L^\spadesuit \cup B^\spadesuit}^* (\beta, j)$; recall that the other direction holds trivially. Can we convert $\spadesuit$ into a free connected caterpillar? This is the goal of the next step.

\medskip

\noindent
\paragraph{\underline{Step 3: Construct a Free Connected Caterpillar}}

\smallskip

\noindent To achieve our goal, we are going to carefully replace each term occurring in $L^\spadesuit \cup B^\spadesuit$ at a certain position $\pi \in \Pi(L^\spadesuit \cup B^\spadesuit)$ with a new constant that only depends on the equivalence class of $\pi$ w.r.t.~the equivalence relation $\simeq_{L^\spadesuit \cup B^\spadesuit}^*$.
As usual, we write $[\pi]_{\simeq_{L^\spadesuit \cup B^\spadesuit}^*}$ for the equivalence class of $\pi$ w.r.t.~$\simeq_{L^\spadesuit \cup B^\spadesuit}^*$.
Let $\bar h$ be a function that maps each atom $\alpha = R(t_1,\ldots,t_n) \in L^\spadesuit \cup B^\spadesuit$ to the atom
\[
R\left(c_{[(\alpha,1)]_{\simeq_{L^\spadesuit \cup B^\spadesuit}^*}},\ldots,c_{[(\alpha,n)]_{\simeq_{L^\spadesuit \cup B^\spadesuit}^*}}\right),
\]
where, for each $1 \leq i \leq n$, $c_{[(\alpha,i)]_{\simeq_{L^\spadesuit \cup B^\spadesuit}^*}}$ is a constant from $\ins{C}$.

Having the function $\bar h$ in place, it is not difficult to see how $\spadesuit$ can be converted into the desired free connected caterpillar $\heartsuit$. In particular, $\heartsuit = (L^\heartsuit,B^\heartsuit,T^\heartsuit,G^\heartsuit)$, where
\begin{itemize}
\item $L^\heartsuit = \{\bar h(\alpha) : \alpha \in L^\spadesuit\}$,

\item $B^\heartsuit = (\alpha_i^\heartsuit)_{i \geq 0}$ with $\alpha^\heartsuit_i = \bar h(\alpha^\spadesuit_i)$ for each $i \geq 0$,

\item $T^\heartsuit = (\sigma_i^\heartsuit,h_i^\heartsuit)_{i > 0}$ with $\sigma_i^\heartsuit = \sigma_i^\spadesuit$ and $h_i^\heartsuit = \bar h \circ h_i^\spadesuit$, for $i > 0$,

\item $G^\heartsuit = (\gamma_i^\heartsuit)_{i > 0}$ with $\gamma_i^\heartsuit = \gamma_i^\spadesuit$ for each $i > 0$.
\end{itemize}
Stickiness allows us to show the following, which concludes the proof of the first implication; for the details see the appendix:

\begin{lemma}\label{lem-step-3}
$\heartsuit$ is a free connected caterpillar for $\dep$.
\end{lemma}

\subsection{Implication 2: From a Free Connected Caterpillar to a Uniformly Connected One}

Let us now concentrate on the second implication. The proof relies on the fact that we can check whether a free connected caterpillar exists via a deterministic B\"{u}chi automaton. As usual, for an automaton $\mathcal A$, we write $L({\mathcal A})$ for its language. Then:

\begin{lemma}\label{lem:buchi-automaton}
We can construct a deterministic B\"{u}chi automaton ${\mathcal A}_\dep$ where $L({\mathcal A}_\dep) \neq \emptyset$ iff there is a free connected caterpillar for $\dep$.
\end{lemma}

Let us stress that the purpose of the automaton ${\mathcal A}_\dep$ provided by Lemma~\ref{lem:buchi-automaton} is twofold: it is used here, together with a pumping argument, for establishing the second implication, and it will be also used in Section~\ref{sec:decide-via-buchi} for showing that the problem of deciding whether a finitary caterpillar exists is decidable in elementary time. We proceed to give some details concerning ${\mathcal A}_\dep$ that allow us to intuitively explain how we get the second implication, while the detailed construction can be found in the appendix. It should not be surprising that this is the place where freeness plays a role.


The automaton ${\mathcal A}_\dep$ operates on what we call caterpillar words over a finite alphabet $\Lambda_\dep$ consisting of triples of the form $(\sigma,\gamma,P)$, where $\sigma \in \dep$, $\gamma \in \body{\sigma}$, and, with $R$ being the predicate of $\head{\sigma}$, $P \subseteq [\ar{R}]$.
%
Intuitively, a caterpillar word $\mathbf{w} = w_1w_2, \cdots$, with $w_i = (\sigma_i,\gamma_i,P_i)$, is a candidate symbolic representation of a free connected caterpillar, where $w_i$ marks a pass-on point iff $P_i$ is non-empty. In fact, $P_i$ indicates at which positions of $\head{\sigma_i}$ the new relay term appears.
Roughly, ${\mathcal A}_\dep$ accepts $\mathbf{w}$ iff $\mathbf{w}$ encodes a free connected caterpillar, while it enters an accepting state only when it reads a symbol $w_i$ that marks a pass-on point.


We get the second implication from the following observation, which can be shown by an obvious pumping argument:

\begin{observation}\label{obs:pumping}
Let $\mathcal A$ be a deterministic B\"{u}chi automaton with $n_{\mathcal A}$ states. If $L(\mathcal A) \neq \emptyset$, then there is $\mathbf{w} \in L(\mathcal A)$ s.t. among each $n_{\mathcal A}$ consecutive states visited by $\mathcal A$ on input $\mathbf{w}$, at least one is accepting.
\end{observation}

Suppose now  that there exists a free connected caterpillar. By Lemma~\ref{lem:buchi-automaton}, $L({\mathcal A}_\dep) \neq \emptyset$. By applying the above observation to the automaton ${\mathcal A}_\dep$, we get a word $\mathbf{w}$ that encodes a free connected caterpillar $\diamondsuit$ such that the distance between two consecutive pass-on points is bounded by the number of states of ${\mathcal A}_\dep$. Thus, $\diamondsuit$ is a free uniformly connected caterpillar, as needed.

\subsection{Implication 3: From a Free Uniformly Connected Caterpillar to a Finitary One}

We now proceed with the last implication.
%
Consider a free uniformly connected caterpillar $\diamondsuit = (L^\diamondsuit,B^\diamondsuit,T^\diamondsuit,G^\diamondsuit)$, where $B^\diamondsuit = (\alpha_i^\diamondsuit)_{i \geq 0}$, $T^\diamondsuit = (\sigma_i^\diamondsuit,h_i^\diamondsuit)_{i > 0}$, and $G^\diamondsuit = (\gamma_i^\diamondsuit)_{i > 0})$.
%
%
Our intention is to obtain from $\diamondsuit$ a finitary caterpillar by unifying some terms of $\adom{L^\diamondsuit}$ in order to make $L^\diamondsuit$ finite, while the rest remains a valid caterpillar. This can be done via what we call a unifying function.

A {\em unifying function} for $\diamondsuit$ is a function $h : {\mathfrak V} \rightarrow {\mathfrak T}$, where ${\mathfrak V} \subseteq \adom{L^\diamondsuit}$, and ${\mathfrak T}$ a set of new terms not occurring in $L^\diamondsuit \cup B^\diamondsuit$; it is called unifying since it essentially unifies the terms of ${\mathfrak V}$.
Let $h(\diamondsuit) = \left(h(L^\diamondsuit), (h(\alpha_i^\diamondsuit))_{i \geq 0}, (\sigma_i,h \circ h_i^\diamondsuit)_{i > 0}, (\gamma_i^\diamondsuit)_{i > 0}\right)$. Then:

\begin{lemma}\label{lem:main-unifying-function}
There exists a unifying function $h$ for $\diamondsuit$ such that $h(\diamondsuit)$ is a finitary caterpillar for $\dep$.
\end{lemma}

It is not difficult to show that no matter how a unifying function $h$ for $\diamondsuit$ is defined, $h(\diamondsuit)$ is a proto-caterpillar that satisfies condition (1) of Definition~\ref{caterpillar}.
The non-trivial task is to define $h$ is such a way that $h(L^\diamondsuit)$ is finite, and $h(\diamondsuit)$ satisfies condition (2) of Definition~\ref{caterpillar}. The key here is, by exploiting uniformity, which provides a bound on the distance between two consecutive pass-on points, to define a sufficiently large {\em finite} set of new terms to which infinitely many carefully chosen terms of $\adom{L^\diamondsuit}$ are mapped to; the details can be found in the appendix. This completes the proof of Theorem~\ref{the:finitary-caterpillar}.

\subsection{Deciding $\rctaapr(\class{S})$ via B\"{u}chi Automata}\label{sec:decide-via-buchi}\label{sec:decide-via-buchi}

By Theorems~\ref{the:fairness-lemma} and~\ref{the:finitary-caterpillar}, given a set $\dep \in \class{S}$, deciding whether $\dep \not\in \rctaa$ is equivalent to the problem of checking whether there exists a finitary caterpillar for $\dep$. By exploiting the B\"{u}chi automaton provided by Lemma~\ref{lem:buchi-automaton}, we can easily show that:

\begin{lemma}\label{lem:finitary-decidable}
The problem of deciding whether there exists a finitary caterpillar for $\dep$ is decidable in elementary time.
\end{lemma}

Since the emptiness problem of deterministic B\"{u}chi automata is feasible in linear time in the size of the automaton, and since the automaton provided by Lemma~\ref{lem:buchi-automaton} can be constructed in elementary time, checking whether a free connected caterpillar for $\dep$ exists is feasible in elementary time.
Now, observe that the three implications established above, together with the $(2) \Rightarrow (1)$ direction of Theorem~\ref{the:finitary-caterpillar}, imply that there exists a free connected caterpillar iff there exists a finitary caterpillar, and Lemma~\ref{lem:finitary-decidable} follows.
\section{Future Work}\label{sec:conclusions}
%


Here are some non-trivial questions that beg for an answer:
(1) What about the exact complexity of our problems?
%
%
(2) What about restricted chase termination for guarded or sticky sets of multi-head TGDs?
%
(3) What about the more liberal version of the problem that asks whether there is a finite restricted chase derivation? 

%
%
%
%
%

\bibliographystyle{ACM-Reference-Format}


\begin{thebibliography}{24}


\ifx \showCODEN    \undefined \def \showCODEN     #1{\unskip}     \fi
\ifx \showDOI      \undefined \def \showDOI       #1{#1}\fi
\ifx \showISBNx    \undefined \def \showISBNx     #1{\unskip}     \fi
\ifx \showISBNxiii \undefined \def \showISBNxiii  #1{\unskip}     \fi
\ifx \showISSN     \undefined \def \showISSN      #1{\unskip}     \fi
\ifx \showLCCN     \undefined \def \showLCCN      #1{\unskip}     \fi
\ifx \shownote     \undefined \def \shownote      #1{#1}          \fi
\ifx \showarticletitle \undefined \def \showarticletitle #1{#1}   \fi
\ifx \showURL      \undefined \def \showURL       {\relax}        \fi
\providecommand\bibfield[2]{#2}
\providecommand\bibinfo[2]{#2}
\providecommand\natexlab[1]{#1}
\providecommand\showeprint[2][]{arXiv:#2}

\bibitem[\protect\citeauthoryear{Aho, Sagiv, and Ullman}{Aho
  et~al\mbox{.}}{1979}]%
        {AhSU79}
\bibfield{author}{\bibinfo{person}{Alfred.~V. Aho}, \bibinfo{person}{Yehoshua
  Sagiv}, {and} \bibinfo{person}{Jeffrey~D. Ullman}.}
  \bibinfo{year}{1979}\natexlab{}.
\newblock \showarticletitle{Efficient Optimization of a Class of Relational
  Expressions}.
\newblock \bibinfo{journal}{\emph{{ACM} Trans. Database Syst.}}
  \bibinfo{volume}{4}, \bibinfo{number}{4} (\bibinfo{year}{1979}).
\newblock


\bibitem[\protect\citeauthoryear{Baget, Lecl{\`e}re, Mugnier, and Salvat}{Baget
  et~al\mbox{.}}{2011}]%
        {BLMS11}
\bibfield{author}{\bibinfo{person}{Jean-Fran\c{c}ois Baget},
  \bibinfo{person}{Michel Lecl{\`e}re}, \bibinfo{person}{Marie-Laure Mugnier},
  {and} \bibinfo{person}{Eric Salvat}.} \bibinfo{year}{2011}\natexlab{}.
\newblock \showarticletitle{On rules with existential variables: {W}alking the
  decidability line}.
\newblock \bibinfo{journal}{\emph{Artif. Intell.}} \bibinfo{volume}{175},
  \bibinfo{number}{9-10} (\bibinfo{year}{2011}), \bibinfo{pages}{1620--1654}.
\newblock


\bibitem[\protect\citeauthoryear{Barcel{\'{o}}, Gottlob, and
  Pieris}{Barcel{\'{o}} et~al\mbox{.}}{2016}]%
        {BaGP16}
\bibfield{author}{\bibinfo{person}{Pablo Barcel{\'{o}}}, \bibinfo{person}{Georg
  Gottlob}, {and} \bibinfo{person}{Andreas Pieris}.}
  \bibinfo{year}{2016}\natexlab{}.
\newblock \showarticletitle{Semantic Acyclicity Under Constraints}. In
  \bibinfo{booktitle}{\emph{PODS}}. \bibinfo{pages}{343--354}.
\newblock


\bibitem[\protect\citeauthoryear{Benedikt, Konstantinidis, Mecca, Motik,
  Papotti, Santoro, and Tsamoura}{Benedikt et~al\mbox{.}}{2017}]%
        {BKMMPST17}
\bibfield{author}{\bibinfo{person}{Michael Benedikt}, \bibinfo{person}{George
  Konstantinidis}, \bibinfo{person}{Giansalvatore Mecca},
  \bibinfo{person}{Boris Motik}, \bibinfo{person}{Paolo Papotti},
  \bibinfo{person}{Donatello Santoro}, {and} \bibinfo{person}{Efthymia
  Tsamoura}.} \bibinfo{year}{2017}\natexlab{}.
\newblock \showarticletitle{Benchmarking the Chase}. In
  \bibinfo{booktitle}{\emph{PODS}}. \bibinfo{pages}{37--52}.
\newblock


\bibitem[\protect\citeauthoryear{Calautti, Gottlob, and Pieris}{Calautti
  et~al\mbox{.}}{2015}]%
        {CaGP15}
\bibfield{author}{\bibinfo{person}{Marco Calautti}, \bibinfo{person}{Georg
  Gottlob}, {and} \bibinfo{person}{Andreas Pieris}.}
  \bibinfo{year}{2015}\natexlab{}.
\newblock \showarticletitle{Chase Termination for Guarded Existential Rules}.
  In \bibinfo{booktitle}{\emph{PODS}}. \bibinfo{pages}{91--103}.
\newblock


\bibitem[\protect\citeauthoryear{Calautti and Pieris}{Calautti and
  Pieris}{2019}]%
        {CaPi19}
\bibfield{author}{\bibinfo{person}{Marco Calautti} {and}
  \bibinfo{person}{Andreas Pieris}.} \bibinfo{year}{2019}\natexlab{}.
\newblock \showarticletitle{Oblivious Chase Termination: The Sticky Case}. In
  \bibinfo{booktitle}{\emph{ICDT}}. \bibinfo{pages}{17:1--17:18}.
\newblock


\bibitem[\protect\citeauthoryear{Cal\`{\i}, Gottlob, and Kifer}{Cal\`{\i}
  et~al\mbox{.}}{2013}]%
        {CaGK13}
\bibfield{author}{\bibinfo{person}{Andrea Cal\`{\i}}, \bibinfo{person}{Georg
  Gottlob}, {and} \bibinfo{person}{Michael Kifer}.}
  \bibinfo{year}{2013}\natexlab{}.
\newblock \showarticletitle{Taming the Infinite Chase: Query Answering under
  Expressive Relational Constraints}.
\newblock \bibinfo{journal}{\emph{J. Artif. Intell. Res.}}
  \bibinfo{volume}{48} (\bibinfo{year}{2013}), \bibinfo{pages}{115--174}.
\newblock


\bibitem[\protect\citeauthoryear{Cal\`{\i}, Gottlob, and Lukasiewicz}{Cal\`{\i}
  et~al\mbox{.}}{2012a}]%
        {CaGL12}
\bibfield{author}{\bibinfo{person}{Andrea Cal\`{\i}}, \bibinfo{person}{Georg
  Gottlob}, {and} \bibinfo{person}{Thomas Lukasiewicz}.}
  \bibinfo{year}{2012}\natexlab{a}.
\newblock \showarticletitle{A general {D}atalog-based framework for tractable
  query answering over ontologies}.
\newblock \bibinfo{journal}{\emph{J. Web Sem.}}  \bibinfo{volume}{14}
  (\bibinfo{year}{2012}), \bibinfo{pages}{57--83}.
\newblock


\bibitem[\protect\citeauthoryear{Cal\`{\i}, Gottlob, and Pieris}{Cal\`{\i}
  et~al\mbox{.}}{2012b}]%
        {CaGP12}
\bibfield{author}{\bibinfo{person}{Andrea Cal\`{\i}}, \bibinfo{person}{Georg
  Gottlob}, {and} \bibinfo{person}{Andreas Pieris}.}
  \bibinfo{year}{2012}\natexlab{b}.
\newblock \showarticletitle{Towards more expressive ontology languages: {T}he
  query answering problem}.
\newblock \bibinfo{journal}{\emph{Artif. Intell.}}  \bibinfo{volume}{193}
  (\bibinfo{year}{2012}), \bibinfo{pages}{87--128}.
\newblock


\bibitem[\protect\citeauthoryear{den Bussche}{den Bussche}{2015}]%
        {Jan}
\bibfield{author}{\bibinfo{person}{Jan~Van den Bussche}.}
  \bibinfo{year}{2015}\natexlab{}.
\newblock
\newblock
\shownote{Personal Communication.}


\bibitem[\protect\citeauthoryear{Deutsch, Nash, and Remmel}{Deutsch
  et~al\mbox{.}}{2008}]%
        {DeNR08}
\bibfield{author}{\bibinfo{person}{Alin Deutsch}, \bibinfo{person}{Alan Nash},
  {and} \bibinfo{person}{Jeff~B. Remmel}.} \bibinfo{year}{2008}\natexlab{}.
\newblock \showarticletitle{The Chase Revisisted}. In
  \bibinfo{booktitle}{\emph{PODS}}. \bibinfo{pages}{149--158}.
\newblock


\bibitem[\protect\citeauthoryear{Deutsch and Tannen}{Deutsch and
  Tannen}{2003}]%
        {DeTa03}
\bibfield{author}{\bibinfo{person}{Alin Deutsch} {and} \bibinfo{person}{Val
  Tannen}.} \bibinfo{year}{2003}\natexlab{}.
\newblock \showarticletitle{Reformulation of {XML} Queries and Constraints}. In
  \bibinfo{booktitle}{\emph{ICDT}}. \bibinfo{pages}{225--241}.
\newblock


\bibitem[\protect\citeauthoryear{Fagin, Kolaitis, Miller, and Popa}{Fagin
  et~al\mbox{.}}{2005}]%
        {FKMP05}
\bibfield{author}{\bibinfo{person}{Ronald Fagin}, \bibinfo{person}{Phokion~G.
  Kolaitis}, \bibinfo{person}{Ren{\'{e}}e~J. Miller}, {and}
  \bibinfo{person}{Lucian Popa}.} \bibinfo{year}{2005}\natexlab{}.
\newblock \showarticletitle{Data exchange: semantics and query answering}.
\newblock \bibinfo{journal}{\emph{Theor. Comput. Sci.}} \bibinfo{volume}{336},
  \bibinfo{number}{1} (\bibinfo{year}{2005}), \bibinfo{pages}{89--124}.
\newblock


\bibitem[\protect\citeauthoryear{Gogacz and Marcinkowski}{Gogacz and
  Marcinkowski}{2014}]%
        {GoMa14}
\bibfield{author}{\bibinfo{person}{Tomasz Gogacz} {and} \bibinfo{person}{Jerzy
  Marcinkowski}.} \bibinfo{year}{2014}\natexlab{}.
\newblock \showarticletitle{All-Instances Termination of Chase is Undecidable}.
  In \bibinfo{booktitle}{\emph{ICALP}}. \bibinfo{pages}{293--304}.
\newblock


\bibitem[\protect\citeauthoryear{Grahne and Onet}{Grahne and Onet}{2018}]%
        {GrOn18}
\bibfield{author}{\bibinfo{person}{G{\"{o}}sta Grahne} {and}
  \bibinfo{person}{Adrian Onet}.} \bibinfo{year}{2018}\natexlab{}.
\newblock \showarticletitle{Anatomy of the Chase}.
\newblock \bibinfo{journal}{\emph{Fundam. Inform.}} \bibinfo{volume}{157},
  \bibinfo{number}{3} (\bibinfo{year}{2018}), \bibinfo{pages}{221--270}.
\newblock


\bibitem[\protect\citeauthoryear{Grau, Horrocks, Kr{\"{o}}tzsch, Kupke, Magka,
  Motik, and Wang}{Grau et~al\mbox{.}}{2013}]%
        {GHKK+13}
\bibfield{author}{\bibinfo{person}{Bernardo~Cuenca Grau}, \bibinfo{person}{Ian
  Horrocks}, \bibinfo{person}{Markus Kr{\"{o}}tzsch}, \bibinfo{person}{Clemens
  Kupke}, \bibinfo{person}{Despoina Magka}, \bibinfo{person}{Boris Motik},
  {and} \bibinfo{person}{Zhe Wang}.} \bibinfo{year}{2013}\natexlab{}.
\newblock \showarticletitle{Acyclicity Notions for Existential Rules and Their
  Application to Query Answering in Ontologies}.
\newblock \bibinfo{journal}{\emph{J. Artif. Intell. Res.}}
  \bibinfo{volume}{47} (\bibinfo{year}{2013}), \bibinfo{pages}{741--808}.
\newblock


\bibitem[\protect\citeauthoryear{Greco, Molinaro, and Spezzano}{Greco
  et~al\mbox{.}}{2012}]%
        {GrMS12}
\bibfield{author}{\bibinfo{person}{Sergio Greco}, \bibinfo{person}{Cristian
  Molinaro}, {and} \bibinfo{person}{Francesca Spezzano}.}
  \bibinfo{year}{2012}\natexlab{}.
\newblock \bibinfo{booktitle}{\emph{Incomplete Data and Data Dependencies in
  Relational Databases}}.
\newblock \bibinfo{publisher}{Morgan {\&} Claypool Publishers}.
\newblock


\bibitem[\protect\citeauthoryear{Greco, Spezzano, and Trubitsyna}{Greco
  et~al\mbox{.}}{2011}]%
        {GrST11}
\bibfield{author}{\bibinfo{person}{Sergio Greco}, \bibinfo{person}{Francesca
  Spezzano}, {and} \bibinfo{person}{Irina Trubitsyna}.}
  \bibinfo{year}{2011}\natexlab{}.
\newblock \showarticletitle{Stratification Criteria and Rewriting Techniques
  for Checking Chase Termination}.
\newblock \bibinfo{journal}{\emph{{PVLDB}}} \bibinfo{volume}{4},
  \bibinfo{number}{11} (\bibinfo{year}{2011}), \bibinfo{pages}{1158--1168}.
\newblock


\bibitem[\protect\citeauthoryear{Kr{\"{o}}tzsch, Marx, and
  Rudolph}{Kr{\"{o}}tzsch et~al\mbox{.}}{2019}]%
        {KrMR19}
\bibfield{author}{\bibinfo{person}{Markus Kr{\"{o}}tzsch},
  \bibinfo{person}{Maximilian Marx}, {and} \bibinfo{person}{Sebastian
  Rudolph}.} \bibinfo{year}{2019}\natexlab{}.
\newblock \showarticletitle{The Power of the Terminating Chase (Invited Talk)}.
  In \bibinfo{booktitle}{\emph{ICDT}}. \bibinfo{pages}{3:1--3:17}.
\newblock


\bibitem[\protect\citeauthoryear{Lecl{\`{e}}re, Mugnier, Thomazo, and
  Ulliana}{Lecl{\`{e}}re et~al\mbox{.}}{2019}]%
        {LMTU19}
\bibfield{author}{\bibinfo{person}{Michel Lecl{\`{e}}re},
  \bibinfo{person}{Marie{-}Laure Mugnier}, \bibinfo{person}{Micha{\"{e}}l
  Thomazo}, {and} \bibinfo{person}{Federico Ulliana}.}
  \bibinfo{year}{2019}\natexlab{}.
\newblock \showarticletitle{A Single Approach to Decide Chase Termination on
  Linear Existential Rules}. In \bibinfo{booktitle}{\emph{ICDT}}.
  \bibinfo{pages}{18:1--18:19}.
\newblock


\bibitem[\protect\citeauthoryear{Marnette}{Marnette}{2009}]%
        {Marn09}
\bibfield{author}{\bibinfo{person}{Bruno Marnette}.}
  \bibinfo{year}{2009}\natexlab{}.
\newblock \showarticletitle{Generalized schema-mappings: from termination to
  tractability}. In \bibinfo{booktitle}{\emph{PODS}}. \bibinfo{pages}{13--22}.
\newblock


\bibitem[\protect\citeauthoryear{Meier, Schmidt, and Lausen}{Meier
  et~al\mbox{.}}{2009}]%
        {MeSL09}
\bibfield{author}{\bibinfo{person}{Michael Meier}, \bibinfo{person}{Michael
  Schmidt}, {and} \bibinfo{person}{Georg Lausen}.}
  \bibinfo{year}{2009}\natexlab{}.
\newblock \showarticletitle{On Chase Termination Beyond Stratification}.
\newblock \bibinfo{journal}{\emph{PVLDB}} \bibinfo{volume}{2},
  \bibinfo{number}{1} (\bibinfo{year}{2009}), \bibinfo{pages}{970--981}.
\newblock


\bibitem[\protect\citeauthoryear{Nenov, Piro, Motik, Horrocks, Wu, and
  Banerjee}{Nenov et~al\mbox{.}}{2015}]%
        {NPMHWB15}
\bibfield{author}{\bibinfo{person}{Yavor Nenov}, \bibinfo{person}{Robert Piro},
  \bibinfo{person}{Boris Motik}, \bibinfo{person}{Ian Horrocks},
  \bibinfo{person}{Zhe Wu}, {and} \bibinfo{person}{Jay Banerjee}.}
  \bibinfo{year}{2015}\natexlab{}.
\newblock \showarticletitle{RDFox: {A} Highly-Scalable {RDF} Store}. In
  \bibinfo{booktitle}{\emph{ISWC}}. \bibinfo{pages}{3--20}.
\newblock


\bibitem[\protect\citeauthoryear{Urbani, Kr{\"{o}}tzsch, Jacobs, Dragoste, and
  Carral}{Urbani et~al\mbox{.}}{2018}]%
        {UKJDC18}
\bibfield{author}{\bibinfo{person}{Jacopo Urbani}, \bibinfo{person}{Markus
  Kr{\"{o}}tzsch}, \bibinfo{person}{Ceriel J.~H. Jacobs},
  \bibinfo{person}{Irina Dragoste}, {and} \bibinfo{person}{David Carral}.}
  \bibinfo{year}{2018}\natexlab{}.
\newblock \showarticletitle{Efficient Model Construction for Horn Logic with
  VLog - System Description}. In \bibinfo{booktitle}{\emph{IJCAR}}.
  \bibinfo{pages}{680--688}.
\newblock


\end{thebibliography}


\newpage

\appendix
\section{Additional Notions}
%

%
An {\em equality type} over a schema $\ins{S}$ is a pair $(R,E)$, where $R \in \ins{S}$ and $E$ is a partition of $\{1,\ldots,\ar{R}\}$ (which we see as a set of subsets of $\{1,\ldots,\ar{R}\}$).
Let $\etypes{\ins{S}}$ be the set of all possible equality types over $\ins{S}$, which is clearly finite.
Given an atom $\alpha = R(t_1,\ldots,t_n)$ over $\ins{S}$, its equality type, denoted $\et{\alpha}$, is the equality type $(R,E) \in \etypes{\ins{S}}$ such that $t_i = t_j$ iff $i,j$ coexist in a set of $E$.
A homomorphism from a set of atoms $A$ to a set of atoms $B$ is an {\em isomorphism} from $A$ to $B$ if it is 1-1, and its inverse $h^{-1}$ is a homomorphism from $B$ to $A$.

\section{Proofs from Section~\ref{sec:fairness-lemma}}
%


\subsection{Fairness Theorem and Multi-head TGDs}

Here is an example showing that the Fairness Theorem (Theorem~\ref{the:fairness-lemma}) does not hold for multi-head TGDs:

\begin{example}
Consider the set $\dep$ of TGDs consisting of
\[
R(x,y,y)\ \ra\ \exists z \, R(x,z,y),R(z,y,y) \qquad R(x,y,z)\ \ra\ R(z,z,z).
\]
It should be clear that there exists an infinite restricted chase derivation of $\{R(a,b,b)\}$ w.r.t.~$\dep$; apply only the first TGD. However, every valid restricted chase derivation of $\{R(a,b,b)\}$ w.r.t.~$\dep$ is finite. \hfill\markfull
\end{example}

Let us say that the recent paper~\cite{LMTU19}, which studies the restricted chase termination problem for linear TGDs, provides an example that refutes Theorem~\ref{the:fairness-lemma} even if we use only binary predicates.

\subsection{Proof of Lemma~\ref{lem:diagonal-property}}

Clearly, $J_{0}^{0} = D$ since $(J_{i}^{0})_{i \geq 0}$ is a chase derivation of $D$ w.r.t.~$\dep$. It remains to show that, for $k \geq 0$, there is an active trigger $(\sigma,h)$ for $\dep$ on $J_{k}^{k}$ such that $J_{k}^{k} \app{\sigma}{h} J_{k+1}^{k+1}$. Note that $J_{k}^{k} = J_{k}^{k+1}$ by the diagonal property. Thus, it suffices to show that there exists an active trigger $(\sigma,h)$ for $\dep$ on $J_{k}^{k+1}$ such that $J_{k}^{k+1} \app{\sigma}{h} J_{k+1}^{k+1}$. This holds since  $(J_{i}^{k+1})_{i \geq 0}$ is a chase derivation of $D$ w.r.t.~$\dep$.

\subsection{Proof of Lemma~\ref{lem:A-finite}}

Let $\mathcal{I}^n = \bigcup_{i \geq 0} I_{i}^{n}$. Since $\dep$ is finite, it suffices to show that, for each TGD $\hat{\sigma} \in \dep$, the set of atoms $B_{\hat{\sigma}}$ that collects all the atoms of $\mathcal{I}^n$ that are stopped by $\result{\sigma,h}$ and are generated by a trigger that involves $\hat{\sigma}$, is finite. Indeed, $B_{\hat{\sigma}}$ is finite implies $A$ is finite, since the cardinality of $B_{\hat{\sigma}}$ coincides with the cardinality of $\{i \geq 0 : \result{\sigma,h} \crel{s} \result{\sigma_i,h_i} \text{ and } \sigma_i = \hat{\sigma}\}$.

Since all the atoms of $B_{\hat{\sigma}}$ have been created by the same TGD, 
and they are all stopped by $\result{\sigma,h}$, we can conclude that they are equal when restricted to their frontier.
Towards a contradiction, assume that two atoms $\alpha,\beta \in B_{\hat{\sigma}}$ have the same equality type. This implies that $\alpha \crel{s} \beta$ and $\beta \crel{s} \alpha$, which contradicts the fact that $\alpha$ and $\beta$ belong to the result of a restricted chase derivation. Thus, for every two distinct atoms $\alpha,\beta \in B_{\hat{\sigma}}$, $\alpha$ and $\beta$ have different equality types. Since there are only finitely many equality types over $\sch{\dep}$, we conclude that $B_{\hat{\sigma}}$ is finite, as needed.

\subsection{Proof of Lemma~\ref{lem:correctness}}

By construction, $I_{0}^{n+1} = I_{0}^{n} = D$. It remains to show that, for $i \geq 0$,
there is an active trigger $(\sigma_i,h_i)$ for $\dep$ on $I_{i}^{n+1}$ such that $I_{i}^{n+1} \app{\sigma_i}{h_i} I_{i+1}^{n+1}$. We proceed by considering the following cases:

\medskip

\paragraph{Case 1.} For $0 \leq i \leq \ell$, the claim is trivial since $(I_{i}^{n+1})_{0 \leq i \leq \ell} = (I_{i}^{n})_{0 \leq i \leq \ell}$, while $(I_{i}^{n})_{i \geq 0}$ is a chase derivation of $D$ w.r.t.~$\dep$.

\medskip

\paragraph{Case 2.} For $i = \ell + 1$, the claim holds since $(\sigma,h)$
is an active trigger for $\dep$ on $I_{\ell}^{n}=I_{\ell}^{n+1}$ and $I_{i}^{n+1}= I_{i-1}^{n+1} \cup \{\result{\sigma,h}\}$.

\medskip

\paragraph{Case 3.} Finally, assume that $i \geq \ell + 2$. Recall that $I_{i-1}^{n}$ is obtained from $I_{i-2}^{n}$ by applying the active trigger $(\sigma_{i-2},h_{i-2})$. We are going to show that $I_{i-1}^{n+1} \app{\sigma_{i-2}}{h_{i-2}}  I_{i}^{n+1}$.
Clearly,  $I_{i}^{n+1} = I_{i-1}^{n+1}\cup \{\result{\sigma_{i-2},h_{i-2}}\}$ since $I_{i-1}^{n} = I_{i-2}^{n}\cup \{\result{\sigma_{i-2},h_{i-2}}\}$. It is also clear that $(\sigma_{i-2},h_{i-2})$ is a trigger for $\dep$ on $I_{i-1}^{n+1}$. It remains to show that it is also active.
Assume that this is not the case. Fact~\ref{fact:stop-relation} implies that there is an atom $\alpha \in I_{i-1}^{n+1}$ such that $\alpha \crel{s} \result{\sigma_{i-2},h_{i-2}}$. Recall that $I_{i-1}^{n+1}= I_{i-2}^{n} \cup \{\result{\sigma,h}\}$. Since $(\sigma_{i-2},h_{i-2})$ is an active trigger for $\dep$ on $I_{i-2}^{n}$, we conclude that $\alpha \not\in I_{i-2}^{n}$. Moreover, since $i-2$ is greater that all the elements of $A$, we get that $\result{\sigma,h}$ does not stop $\result{\sigma_{i-2},h_{i-1}}$, which implies that $\alpha \neq \result{\sigma_{i-2},h_{i-1}}$. Hence, $\alpha \not\in I_{i-1}^{n+1}$, which is a contradiction.


\section{Proofs from Section~\ref{sec:guardedness}}
%


\subsection{Proof of Theorem~\ref{j-chaseable-lemma}}

The fact that $(1) \Rightarrow (2)$ is easy: simply define $A$ as the set $\bigcup_{i \geq 0} I_i$, where $(I_i)_{i \geq 0}$ is the infinite restricted chase derivation of $D$ w.r.t.$\dep$ that exists by hypothesis.
For the other direction, by exploiting $A$, we are going to inductively construct an infinite restricted chase derivation $(I_i)_{i \geq 0}$ of $D$ w.r.t~$\dep$.
Clearly, $I_0$ is defined as $D$.
Suppose that we have already constructed $(I_i)_{1 \leq i \leq n-1}$, for some natural number $n > 1$. Due to condition (1) of Definition~\ref{j-chaseable-def}, there exists an atom $\alpha \in A \setminus B$, where $B = \bigcup_{0 \leq i < n} I_i$, that is minimal w.r.t.~$\crel{b}$, i.e., for every $\beta \in A \setminus B$, $\alpha \crel{b} \beta$. We define $I_n$ as the instance $B \cup \{\alpha\}$. It remains to show that there exists an active trigger $(\sigma,h)$ for $\dep$ on $B$ such that $\alpha = \result{\sigma,h}$. By condition (2) of Definition~\ref{j-chaseable-def}, we get that all the parents of $\alpha$ occur in $B$, and thus $(\sigma,h)$ is a trigger for $\dep$ on $B$.
To show that $(\sigma,h)$ is active, by Fact~\ref{fact:stop-relation}, we need to show that there is no $\beta \in B$ such that $\beta \crel{s} \alpha$. Towards a contradiction, assume that such $\beta$ exists. This implies that $\alpha \crel{b} \beta$ (recall that $\crel{s}^{-1} \subseteq \crel{b})$. But this implies that $\crel{b}$ over $B$ contains a cycle, which is a contradiction due to the third condition of Definition~\ref{j-chaseable-def}, and the claim follows.


\subsection{Proof of Theorem~\ref{the:treefication}}

In the rest of the subsection, let $\dep$ be a set of single-head guarded TGDs. As it is common when studying guarded TGDs, we need a refined version of the parent relation over the real oblivious chase that distinguishes between guard- and side-parents.

\medskip

\noindent
\paragraph{Guard- and Side-Parent Relation.} Consider the real oblivious chase $\chase{D}{\dep} = \langle V, \crel{p}, \lambda, \tau \rangle$ of a database $D$ w.r.t.~$\dep$. We can naturally define the {\em guard-parent} relation $\crel{\mathit{gp}}$ over $V$ as the subrelation of $\crel{p}$ by keeping only the pairs of nodes $(v,u)$ where $v$ corresponds to the guard atom of the TGD in $\tau(u)$. Formally, the guard-parent relation $\crel{\mi{gp}}$ (over $V$) is defined as
\[
\{\langle v,u \rangle\ :\ v \crel{p} u \text{ and, with } \tau(u) = (\sigma,h),\ h(\guard{\sigma}) = \lambda(v)\}.
\]
For a node $u$, we may write $\mi{gp}(u)$ for its guard-parent, i.e., if $v \crel{\mi{gp}} u$, then $\mi{gp}(u) = v$. We denote by $\ccrel{\mi{gp}}$ the transitive closure of $\crel{\mi{gp}}$. Observe that, due to guardedness, $\chase{D}{\dep}$ can be seen as a forest w.r.t.~$\crel{\mi{gp}}$, with the nodes of $V$ labeled with atoms of $D$ being the roots of the trees, and all the other nodes are the non-root nodes. It would be conceptually useful to have this forest in mind.

Regarding the side-parents, it is not enough to simply keep the pairs $v \crel{p} u$ where $v$ corresponds to a side atom (i.e., an atom different than the guard) of the TGD in $\tau(u)$. In addition, we need to know which terms of the atom $\lambda(\mi{gp}(u))$ occur in $\lambda(v)$ and at which positions.\footnote{For the discussion in the main body of the paper, the simple side-parent relation $\crel{sp}$ was enough. However, for the formal proof we need this additional information.}
This can be achieved via the notion of sideatom type.\footnote{Note that this notion has been also introduced and used in Section~\ref{sec:msol-sentence} where we talk about chaseable abstract join trees. We repeat it here for the sake of readability.}
A {\em sideatom type} $\pi$ (w.r.t~$\dep$) is a triple $\langle P, m, \xi \rangle$, where $P/n \in \sch{\dep}$, $m \leq \ar{\dep}$ is a natural number, called the arity of $\pi$, and $\xi : [n] \ra [m]$.
%
Given two atoms $\beta$ and $\gamma$, we say that $\beta$ is a {\em $\pi$-sideatom} of $\gamma$, denoted $\beta \subseteq_\pi \gamma$, if the predicate of $\beta$ is $P$, the predicate of $\gamma$ has arity $m$, and $\beta[i] = \gamma(\xi(i))$ for each $i \in [n]$.
For example, the atom $\beta = P(a,b,c)$ is a $\pi$-sideatom of $\gamma = R(a,d,c,b)$ with $\pi = \langle P, 4, \{1 \mapsto 1, 2 \mapsto 4, 3 \mapsto 3\} \rangle$.
%
%
Consider now a node $u \in V$ such that $v \crel{p} u$, $v_1 \crel{p} u$, \ldots, $v_m \crel{p} u$, $\tau(u) = (\sigma,h)$, where $\body{\sigma} = \gamma,\gamma_1,\ldots,\gamma_m$ with $\gamma = \guard{\sigma}$, and $h(\gamma) = \lambda(v)$, $h(\gamma_1) = \lambda(v_1)$, \ldots, $h(\gamma_m) = \lambda(v_m)$. Let $\pi_1,\ldots,\pi_m$ be sideatom types such that, for each $i \in [m]$, $\lambda(v_i) \subseteq_{\pi_i} \lambda(v)$ (or $\lambda(v_i) \subseteq_{\pi_i} \lambda(\mi{gp}(u))$). Then, for each $i \in [m]$, we say that $v_i$ is a {\em $\pi_i$-side-parent} of $u$, written $v_i \crel{\mi{sp}}^{\pi_i} u$.


As for the relation $\crel{p}$, notice that, strictly speaking, $\crel{\mi{gp}}$ and $\crel{\mi{sp}}^\pi$, for some sideatom type $\pi$, are relations over the node set $V$ of $\chase{D}{\dep}$. However, for convenience, we will usually see these relations as relations over the set multiset consisting of the atoms of $\chase{D}{\dep}$. Thus, we will directly refer to the guard-parent of an atom $\alpha$ of $\chase{D}{\dep}$ and write $\mi{gp}(\alpha)$.

\medskip

Let us now proceed with the proof of Theorem~\ref{the:treefication}. By hypothesis, there exists an infinite restricted chase derivation $(I_i)_{i \geq 0}$ of $D$ w.r.t.~$\dep$. The proof proceeds in three main steps:
\begin{enumerate}
\item We first construct from $D$ an acyclic database $D_{\mi{ac}}$. In fact, we explicitly construct a join tree $(T_{\mi{ac}},\lambda)$, where $T_{\mi{ac}} = (V,E)$, and the database $D_{\mi{ac}}$ is defined as $\{\lambda(v) : v \in V\}$.

\item We then show that there is an auxiliary infinite sequence of instances $(K_i)_{i \geq 0}$, where $K_0 = D_{\mi{ac}}$, which somehow mimics the infinite restricted chase derivation $(I_i)_{i \geq 0}$ of $D$ w.r.t.~$\dep$.

\item Finally, by exploiting the sequence $(K_i)_{i \geq 0}$, we construct an infinite restricted chase derivation $(J_i)_{i \geq 0}$ of $D_{\mi{ac}}$ w.r.t.~$\dep$.
\end{enumerate}

We proceed to give more details for each of the above steps. But first we need to fix some notation. Let $\mathcal{I} = \bigcup_{i \geq 0} I_i$. We write $(\sigma_i^\mathcal{I}, h_i^\mathcal{I})$ for the trigger such that $I_i \langle \sigma_i^\mathcal{I}, h_i^\mathcal{I} \rangle I_{i+1}$. For brevity, we write $\beta^\mathcal{I}_i$ for the atom $\result{\sigma_i^\mathcal{I}, h_i^\mathcal{I}}$, and $\gamma^\mathcal{I}_i$ for $\guard{\sigma^\mathcal{I}_i}$. Given an atom $\beta \in \mathcal{I}$, we define $\mathcal{I}_\beta$ as the set $\{\alpha \in \mathcal{I} : \beta \ccrel{\mi{gp}} \alpha\}$.

\medskip

\noindent
\paragraph{\underline{Step 1: The Acyclic Database $D_{\mi{ac}}$}}

\smallskip

\noindent Since $D$ is finite, while $\mathcal{I}$ is infinite, we can conclude that there exists an atom $\alpha^\infty \in D$ such that the set  $\mathcal{I}_{\alpha^\infty}$ is infinite. One may think that the acyclic database $D_{\mi{ac}}$ consists of the atom $\alpha^\infty$ together with the atoms of $D$ that can serve as its side atoms, i.e., the database
\[
\{\alpha^\infty\}\ \cup\ \{\beta \in D : \beta \subseteq_{\pi} \alpha^\infty \text{ for some sideatom type } \pi\}.
\]
However, as explained in the main body of the paper (see Example~\ref{exa:acyclic-db-counterexample}), this is not the case due to what we call remote side-parents:

\begin{definition}\label{def:remote-side-parent}
Consider two distinct atoms $\alpha,\beta \in D$, and two atoms $\alpha', \beta'\in \mathcal{I}$.
The tuple $\langle \alpha,\alpha',\beta,\beta' \rangle$ is a {\em remote-side-parent situation} if $\alpha \ccrel{\mi{gp}} \alpha'$, $\beta \ccrel{\mi{gp}} \beta'$, and $\beta' \crel{\mi{sp}}^{\pi} \alpha'$ for some sideatom type $\pi$. If this is the case, then we say that $\alpha$ {\em longs for} $\beta$. \hfill\markfull
\end{definition}

The following easy lemma collects a couple of useful facts about the notion of remote-side-parent situation, which would be crucial for the construction of the acyclic database $D_{\mi{ac}}$.

\begin{lemma}\label{j-20}
\begin{enumerate}
\item If $\langle \alpha,\alpha',\beta,\beta' \rangle$ is a remote-side-parent situation, then $\beta' \subseteq_{\pi} \alpha$ and $\beta' \subseteq_{\pi'} \beta$ for some types $\pi$ and $\pi'$.

\item There exists a natural number $\ell_\infty$ such that, if $\langle \alpha^\infty,\alpha',\beta,\beta' \rangle$ is a remote-side-parent situation, then $\beta'\in I_{\ell_\infty}$.
\end{enumerate}
\end{lemma}

\begin{proof}
It is easy to verify that claim (1) holds due to guardedness. For claim (2) it suffices to observe that the following holds, which is a consequence of (1): for an atom $\alpha \in D$, there are only finitely many pairs of atoms $\beta,\beta'$ such that, for some atom $\alpha'$ with $\alpha \ccrel{\mi{gp}} \alpha'$, $\langle \alpha,\alpha',\beta,\beta' \rangle$ is a remote-side-parent situation.
\end{proof}

\noindent
\paragraph{The Construction of $D_{\mi{ac}}$.} Let us now formally define the acyclic database $D_{\mi{ac}}$. We will construct, via simultaneous induction:
\begin{enumerate}
\item A labeled tree $(T_{\mi{ac}},\lambda)$, where $T_{\mi{ac}} = (V,E)$ and $\lambda$ is a labeling function from $V$ to $\{R([t_1]_v,\ldots,[t_n]_v) : R/n \in \sch{\dep}, t_i \in \adom{D} \text{ and } v \in V\}$, i.e., the set of atoms that can be formed using predicates of $\sch{\dep}$
and constants from the set $\{[t]_v : t \in \adom{D} \text{ and } v \in V\}$.

\item A mapping $h_{\mi{ac}}$ from $\{ \lambda(v) : v \in V\}$ to $D$.

\item A function $\mathsf{depth}$ from $\{ \lambda(v) : v \in V\}$ to $\mathbb N$.
\end{enumerate}

The constants of the form $[t]_v$ used above provide us with a simple mechanism for uniformly renaming a constant $t \in \adom{D}$ into a fresh constant, while this renaming step is performed with respect to a certain node $v$ of $T_{\mi{ac}}$. This allows us to break the connection among occurrences of the same constant that are semantically different; this will be made clear in a while. The construction follows:

\medskip

\noindent \underline{Base Case.} Let $v \in V$ be the root node of $T_{\mi{ac}}$. Then, $\lambda(v) = \alpha^\infty$, $h_{\mi{ac}}(\lambda(v)) = \alpha^\infty$, and $\dept{\lambda(v)} = 0$.

\medskip

\noindent \underline{Inductive Step.}
Assume that $v \in V$ is such that $h_{\mi{ac}}(\lambda(v)) = \alpha$, for some $\alpha \in D$, with $\dept{\lambda(v)} < \ell_\infty$. Then, for each $\beta \in D$ such that $\alpha$ longs for $\beta$, we add a new node $u$ to $V$, and  the edge $(v,u)$ to $E$, in such a way that:

\begin{itemize}
\item the atom $\lambda(u)$ is of the following form:
\begin{itemize}
\item it has the same predicate as the atom $\beta$,
\item $\lambda(u)[i] = \lambda(u)[j]$ iff $\beta[i] = \beta[j]$,
\item $\lambda(u)[i] = \lambda(v)[j]$ iff $\beta[i] = \alpha[j]$,
\item if $\lambda(u)[i]$ does not occur in $\lambda(v)$, then $\lambda(u)[i] = [\beta[i]]_u$.
\end{itemize}
\item $h_{\mi{ac}}(\lambda(u)) = \beta$, and
\item $\dept{\lambda(u)} = \dept{\lambda(v)} + 1$.
\end{itemize}
This completes the construction of $(T_{\mi{ac}},\lambda)$, $h_{\mi{ac}}$ and $\mathsf{depth}$. Having $(T_{\mi{ac}},\lambda)$ in place, we define $D_{\mi{ac}}$ as $\{\lambda(v) : v \in V\}$.


Before we proceed any further, it is important to observe that different nodes $v,u$ of $T_{\mi{ac}}$ (possibly of different depths)
may have the same label, i.e. it may happen that $\lambda(v) = \lambda(u)$. In this case, we  treat them as two different atoms since,
although  syntactically the same, they are present in $T_{\mi{ac}}$ for different reasons. Therefore, strictly speaking,
$D_{\mi{ac}}$ is a {\em multiset database}, i.e., it can hold many occurrences of the same atom, which are treated as different atoms. Let us clarify that the notion of acyclicity given in Definition~\ref{def:acyclic-dbs} can be directly applied to multiset instances, i.e., a multiset instance is acyclic iff it admits a join tree.

\begin{lemma}\label{lem:acyclic-database-properties}
\begin{enumerate}
\item $D_{\mi{ac}}$ is an acyclic multiset database.
\item The mapping $h_{\mi{ac}}$ is a homomorphism from $D_{\mi{ac}}$ to $D$.
\item For each two vertices $u,v\in T_{\mi{ac}}$, the mapping $h_{\mi{ac}}$ is an isomorphism from $\{\lambda(u),\lambda(v) \}$ to
$\{h_{\mi{ac}}(\lambda(u)),h_{\mi{ac}}(\lambda(v)) \}$
\end{enumerate}
\end{lemma}

\begin{proof}
For (1) it suffices to show that $T_{\mi{ac}}$ is finite, and that it enjoys the connectedness condition (condition (2) of Definition~\ref{def:acyclic-dbs}). By Lemma~\ref{j-20}(1), for $\alpha \in D$, there are only finitely many pairs of atoms $\beta,\beta'$ such that $\langle \alpha,\alpha',\beta,\beta' \rangle$ is a remote-side-parent situation for some $\alpha'$. Thus, by construction, the branching degree of $T_{\mi{ac}}$ is finite. Since the depth of $T_{\mi{ac}}$ is bounded by $\ell_\infty$, $T_{\mi{ac}}$ is finite. The fact that $T_{\mi{ac}}$ enjoys the connectedness condition follows by construction; here, the renaming of the constants $t$ of $\adom{D}$ to $[t]_v$ is crucial. Claims (2) and (3) also follow by construction.
\end{proof}

It remains to show that there exists an infinite restricted chase derivation of $D_{\mi{ac}}$ w.r.t.~$\dep$.
Indeed, by showing the latter statement for the multiset database $D_{\mi{ac}}$, we can conclude that there exists an infinite restricted chase derivation of the acyclic database obtained from $D_{\mi{ac}}$ by keeping only one occurrence of each atom w.r.t.~$\dep$.

\medskip

\noindent
\paragraph{\underline{Step 2: An Auxiliary Infinite Sequence of Instances}}

\smallskip

\noindent As we already explained, $D_{\mi{ac}}$ consists of several (slightly modified) copies of atoms of $D$. It is like seeing the atoms of $D$ through several distorting mirrors, where the mirror images are atoms of $D_{\mi{ac}}$. Imagine now that we watch the restricted chase derivation $(I_i)_{i \geq 0}$ through those mirrors. Although during a restricted chase step only one atom,
let us say $\alpha$, is generated, in the mirrors we see the generation of several atoms, which are the distorted images of $\alpha$. In order to formalize this phenomenon, we  define a variant of the restricted chase, called {\em weakly restricted chase}.

\medskip

\noindent
\paragraph{Weakly Restricted Chase.} Our intention is to define a variant of chase that allows us to apply several active triggers at the same time, and operates on multiset instances, i.e., multisets of atoms. The reason why we need to consider multisets is because two different mirror images may be syntactically the same.

\begin{definition}\label{def:weakly-restricted-chase}
Consider a multiset instance $K$, and let $S$ be a set of active triggers for $\dep$ on $K$. An application of $S$ to $K$, called {\em weakly restricted chase step}, returns the multiset instance
\[
K'\ =\ K\ \cup\ \{\result{\sigma,h} : (\sigma,h) \in S\},
\]
and is denoted as $K \langle S \rangle K'$.
A sequence of multiset instances $(K_i)_{i \geq 0}$, where $K_0$ is the database $D'$, is a {\em weakly restricted chase derivation} of $D'$ w.r.t.~$\dep$ if, for each $i \geq 0$, there exists a set $S$ of active triggers for $\dep$ on $K_i$ such that $K_i \langle S \rangle K_{i+1}$. \hfill\markfull
\end{definition}

The auxiliary infinite sequence of instances that we are looking for, which will eventually lead to an infinite restricted chase derivation of $D_{\mi{ac}}$ w.r.t.~$\dep$, is an infinite weakly restricted chase derivation of $D_{\mi{ac}}$ w.r.t.~$\dep$. This is essentially (modulo some condition, called the {\em depth condition}, given below) the infinite restricted chase derivation $(I_i)_{i \geq 0}$ seen through the mirrors discussed above.

\medskip

\noindent
\paragraph{The Auxiliary Sequence $(K_i)_{i \geq 0}$.} We now inductively construct a sequence of multiset instances $(K_i)_{i \geq 0}$, together with a mapping $\bar h$ from $\bigcup_{i \geq 0} K_i$ to $\mathcal{I}$:

\medskip

\noindent \underline{Base Case.} Let $K_0 = D_{\mi{ac}}$, and for each $\alpha \in K_0$, $\bar h(\alpha) = h_{\mi{ac}}(\alpha)$. Recall that $h_{\mi{ac}}$ is the mapping from $D_{\mi{ac}}$ to $D$ provided by Lemma~\ref{lem:acyclic-database-properties}.

\medskip

\noindent \underline{Inductive Step.} Suppose now that $K_i$ and $\bar h : K_i\rightarrow \mathcal{I}$ have been already defined, for $i > 0$. Let $S_i$ be the set of all active triggers for $\dep$ on $K_i$ of the form $(\sigma^\mathcal{I}_{i},h)$, i.e., they use the same TGD $\sigma^{\mathcal{I}}_{i}$ that has been used in $(I_i)_{i \geq 0}$ to generate the atom $\beta^\mathcal{I}_{i}$, such that:
\begin{enumerate}
\item $\bar h(h(\gamma_{i}^\mathcal{I})) = \mi{gp}(\beta^\mathcal{I}_{i})$, which simply states that the atom of $K_i$ that is now
about to become a guard-parent must be a mirror image of the guard-parent of the atom $\beta^\mathcal{I}_{i}$ in $\mathcal{I}$.

\item (\textbf{Depth Condition}) $\alpha \ccrel{\mi{gp}} h(\gamma_{i}^\mathcal{I})$, for some atom $\alpha \in D_{\mi{ac}}$ such that $\alpha = \alpha^{\infty}$, or $\dept{\alpha} < \ell_\infty - i$.
\end{enumerate}
We define $K_{i+1}$ as the multiset instance
\[
K_i\ \cup\ \{\result{\sigma_{i}^{\mathcal{I}},h} : (\sigma_{i}^{\mathcal{I}},h) \in S_i\},
\]
i.e., $K_i \langle S_i \rangle K_{i+1}$.
Furthermore, for each $\alpha \in K_{i+1} \setminus K_i$, let $\bar h(\alpha) = \beta^\mathcal{I}_{i}$. This completes the definition of $(K_i)_{i \geq 0}$.

\medskip

\noindent
\paragraph{The Structure of the Auxiliary Sequence.} By construction, $(K_i)_{i \geq 0}$ is a weakly restricted chase derivation of $D_{\mi{ac}}$ w.r.t.~$\dep$. What is not immediately clear is that $(K_i)_{i \geq 0}$ is infinite.
Our goal, in the rest of this subsection, is to understand how $\mathcal{K} = \bigcup_{i \geq 0} K_i$ relates to $\mathcal I$. This analysis will give us useful information about the structure of $\mathcal{K}$, which will be crucial later, and also it will allow us to conclude that $\mathcal K$ is infinite.

For an atom $\beta \in \mathcal{K}$, we define $\mathcal{K}_\beta$ as the set $\{\alpha \in \mathcal{K} : \beta \ccrel{\mi{gp}} \alpha\}$.
The main technical lemma that we need to show follows:

\begin{lemma}\label{k-structure}
For each $i \geq 0$, the following statements hold:
\begin{enumerate}
\item For each $\alpha \in  D_{\mi{ac}}$ such that $\alpha \neq {\alpha^\infty}$, $\bar h$ is an isomorphism from $\mathcal{K}_{\alpha}\cap K_i$ to $\mathcal{I}_{\bar h(\alpha)}\cap I_k$, where $k= \min\{i,\ell_\infty-\dept{\alpha}\}$.

\item The mapping $\bar h$ is an isomorphism from $\mathcal{K}_{\alpha^\infty}\cap K_i$ to $\mathcal{I}_{\alpha^\infty}\cap I_i$.
\end{enumerate}
\end{lemma}

\begin{proof}
For $\alpha\in D_{\mi{ac}}$, we define $T^K_i(\alpha)={\mathcal K}_\alpha\cap K_i$ and $T^I_i(\alpha)={\mathcal I}_{\bar h(\alpha)}\cap I_k$, where $k= \min\{i,\ell_\infty-\dept{\alpha}\}$ if $\dept{\alpha}>0$, and $k = i$ otherwise. Let also ${\alpha^I_i}$ be an atom of $D$ such that $\alpha^I_i \ccrel{gp} \beta^I_i $.

The lemma says that, for $i \geq 0$ and $\alpha \in D_{\mi{ac}}$, $\bar h$ is an isomorphism from $T^K_i(\alpha)$ to $T^I_i(\alpha)$.
We proceed by induction on $i$. The lemma holds for $i=0$, since $I_0=D$ and $K_0=D_{\mi{ac}}$. Assume now that it is also true for some $i \geq 0$.

First observe that, if $\dept{\alpha}>0$ and $\dept{\alpha}\geq \ell_\infty-i$, for $\alpha\in D_{\mi{ac}}$, then $T^K_i(\alpha)=T^K_{i+1}(\alpha)$ (this follows from the Depth Condition), and $T^I_i(\alpha)= T^I_{i+1}(\alpha)$. Thus, for such $\alpha$, the claim directly follows from the hypothesis.

It of course follows from the hypothesis (the part where it tells us something about $\bar h$) that, if $\beta\ccrel{gp} \beta'$, for some $\beta,\beta'\in K_i$, then also $\bar h(\beta)\ccrel{gp} \bar h(\beta')$. In other words, if $\beta\in T_i^K(\alpha)$, then $\bar h(\beta)\in T_i^K(\bar h (\alpha))$. This means that, if $\bar h(\alpha)\neq \alpha^I_i$, then $T^K_i(\alpha)=T^K_{i+1}(\alpha)$. In this case, we also have that $T^I_i(\alpha)= T^I_{i+1}(\alpha)$. Hence, again, for such $\alpha$ our claim directly follows from the hypothesis.

Let us now concentrate on the only interesting case, where $\alpha$ is not too deep and is a mirror image of $\alpha^I_i$ (and, therefore, ${\mathcal K}_\alpha$ is a mirror image of ${\mathcal I}_{\alpha^I_i}$), which, formally speaking, means that $\bar h(\alpha)= \alpha^I_i$ and $\dept{\alpha}=0$ or $\dept{\alpha} < \ell_\infty-i$.

Clearly, for such $\alpha$, there exists exactly one atom which is in $T^I_{i+1}(\alpha)$ but not in $ T^I_{i}(\alpha)$. This
atom is $\beta^I_i$. It is also easy to see that $gp(\beta^I_i)$ (or, $\beta_{gp}$, for short) is somewhere in $ T^I_i(\alpha)$, and, by hypothesis, there is an atom $\kappa_{gp}$ somewhere in $T^K_i(\alpha)$ such that $\bar h(\kappa_{gp})=\beta_{gp}$.

Now, suppose there is an active trigger $(\sigma^I_i,h)$ on $K_i$ such that $h(\gamma_i)=\kappa_{gp}$. Then the atom, call it $\kappa_{new}$, will
appear in $K_{i+1}$ as the result of this trigger, with $\bar h(\kappa_{new})=\beta^I_i$, and it is an easy exercise to verify that the new function $\bar h$ will be indeed an isomorphism between $T^K_{i+1}(\alpha)$ and $T^I_{i+1}(\alpha)$.

Thus, the only thing that remains to be shown is that such an active trigger indeed exists. For that we need to show:

\smallskip


(A) All the sideatoms of $\kappa_{gp}$ required by $\sigma_{i}^\mathcal{I}$ occur in ${K}_i$.


(B) The trigger $(\sigma_i^\mathcal{I},h)$ for $\dep$ on $K_i$ is active.

\smallskip

For (A), suppose that $\pi$ is a sideatom type of $\gamma_{i}^\mathcal{I}$ required by $\sigma_{i}^\mathcal{I}$. We know that there is $\beta^\pi \in {I}_i$ such that $\beta^\pi \crel{sp}^\pi \beta_{gp}$.
If $\beta^\pi \in {\mathcal I}_{\alpha^I_i}= {\mathcal I}_{\bar h(\alpha)}$, then, by induction hypothesis,
there is $\kappa^\pi \in {\mathcal K}_\alpha \cap K_i$ such that $\kappa^\pi \crel{sp}^\pi \kappa$.
But what if $\beta^\pi$ is a remote side-parent?

Here is where the essence of the construction of $D_{\mi{ac}}$, and of the Depth Condition, reveals itself. If $\beta^\pi \not\in {\mathcal I}_{\alpha^I_i}$, then there exists $\alpha'\in D$ such that $\langle \alpha^I_i, \beta^I_i, \alpha', \beta^\pi \rangle$ is a remote-side-parent situation, and $\beta^\pi  = \beta^\mathcal{I}_j$ for some $j < i$.
Thus, in $(T_{\mi{ac}},\lambda)$ there is an edge $(v,u)$ such that $\lambda(v) = \alpha$ and $\lambda(u) = \kappa'$, for some $\kappa'$ with $\bar h(\kappa')=\alpha'$, as postulated in the Inductive Step of the construction of $D_{\mi{ac}}$.

We know that either $\dept{\alpha}\leq \ell_\infty-i$ or $\dept{\alpha}> \ell_\infty-i$ but $\alpha=\alpha^\infty$. In both cases, by induction hypothesis, $\bar h$ is an isomorphism from $T^K_{i}(\kappa')$ to $T^I_{i}(\kappa')$. Let now $\kappa^\pi$ be an element of $T^K_i(\kappa')$ such that $\bar h(\kappa^\pi)=\beta^\pi$.
It follows from the fact that $\bar h$ is an isomorphism from $T^K_{i}(\alpha)$ and $T^I_{i}(\alpha)$, and from Lemma \ref{lem:acyclic-database-properties}(3), that $\bar h$ is also an isomorphism from $T^K_{i}(\alpha)\cup T^K_{i}(\kappa')$ to $T^I_{i}(\alpha)\cup T^I_{i}(\kappa')$ (notice that guardedness is crucial here). Hence, $\kappa^\pi \crel{sp}^\pi \kappa_{gp}$.

For (B), assume that the trigger $(\sigma_i^\mathcal{I},h)$ for $\dep$ on $K_i$ is not active. Thus, there is $\alpha_{\mi{bad}} \in K_i$ such that $\alpha_{\mi{bad}} \crel{s} \result{\sigma_i^\mathcal{I},h}$. We can then conclude that $\bar h(\alpha_{\mi{bad}}) \crel{s} \bar h(\result{\sigma_i^\mathcal{I},h})$. This follows from the fact that, by claim (2) of  Lemma~\ref{lem:acyclic-database-properties}, $\bar h$ is a homomorphism from $\mathcal{K}$ to $\mathcal{I}$, and thus, if terms are equal in $\alpha_{\mi{bad}}$ and $\result{\sigma_i^\mathcal{I},h}$, then they are not less equal in $\bar h(\alpha_{\mi{bad}})$ and $\bar h(\result{\sigma_i^\mathcal{I},h})$. But then $(\sigma_i^\mathcal{I},h)$ for $\dep$ on $I_i$ is not active due to the atom $\bar h(\alpha_{\mi{bad}}) \in I_i$, which is a contradiction. This concludes the proof of Lemma \ref{k-structure}.
\end{proof}

Let us try to intuitively explain the above complicated lemma. Clearly, both $\mathcal I$ and $\mathcal K$ are forests (with $\crel{\mi{gp}}$ being the tree relation). The roots of the trees in $\mathcal I$ are atoms of $D$, while the roots of the trees in $\mathcal K$ are atoms of $D_{\mi{ac}}$. Each atom in $D_{\mi{ac}}$ has its original atom in $D$, and $\bar h$ tells us which is this atom.
Now, the second claim of the lemma (which looks simpler) states the following: at every stage of the construction of $\mathcal K$, the tree that has been constructed up to this point over the root $\alpha^\infty \in D_{\mi{ac}}$, it is isomorphic to the tree that has been built over $\alpha^\infty$ up to the same point of the construction of $\mathcal I$.
This is actually expected since the construction of $\mathcal K$ is exactly the construction of $\mathcal I$, but seen in a room full of distorting mirrors, and imagining that $\alpha^\infty$ is {\em the only} element of $D_{\mi{ac}}$ that is not a mirror image, but the real atom.
Regarding the first claim, as long as $i$ is small enough, the situation is similar to the one in (2). The tree constructed in $\mathcal K$, until stage $i$, over the root $\alpha\in D_{\mi{ac}}$ is isomorphic to the tree constructed in $\mathcal I$ until the same point in time over the root $\bar h(\alpha)\in D$. For some time we can see a faithful image of the construction, despite the fact that many mirror reflections are needed. But, when $i$ is too large (compared to $\dept{\alpha}$) we can no longer see anything new. Notice that, in particular, if $\dept{\alpha} = \ell_\infty$, the lemma states that no tree at all will be built over the root $\alpha$.

Let us now state a useful corollary, which directly follows from Lemma~\ref{k-structure}; for the proof of claim (2)
recall that ${\mathcal I}_{\alpha^\infty}$ is infinite, while the proof of claim (5) uses claims (3) and (4).

\begin{corollary}\label{k-take-away}
\begin{enumerate}
\item
$\bar h$ is an isomorphism from $\mathcal{K}_{\alpha^{\infty}}$ to $\mathcal{I}_{\alpha^\infty}$.

\item
The weakly restricted chase derivation $(K_i)_{i \geq 0}$ is infinite.

\item
For each atom $\alpha \in  D_{\mi{ac}}$ such that $\alpha \neq {\alpha^\infty}$, the mapping $\bar h$ is an isomorphism from $\mathcal{K}_{\alpha}$ to $\mathcal{I}_{\bar h(\alpha)}\cap I_{\ell_\infty - \dept{\alpha}}$.

\item
For $\alpha, \alpha' \in D_{\mi{ac}}$ with $\bar h(\alpha)= \bar h(\alpha')$ and $\dept{\alpha} \leq \dept{\alpha'}$, there is a 1-1 homomorphism $g$ from ${\mathcal K}_{\alpha'}$ to ${\mathcal K}_\alpha$ and $g(\alpha) = \alpha'$.

\item
Let $\beta,\beta' \in D_{\mi{ac}}$ such that $\dept{\beta}\leq \dept{\beta'}$. For each $\alpha\in {\mathcal K}_\beta$ and $ \alpha' \in {\mathcal K}_{\beta'} $ such that $\bar h(\alpha)= \bar h(\alpha')$, there
exists a 1-1 homomorphism $g$ from ${\mathcal K}_{\alpha'}$ to ${\mathcal K}_\alpha$ and $g(\alpha)=\alpha'$.
\end{enumerate}
\end{corollary}

\noindent \paragraph{\underline{Step 3: An Infinite Restricted Chase Derivation}}

\smallskip

\noindent In this last step of the proof of the Treeification Theorem, our task is to extract from the infinite weakly restricted chase derivation of $D_{\mi{ac}}$ w.r.t.~$\dep$ constructed above, an infinite restricted chase derivation $(J_i)_{i \geq 0}$ of $D_{\mi{ac}}$ w.r.t.~$\dep$.
For this, we first need to a fix a notation allowing us to directly address the atoms of $\mathcal{K}$.

Let $\mathcal N$ be the set of pairs of natural numbers defined as
\[
\{(i,j) : i \geq 0 \text{ and } 0 \leq j < |K_{i+1}\setminus K_{i}|\}.
\]
By $\leq$ and $<$ we denote the lexicographic ordering on $\mathcal N$. Note that $\langle {\mathcal N}, <\rangle $ and $\langle {\mathbb N}, < \rangle $ are isomorphic. Now, let $(\kappa_w)_{w\in {\mathcal N}}$ be an enumeration of all the atoms of $\mathcal{K}$ such that:
\begin{itemize}
\item $\kappa_{[i,j]}\in K_{i+1} \setminus K_{i}$, and

\item if $ \kappa_{[i,j]} \in \mathcal{K}_\alpha$ and $\kappa_{[i,j']}\in \mathcal{K}_{\beta}$, for some $j \leq j'$ and $\alpha,\beta \in D_\mi{ac}$, then $\dept{\alpha} \leq \dept{\beta}$.
\end{itemize}

\begin{algorithm}[t]
    $\mi{Pending} := \mathcal{K} \setminus K_0$;\\
    $\mi{Born} := K_0$;\\
    $\mi{Stopped} := \emptyset$;\\
    $m := 0$;\\
    $J_0 := K_0$;\\
    \While{$\mi{Pending} \neq \emptyset$}
    {
            let $\kappa$ be the $\leq$-smallest element of $\mi{Pending}$;\\
            $\mi{Pending} := \mi{Pending} \setminus \{\kappa\}$;\\
            \eIf{\text{\rm there is an active trigger} $(\sigma,h)$ \textrm{\rm for} $\dep$ \textrm{\rm on} $J_m$ \text{\rm such that} $\kappa = \result{\sigma,h}$}
            {
            $\mi{Born} := \mi{Born} \cup \{\kappa\}$;\\
            $J_{m+1} := J_m \cup \{\kappa\}$;\\
            $m := m+1$;
            }
            {
            $\mi{Stopped} := \mi{Stopped} \cup \{\kappa\}$;\\
            \ForEach{$\beta \in \mi{Pending}$ \text{\rm such that} $\kappa \ccrel{\mi{gp}} \beta$}
            {
            $\mi{Pending} := \mi{Pending} \setminus \{\beta\}$;\\
            $\mi{Stopped} := \mi{Stopped} \cup \{\beta\}$;
            }
            }
    }
    \Return{$(J_i)_{i \geq 0}$}.
\end{algorithm}

We  now present a simple (not necessarily terminating) procedure, dubbed $\mathsf{Extract}(\mathcal{K},\dep)$, that extracts from $\mathcal{K}$ an infinite restricted chase derivation $(J_i)_{i \geq 0}$ of $D_{\mi{ac}}$ w.r.t.~$\dep$. This algorithm is depicted in the box above.
It is clear that each time the while-loop is entered it holds that $J_m = \mi{Born}$. It also follows by construction that:

\begin{lemma}\label{lem:extract-correctness}
The sequence of instances  $(J_i)_{i \geq 0}$ produced by $\mathsf{Extract}(\mathcal{K},\dep)$ is a restricted chase derivation of $D_{\mi{ac}}$ w.r.t.~$\dep$.
\end{lemma}

The crucial question is whether this sequence is infinite. A positive answer to this question will conclude the proof of the treeification theorem. The rest of the section is devoted to showing that:

\begin{lemma}\label{lem:extract-infinite}
The sequence of instances $(J_i)_{i \geq 0}$ produced by $\mathsf{Extract}(\mathcal{K},\dep)$ is infinite.
\end{lemma}

We first show the following {\em loop invariant lemma}, that intuitively states the following: at each point of the execution of our iterative procedure, if an atom is not stopped, then there is a whole tuple of candidates that can act as its side-parents that are also not stopped.
%

\begin{lemma}[\textbf{Loop Invariant}]\label{lem:loop-invariant}
Consider two atoms $\alpha,\beta \in \mathcal{K}$ such that $\beta \crel{sp}^{\pi} \alpha$, for some sideatom type $\pi$. If $\alpha \in \mi{Born} \cup \mi{Pending}$, then there exists $\beta' \in \mi{Born} \cup \mi{Pending}$ such that $\beta' \crel{sp}^{\pi} \alpha$.
\end{lemma}

\begin{proof}
We proceed by induction on the number of iterations of $\mathsf{Extract}(\mathcal{K},\dep)$. Clearly, the loop invariant holds at the beginning of the execution.
Suppose now that it holds at some point of the execution when we enter the while-loop. Let $\kappa_{[i,j]}$ be the current atom, i.e., the $\leq$-smallest atom of the set $\mi{Pending}$.

It is easy to see that the guard-parent of $\kappa_{[i,j]}$ necessarily belongs to $\mi{Born}$: it cannot be in $\mi{Pending}$ because then $\kappa_{[i,j]}$ would not be minimal in $\mi{Pending}$, and it cannot be in $\mi{Stopped}$ because in such a case $\kappa_{[i,j]}$ would be in $\mi{Stopped} $ too.
It is clear that, if there exists an active trigger for $\dep$ on $J_m$ that produces $\kappa_{[i,j]}$, then the set $\mi{Born} \cup \mi{Pending}$ remains unchanged since the algorithm will simply remove $\kappa_{[i,j]}$ from $\mi{Pending}$ and add it to $\mi{Born}$. Thus, in this case the loop invariant holds.

Assume now that there is no such an active trigger. There are only two cases in which this can happen:

\medskip


\paragraph{Case 1.} Some of the side-parents $\kappa_{[i,j]}$ needs are not in $\mi{Born}$. Assume that $\pi'$ is a sideatom type required by the TGD $\sigma_{i}^{\mathcal{I}}$, due to which $\kappa_{[i,j]}$ has been generated in $\mathcal K$, and let $\kappa_{[i',j']}$ be any candidate from $\mathcal{K}$ such that $\kappa_{[i',j']}\crel{sp}^{\pi'} \kappa_{[i,j]}$.
It should be clear that $i' < i$. By construction, $\kappa_{[i',j']} \in \mi{Born} \cup \mi{Stopped}$ (since $\kappa_{[i,j]}$ is minimal in $\mi{Pending}$). By induction hypothesis, there is, among these candidates, at least one that belongs to $\mi{Born} \cup \mi{Pending}$. Hence, there must be at least one of those candidates in $\mi{Born}$. This implies that this first case does not apply.

\medskip

\paragraph{Case 2.} There is an atom $\kappa_{[i',j']} \in J_m$ such that $\kappa_{[i',j']} \crel{s} \kappa_{[i,j]}$.
By an argument similar to that for statement (B) in the proof of Lemma~\ref{k-structure} above, we can show that in such a case it would be
$\bar h(\kappa_{[i',j']}) \crel{s} \bar h(\kappa_{[i,j]})$. Since $(I_i)_{i \geq 0}$ is a restricted chase derivation, we get that $i' \geq i$. But since $\kappa_{[i',j']} \in J_m$, we get that $i' = i$, and thus, $j' < j$.
This means that $\bar h(\kappa_{[i,j]}) = \bar h(\kappa_{[i',j']})$, and $\kappa_{[i,j]}\in {\mathcal K}_\beta$ and $\kappa_{[i',j']}\in {\mathcal K}_{\beta'}$ for some $\beta, \beta'\in D_{\mi{ac}}$ such that $\dept{\beta'} \leq \dept{\beta}$.
By Corollary~\ref{k-take-away}, ${\mathcal K}_{\kappa_{[i,j]}}$ is isomorphic to a subset of ${\mathcal K}_{\kappa_{[i',j']}}$ via an isomorphism, let us say $g$, such that $g(\kappa_{[i,j]}) = \kappa_{[i',j']}$.
Assume now that there are $\kappa,\kappa_\pi \in \mi{Born} \cup \mi{Pending}$ such that $\kappa_\pi \crel{sp}^\pi \kappa$. We need to show that, after $\kappa_{[i,j]}$ gets stopped, together with all its $\ccrel{\mi{gp}}$-descendants, the loop invariant will still hold.
Observe that if $\kappa_\pi \not\in {\mathcal K}_{\kappa_{[i,j]}}$, then it is not affected by the removal of atoms of ${\mathcal K}_{\kappa_{[i,j]}}$. Moreover, if $\kappa \in {\mathcal K}_{\kappa_{[i,j]}}$, then it gets stopped, and there is nothing to show about its side-parents.
Thus, the only case that we need to worry about is when $\kappa \not\in {\mathcal K}_{\kappa_{[i,j]}}$ and $\kappa_\pi \in {\mathcal K}_{\kappa_{[i,j]}}$.
In this case all the terms in $\kappa_\pi$ occur in $\fr{\kappa_{[i,j]}}$. Furthermore, since $\kappa_{[i',j']} \crel{s} \kappa_{[i,j]}$, $\fr{\kappa_{[i,j]}} = \fr{\kappa_{[i',j']}}$.
Hence, the atom $g(\kappa_\pi)$, which is in ${\mathcal K}_{\kappa_{[i,j]}}$, and thus in
$\mi{Born} \cup \mi{Pending}$ after the current iteration, is such that $g(\kappa_\pi)\crel{sp}^\pi \kappa$, and the claim follows.
\end{proof}

In order to understand the meaning of this lemma, recall that $\mathcal K$, since it was produced by a weakly restricted chase, is a multiset, and there can be many atoms $\beta \in \mathcal K$ such that $\beta \crel{sp}^{\pi} \alpha$. This is a phenomenon that never happens in a normal restricted chase.
By exploiting the loop invariant lemma, we can show that none of the atoms of $\mathcal{K}_{\alpha^{\infty}}$ is stopped during our iterative procedure.

\begin{lemma}\label{lem:infinite-non-stopped-atoms}
For each $\alpha \in \mathcal{K}$, if $\alpha \in \mathcal{K}_{\alpha^{\infty}}$, then $\alpha$ occurs in an instance of the sequence $(J_i)_{i\geq 0}$ produced by $\mathsf{Extract}(\mathcal{K},\dep)$.
\end{lemma}

\begin{proof}
We need to show that $\alpha \not\in \mi{Stopped}$ for any $\alpha \in \mathcal{K}_{\alpha^{\infty}}$.
Assume that there exists an atom of $\mathcal{K}_{\alpha^{\infty}}$ that belongs to $\mi{Stopped}$; let $\hat{\alpha}$ be the $<$-smallest such atom. The loop invariant lemma (Lemma~\ref{lem:loop-invariant}) ensures that $\hat{\alpha}$ belongs to $\mi{Stopped}$ not because some of its side-parents are missing, but for a different reason; in fact, for one of the following two reasons:
\begin{enumerate}
\item There exists an atom $\beta \in \mi{Stopped}$ such that $\beta \ccrel{\mi{gp}} \hat{\alpha}$. Clearly, $\beta \in \mathcal{K}_{\alpha^{\infty}}$ and also $\beta < \hat{\alpha}$. But this contradicts the fact that $\hat{\alpha}$ is the $<$-smallest atom of $\mathcal{K}_{\alpha^{\infty}}$ that has been stopped. Thus, this reason does not apply.

\item There is $\beta \in \mi{Born}$, with $\beta < \hat{\alpha}$ and $\beta \crel{s} \hat{\alpha}$. Let $\beta = \kappa_{[i',j']}$ and $\hat{\alpha} = \kappa_{[i,0]}$; it follows from the definition of $(\kappa_w)_{w \in \mathcal{N}}$ that $\kappa_{[i,j]} \in \mathcal{K}_{\alpha^{\infty}}$ implies $j = 0$. Clearly, $[i',j'] < [i,0]$. However, since $\kappa_{[i',j']} \crel{s} \kappa_{[i,0]}$, we get that $i' \geq i$. Thus, $j < 0$, which is not possible. Hence, also this reason does not apply.
\end{enumerate}
Since none of the above cases apply, the claim follows.
\end{proof}

Having Lemma~\ref{lem:infinite-non-stopped-atoms}, it is
clear that Lemma~\ref{lem:extract-infinite} follows. Indeed, since
$\mathcal{K}_{\alpha^{\infty}}$ is infinite (Corollary ~\ref{k-take-away}), and since each restricted chase step generates just one atom, we immediately get that the sequence $(J_i)_{i\geq 0}$ of instances produced by
$\mathsf{Extract}(\mathcal{K},\dep)$ is infinite.
Therefore, $(J_i)_{i\geq 0}$ is an {\em infinite} restricted chase derivation of $D_{\mi{ac}}$ w.r.t.~$\dep$. This completes the proof of the Treeification Theorem.


\subsection{Proof of Lemma~\ref{lem:chaseable-guarded-trees-msol}}

Let us assume, for the moment, that we have available the following auxiliary MSOL formulas (more details are given below); as usual, we use lower-case letters $x,y,\ldots$ for first-order variables, and upper-case letters $A,B,\ldots$ for second-order variables:
\begin{eqnarray*}
\phi_{\mi{fin}}(A) &\equiv& A \text{ is finite }\\
\phi_{\pi}(x,y) &\equiv& x \crel{\mi{sp}}^{\pi} y, \text{ for the sideatom type } \pi\\
\phi_b(x,y) &\equiv& x \crel{b}^{+} y.
\end{eqnarray*}

By exploiting the above formulas, we can easily define $\phi_\dep$ as the conjunction of the following four sentences:
\begin{enumerate}
\item $\phi_{\mi{jt}}$ checks whether $T$ is an abstract join tree. It is easy to verify that all the conditions in the definition of
abstract join  trees (see Definition~\ref{j-guarded-tree}) are first-order expressible, apart from the first one, which states that the set $\{x \in V : \origin{x} = F\}$ is finite. For this check we exploit the MSOL formula $\phi_{\mi{fin}}$.

\item $\phi_1$ checks for the first condition of Definition~\ref{def:chaseable-guarded-tree} as follows:
    \[
    \hspace{10mm} \forall x \forall A \, (\forall y \, (\phi_b(y,x)\ \leftrightarrow\ y \in A)\ \ra\ \phi_{\mi{fin}}(A))
    \]

\item $\phi_2$ checks for the second condition; in what follows, we assume that $\sigma$ has body $\alpha,\pi_1,\ldots,\pi_k$:
    \[
    \hspace{8mm} \forall x \forall y \, \left(x \Yleft y\ \wedge\ \origin{y} = \sigma\ \ra\ \bigwedge_{i \in \{1,\ldots,k\}} \exists z \, \phi_{\pi_i}(z,y)\right)
    \]
    Notice that $\origin{y} = \sigma$ is an abbreviation of a big disjunction that checks, via monadic predicates $M_{\tau}$, where $\tau \in \Lambda_\dep$, whether the label of $y$ is of the form $\langle \cdot, \sigma, \cdot \rangle$.

\item $\phi_3$ checks for the third condition as follows:
    \[
    \hspace{6mm}\forall x \, \neg \phi_b(x,x)
    \]
\end{enumerate}

We proceed to give more details about the auxiliary formulas used in $\phi_\dep$. The formal definitions are omitted since they are long and tedious, but we give enough evidence that the formulas are indeed expressible in MSOL. Note that the following discussion heavily relies on the obvious fact below, which we will silently use:

\begin{fact}\label{fact:connectedness}
Let $T = \langle V, \Yleft \rangle$ be an abstract join tree. For each term $t$ in $\Delta(T)$, $\{x \in V : t \text{ occurs in } \delta(x)\}$ induces a connected subtree of $T$.
\end{fact}

\noindent
\paragraph{Formula $\phi_{\mi{fin}}(A)$.} This formula comes from the general MSOL toolbox. It states that every infinite directed path $B$ in $T$, starting from the root node of $T$, has an infinite directed sub-path, starting from some non-root element of $B$, which is disjoint with $A$.

\medskip

\noindent \paragraph{Formula $\phi_{=}^{i,j}(x,y)$, for each $i,j \in \{1,\ldots,\ar{\dep}\}$.} Notice that these formulas have not been explicitly used above. However, they are needed for defining $\phi_\pi$ and $\phi_b$. The formula $\phi_{=}^{i,j}(x,y)$ says that the term in $\delta(x)$ at position $i$ is equal to the term in $\delta(y)$ at position $j$.
This can be expressed in MSOL as follows: there is a set $A \subseteq V$ such that (i) $A$ is a path with $x$ and $y$ being its ends, i.e., $A$ is finite, $x,y$ have exactly one neighbor in $A$, and any other node in $A$ has exactly two neighbors, and (ii) $A$ is a disjoint union of $A_1,\ldots,A_{\ar{\dep}}$ such that $x \in A_i$, $y \in A_{j}$, and, for all pairs $z,w \in A$ such that $z \Yleft w$, $z \in A_k$, $w \in A_{\ell}$ it holds that $[[f,k],[m,\ell]] \in \eq{w}$.

\medskip

\noindent
\paragraph{Formula $\phi_{\pi}(x,y)$.} The formula says that $\delta(x) \subseteq_\pi \delta(y)$. It should be clear that it can be easily expressed by exploiting the formulas $\phi_{=}^{i,j}$ given above for checking whether terms in atoms are equal.

\medskip

\noindent
\paragraph{Formula $\phi_{b}(x,y)$.} We first devise a formula $\psi_b(x,y)$, which states that $x \crel{b} y$. Such a formula can be defined by using $\phi_\pi$ above, and also the formula $\phi_s(x,y) \equiv x \crel{s} y$, which can be in turn defined by exploiting the formulas $\phi_{=}^{i,j}(x,y)$, for $i,j \in \{1,\ldots,\ar{\dep}\}$.

Having $\psi_b$ we can then devise a formula $\phi_{\mi{cl}}(A)$, which states that $A$ is $\crel{b}$-downward closed, i.e.,
for each $x,y \in V$, with $x \crel{b} y$ and $y \in A$ there is also  $x \in A$.

Finally, $\phi_b(x,y)$ simply says that, for every $\crel{b}$-downward closed set $A$ it holds that $y \in A$ implies $x \in A$.


\section{Proofs from Section~\ref{sec:stickiness}}
%


\subsection{Proof of Lemma~\ref{lem-step-3}}

Let us first establish an auxiliary claim, which essentially states that every free connected proto-caterpillar trivially satisfies condition (2) of Definition~\ref{caterpillar}:

\begin{lemma}\label{proto-smoto}
Consider a free connected proto-caterpillar $\diamondsuit = (L^\diamondsuit,(\alpha_i^\diamondsuit)_{i \geq 0},(\sigma_i^\diamondsuit,h_i^\diamondsuit)_{i > 0},(\gamma_i^\diamondsuit)_{i > 0})$. For $\beta \in L^\diamondsuit$ and $i > 0$, $\beta \not\crel{s} \alpha_i^\diamondsuit$.
\end{lemma}

\begin{proof}
Towards a contradiction, assume that there exists $\beta \in L^\diamondsuit$ and $i > 0$ such that $\beta \crel{s} \alpha_i^\diamondsuit$. This implies that there exists a relay term $\ccc$ of $\heartsuit$ occurring in $\fr{\alpha_i^\diamondsuit}$ that occurs also in $\beta$.
Let $\alpha_j^\diamondsuit$, for $j < i$, be the birth atom of $\ccc$. In the special case where $\ccc$ is the first relay term of $\diamondsuit$, then $j=0$. Assuming that $\ccc'$ is the next relay term of $\diamondsuit$ after $\ccc$, let $k > i$ be such that $\alpha_k^\diamondsuit$ is the birth atom of $\ccc'$.
By connectedness, $\ccc$ occurs in $\fr{\alpha_\ell^\diamondsuit}$ for every $j < \ell \leq k$.
Moreover, we know that there exists $j \leq \ell \leq k$ such that $\beta \crel{p} \alpha_\ell^\diamondsuit$. Since $\beta \neq \alpha_{\ell-1}^\diamondsuit$ and $\diamondsuit$ is free, we can conclude that the TGD $\sigma_\ell^\diamondsuit$ (recall that the trigger $(\sigma_\ell^\diamondsuit,h_\ell^\diamondsuit)$ generates $\alpha_\ell^\diamondsuit$), apart from the atom $\gamma_\ell^\diamondsuit$, which is mapped by $h_\ell^\diamondsuit$ to $\alpha_{\ell-1}^\diamondsuit$, has another atom in its body that is mapped by $h_\ell^\diamondsuit$ to $\beta$, while it shares a variable $x$ with $\gamma_\ell^\diamondsuit$ and $h_\ell^\diamondsuit(x) = \ccc$.
Since $\dep$ is a sticky set of TGDs, we conclude that $\ccc$ occurs at an immortal position, which contradicts the fact that $\diamondsuit$ is connected.
\end{proof}

We are now ready to give the proof of Lemma~\ref{lem-step-3}.

\begin{proof}[Proof of Lemma~\ref{lem-step-3}]
The fact that $\heartsuit$ is a connected proto-caterpillar follows from the fact that $\spadesuit$ is a connected proto-caterpillar (Lemma~\ref{lem-step-2}). In particular, by applying $\bar h$ on the atoms occurring in $\spadesuit$, there is no way to violate the conditions (1) - (3) of Definition~\ref{proto-caterpillar}, or the connectedness condition as defined in Definition~\ref{caterpillar-c}.
Moreover, it follows by construction that, for each $(\alpha,i), (\beta,j) \in \Pi(L^\heartsuit \cup B^\heartsuit)$, $\alpha[i] = \beta[j]$ implies $(\alpha,i) \simeq_{L^\heartsuit \cup B^\heartsuit}^* (\beta, j)$, and thus, $\heartsuit$ is free.
It remains to show that $\heartsuit$ enjoys the two conditions of Definition~\ref{caterpillar}, which we recall here:
\begin{enumerate}
\item for each $\beta \in L^\heartsuit$ and $i > 0$, $\beta \not\crel{s} \alpha_i^\heartsuit$, and

\item for each $0 \leq i < j$, $\alpha_i^\heartsuit \not\crel{s} \alpha_j^\heartsuit$.
\end{enumerate}

Since $\heartsuit$ is a free conected proto-caterpillar, (1) immediately follows from Lemma~\ref{proto-smoto}.
%

For (2), towards a contradiction, assume that $\alpha_i^\heartsuit \crel{s} \alpha_j^\heartsuit$ for some $0 \leq i < j$. Since we know that, if a term in $\alpha_i^\heartsuit$ is equal to a term in $\alpha_j^\heartsuit$, then the terms at the same positions in $\alpha_i^\spadesuit$ and $\alpha_j^\spadesuit$ are also equal, we get that $\alpha_i^\spadesuit \crel{s} \alpha_j^\spadesuit$. This implies that $\alpha_{i+n}^\clubsuit \crel{s} \alpha_{j+n}^\clubsuit$; recall, from the construction of $\spadesuit$, that $n$ is such that $\alpha_{n}^\clubsuit$ is the birth atom of the relay term $\ccc_{i_0}$. Therefore, there are atoms $\beta,\beta' \in \mathcal I$ such that $\beta \crel{p}^+ \beta'$ and $\beta \crel{s} \beta'$. But this contradicts the fact that $\beta \not\crel{s} \beta'$ since $(I_i)_{i \geq 0}$ is a restricted chase derivation.
\end{proof}


\subsection{Proof of Lemma~\ref{lem:buchi-automaton}}

The high-level idea of the construction is as follows.
We first show that for an equality type $e = (R,E)$ from $\etypes{\sch{\dep}}$, and a set of positions $\Pi \subseteq \{1,\ldots,\ar{R}\}$, we can build a deterministic B\"{u}chi automaton ${\mathcal A}_{e,\Pi}$ such that $L({\mathcal A}_{e,\Pi}) \neq \emptyset$ iff there exists a free connected caterpillar such that its body starts with an atom of equality type $e$, and its first relay term occurs at positions $\Pi$ of this atom.
This means that a word $\mathbf {w}$ (which we will call caterpillar word and its over a finite alphabet $\Lambda_\dep$) accepted by ${\mathcal A}_{e,\Pi}$ is actually a symbolic representation of a free connected caterpillar as the one above.
Observe now that there are finitely many pairs $(e,\Pi)$, where $e = (R,E)$ is an equality type of $\etypes{\sch{\dep}}$, and $\Pi \subseteq \{1,\ldots,\ar{R}\}$; let $\mathsf{etp}_\dep$ be the set of all such pairs.
Since B\"{u}chi automata are closed under union, i.e., given two B\"{u}chi automata ${\mathcal A}_1$ and ${\mathcal A}_2$, we can construct a B\"{u}chi automaton, denoted ${\mathcal A}_1 \cup {\mathcal A}_2$, that recognizes the language $L({\mathcal A}_1) \cup L({\mathcal A}_2)$, the desired automaton is defined as the deterministic B\"{u}chi automaton
\[
{\mathcal A}_\dep\ =\ \bigcup_{(e,\Pi)\ \in\ \mathsf{etp}_\dep} \, {\mathcal A}_{e,\Pi}.
\]
Thus, our main task in the remainder of the section is, for a pair $(e,\Pi) \in \mathsf{etp}_\dep$, to construct the B\"{u}chi automaton ${\mathcal A}_{e,\Pi}$.





\medskip

\noindent
\paragraph{\underline{Caterpillar Words and Automata}}

\smallskip

\noindent It is easy to see that a free proto-caterpillar is fully described (up to isomorphism, of course) by the equality type of the first atom $\alpha_0$ of its body, and an infinite sequence of TGD-atom pairs $(\sigma_i, \gamma_i)_{i > 0}$, which tells us which TGD of $\dep$ should be used to produce the next atom of the proto-caterpillar's body, and which atom of the body of this TGD must match with the previous atom of the proto-caterpillar's body. The remaining atoms of the body of the TGD tell us which are the leg atoms of the proto-caterpillar.\footnote{Notice that here we silently assume, w.l.o.g., that the proto-caterpillar is minimal in the sense that all the leg atoms participate in the generation of a body atom.}
Of course, not each such sequence translates to a free proto-caterpillar (as it may happen that some $\gamma_{i+1}$ does not match with the $i$-the atom according to $\alpha_0$ and the sequence $(\sigma_1, \gamma_1),(\sigma_2, \gamma_2),\ldots,(\sigma_i, \gamma_i)$), but if it does then the free proto-caterpillar is unique. However, there is no guarantee that this unique free proto-caterpillar is a connected caterpillar.
In order to fully describe a free connected caterpillar we also need somehow to mark the pass-on points. This brings us to the notion of the caterpillar word (for $\dep$). 

We first define the finite alphabet $\Lambda_\dep$, which consists of triples of the form $(\sigma,\gamma,P)$, where $\sigma \in \dep$, $\gamma \in \body{\sigma}$, and, assuming that $R$ is the predicate of $\head{\sigma}$, $P \subseteq \{1,\ldots,\ar{R}\}$ is such that $P \neq \emptyset$ implies there exists $i \in \{1,\ldots,\ar{R}\}$ with $\head{\sigma}[i] \not\in \fr{\sigma}$ and $P = \{j : \head{\sigma}[i] = \head{\sigma}[j]\}$. Then:

\begin{definition}\label{def:caterpillar-word}
A {\em caterpillar word} (for $\dep$) is an infinite word $\mathbf{w} = w_1 w_2 \cdots$ such that, for each $i \geq 1$, $w_i \in \Lambda_\dep$. \hfill\markfull
\end{definition}

Intuitively, a caterpillar word $\mathbf{w} = w_1w_2, \cdots$, with $w_i = (\sigma_i,\gamma_i,P_i)$, is a candidate symbolic representation of a free connected caterpillar, where $w_i$ marks a pass-on point iff $P_i$ is non-empty. In fact, $P_i$ indicates at which positions of $\head{\sigma_i}$ the new relay term appears.
Now, given a pair $(e_0,\Pi_0) \in \mathsf{etp}_\dep$, we say that $\mathbf{w}$ {\em encodes a free connected caterpillar starting at $(e_0,\Pi_0)$}\footnote{We keep this definition semi-formal as the formal one is very tedious and it does not add any technical value to the proof.} if the sequence of TGD-atom pairs $(\sigma_i,\gamma_i)_{i > 0}$ translates to a free connected caterpillar $\diamondsuit = (L^\diamondsuit,(\alpha_{i}^{\diamondsuit})_{i \geq 0},(\sigma_{i}^\diamondsuit,h_{i}^\diamondsuit)_{i > 0},(\gamma_{i}^{\diamondsuit})_{i > 0})$, where
(i) $\et{\alpha_0^\diamondsuit} = e_0$, and the first relay term of $\diamondsuit$ occurs in $\alpha_0^\diamondsuit$ at positions $\Pi_0$,
%
(ii) $\sigma_i = \sigma_i^\diamondsuit$ and $\gamma_i = \gamma_i^\diamondsuit$, for each $i > 0$, and
%
(iii) assuming that $b_1 < b_2 < \cdots$ are the pass-on points of $\diamondsuit$, $P_i \neq \emptyset$ iff $i \in \{b_1,b_2,\ldots\}$, and the $k$-th relay term of $\diamondsuit$ occurs in $\alpha_{b_k}^\diamondsuit$ at positions $P_k$.

Recall that our goal is to construct a deterministic B\"{u}chi automaton ${\mathcal A}_{e_0,\Pi_0}$, with $\Lambda_\dep$ being its alphabet, such that $L({\mathcal A}_{e_0,\Pi_0})$ is exactly the set of caterpillar words that encode a free connected caterpillar starting at $(e_0,\Pi_0)$. The automaton ${\mathcal A}_{e_0,\Pi_0}$ is defined as the (almost) cartesian product of three automata:
\begin{itemize}
\item ${\mathcal A}_{\mathit{pc}}$ that checks whether a caterpillar word $\mathbf{w}$ encodes a free proto-caterpillar such that $e_0$ is the equality type of the first atom of its body. Note that the set $\Pi_0$ does not play any role here. In fact, this automaton will only read the first two elements of each letter of $\mathbf{w}$.

\item ${\mathcal A}_{\mathit{qc}}$ that checks whether a caterpillar word $\mathbf{w}$ that encodes a free proto-caterpillar $\diamondsuit$ is such that $\diamondsuit$ is a {\em quasi-caterpillar}, i.e., it satisfies condition (2) of Definition~\ref{caterpillar}, that is, assuming that $(\alpha_i^\diamondsuit)_{i \geq 0}$ is the body of $\diamondsuit$, $\alpha_i^\diamondsuit \not\crel{s} \alpha_j^\diamondsuit$ for each $0 \leq i < j$. This automaton is quite involved, and as ${\mathcal A}_{\mathit{pc}}$ above, it will read only the first two elements of each letter of $\mathbf{w}$.

\item ${\mathcal A}_{\mathit{cc}}$ that checks whether a caterpillar word $\mathbf{w}$ that encodes a free quasi-caterpillar $\diamondsuit$ is such that $\diamondsuit$ is connected. Since, by Lemma~\ref{proto-smoto}, a free connected quasi-caterpillar is a caterpillar, ${\mathcal A}_{\mathit{cc}}$ essentially checks whether $\diamondsuit$ is a free connected caterpillar, i.e., condition (1) of Definition~\ref{caterpillar} comes for free.
\end{itemize}

By ``almost cartesian product'' we mean that the states of ${\mathcal A}_{e_0,\Pi_0}$ will be triples, consisting of the states of ${\mathcal A}_{\mathit{pc}}$, ${\mathcal A}_{\mathit{qc}}$, and ${\mathcal A}_{\mathit{cc}}$. The transition functions of ${\mathcal A}_{\mathit{pc}}$ and ${\mathcal A}_{\mathit{cc}}$ will only depend on the current symbol of the word $\bf w$ and of the current state of the respective automaton. However, the transition function of ${\mathcal A}_{\mathit{qc}}$ will also use the current state of ${\mathcal A}_{\mathit{pc}}$ as part of its argument.
Regarding the acceptance, each of the three automata will have a designated $\mathit{reject}$ state. If any of them is ever encountered, then we assume that ${\mathcal A}_{e_0,\Pi_0}$ immediately rejects the input word $\mathbf{w}$. Apart from the $\mathit{reject}$ state, ${\mathcal A}_{cc}$ will have an accepting state. The automaton ${\mathcal A}_{e_0,\Pi_0}$, which, as we said, is a B\"{u}chi automaton, will accept if ${\mathcal A}_{\mathit{cc}}$ will encounter this accepting state infinitely many times while reading the word $\bf w$.
We can now describe the three automata in question.


\medskip

\noindent
\paragraph{\underline{The Automaton ${\mathcal A}_{\mathit{pc}}$}}

\smallskip

\noindent Since we are building a finite automaton, and there are infinitely many atoms in a caterpillar's body, there is no hope the automaton, after reading the symbol $w_i$ of the input caterpillar word ${\bf w} = w_1w_2\cdots$, could ``know'' the $i$-th atom of the body of the proto-caterpillar encoded by   $\bf w$. But it can know its equality type.
We define the function $\delta_\mathsf{et} : \etypes{\sch{\dep}} \times \Lambda_\dep \ra \etypes{\sch{\dep}} \cup \{\mathit{reject}\}$ as follows: for each $e \in \etypes{\sch{\dep}}$ and $(\sigma,\gamma,\cdot) \in \Lambda_\dep$,
\begin{itemize}
\item if there is a homomorphism $h$ that maps $\gamma$ to $R(\star_1,\ldots,\star_n)$, with $\star_i = \star_j$ iff $i,j$ coexist in a set of $E$, then $\delta_\mathsf{et}(e,(\sigma,\gamma,\cdot)) = \et{\bar h(\head{\sigma})}$, where $\bar h$ is an extension of $h$ that maps each existentially quantified variable in $\sigma$ to a distinct symbol;

\item otherwise, $\delta_\mathsf{et}(e,(\sigma,\gamma,\cdot)) = \mathit{reject}$.
\end{itemize}

Let ${\mathcal A}_{\mathit{pc}}$ be a B\"{u}chi automaton with
\begin{itemize}
\item $\etypes{\sch{\dep}} \cup \{\mathit{reject}\}$ its set of states,

\item $\Lambda_\dep$ its alphabet,

\item $\delta_\mathsf{et}$ its transition function, and

\item $e_0$ its initial state.
\end{itemize}
Clearly, for a caterpillar word $\mathbf{w}$, ${\mathcal A}_{\mathit{pc}}$ does not reject $\bf w$ iff $\bf w$ encodes a free proto-caterpillar $\diamondsuit = (\cdot,(\alpha_i^\diamondsuit)_{i \geq 0},\cdot,\cdot)$ such that $\et{\alpha_0^\diamondsuit} = e_0$; and if it does, then $\delta_\mathsf{et}(\et{\alpha_{i-1}^\diamondsuit}, w_i) = \et{\alpha_i^\diamondsuit}$, for each $i > 0$.

\medskip

\noindent
\paragraph{\underline{The Automaton ${\mathcal A}_{\mathit{qc}}$}}

\smallskip

\noindent From now on we assume that an input caterpillar word $\bf w$ encodes a free proto-caterpillar $\diamondsuit = (\cdot,(\alpha_i^\diamondsuit)_{i \geq 0},\cdot,\cdot)$ such that the first atom of its body has equality type $e_0$ (in case it does not encode such a free proto-caterpillar, ${\mathcal A}_{\mathit{pc}}$ will take care of it), and proceed towards checking whether $\diamondsuit$ is a quasi-caterpillar, i.e., whether it satisfies condition (2) of Definition~\ref{caterpillar}, that is, $\alpha_i^\diamondsuit \not\crel{s} \alpha_j^\diamondsuit$ for each $0 \leq i < j$.

%
%

Given a finite set $\mathfrak T$ of terms, a {\em $\mathfrak T$-equality type} over a schema $\ins{S}$ is essentially an equality type $(R,E)$ over $\ins{S}$ where, in addition, some of the sets of $E$ are labeled with distinct terms of $\mathfrak T$ indicating that a term $t \in \mathfrak T$ should occur at certain positions. Formally, a {\em $\mathfrak T$-equality type} over $\ins{S}$ is a triple $(R,E,\lambda)$, where $(R,E) \in \etypes{\ins{S}}$, and $\lambda$ is a partial injective function from $E$ to $\mathfrak T$. It is clear that there are only finitely many $\mathfrak T$-equality types over $\ins{S}$.
The $\mathfrak T$-equality type of an atom $\alpha$, denoted $\eqtt{\mathfrak T}{\alpha}$, as well as the canonical atom of a $\mathfrak T$-equality type $e$, denoted $\can{e}$, are defined in the expected way.

Now, for brevity, let ${\mathfrak T}_j$ be the set of terms occurring in $\alpha_j^\diamondsuit$. We can easily show the following useful lemma:

\begin{lemma}\label{lem:aux-stop}
Suppose $i < j < k$ for some $i,j,k \geq 0$. It holds that $\alpha_i^\diamondsuit \crel{s} \alpha_k^\diamondsuit$ iff $\can{\eqtt{{\mathfrak T}_j}{\alpha_i^\diamondsuit}} \crel{s} \alpha_k^\diamondsuit$.
\end{lemma}

\begin{proof}
First, for each set of terms ${\mathfrak T}$ that contains all the terms occurring both in $\alpha_i^\diamondsuit$ and $\alpha_k^\diamondsuit$, it is easy to show that $\alpha_i^\diamondsuit \crel{s} \alpha_k^\diamondsuit$ iff $\can{\eqtt{{\mathfrak T}}{\alpha_i^\diamondsuit}} \crel{s} \alpha_k^\diamondsuit$.
Now, since $\diamondsuit$ is free, if some term occurs in $\alpha_i^\diamondsuit$ and $\alpha_k^\diamondsuit$, then it must also occur in $\alpha_j^\diamondsuit$. This implies that ${\mathfrak T}_j$ contains all the terms occurring both in $\alpha_i^\diamondsuit$ and $\alpha_k^\diamondsuit$, and the claim follows.
\end{proof}

For each $j \geq 0$, let
%
$\Theta_j=\{ \eqtt{{\mathfrak T}}{\alpha_i^\diamondsuit} :  0 \leq i \leq j\}$.
Of course, $\Theta_j$ is a finite set, for each $j \geq 0$. Moreover, if we just know $\et{\alpha_j^\diamondsuit}$, then the number of possible candidates for $\Theta_j$ is finite, and uniformly bounded, so $\Theta_j$ can be seen as a finite piece of information, or as (part of) a state of a finite automaton.
It is possible to construct $\Theta_{j+1}$ only knowing $\Theta_j$, $\et{\alpha_j^\diamondsuit}$ and the $(j+1)$-th symbol $w_{j+1}$ of $\mathbf{w}$. Furthermore, knowing $\Theta_j$ and $\et{\alpha_j^\diamondsuit}$, we can check whether there is $0 \leq i < j$ such that $\alpha_i^\diamondsuit \crel{s} \alpha_j^\diamondsuit$; the latter is a consequence of Lemma~\ref{lem:aux-stop}.
Thus, we can define a function $\delta_\Theta$ such that:

\medskip

\noindent $\delta_\Theta\left((\Theta_j,\et{\alpha_j^\diamondsuit}),(\sigma_{j+1},\gamma_{j+1},\cdot)\right)$\\
\begin{eqnarray*}
=\ \left\{
\begin{array}{ll}
\mathit{reject} & \text{ if } \alpha_i^\diamondsuit \crel{s} \alpha_j^\diamondsuit \text{ for some } i<j,\\
&\\
(\Theta_{j+1},\et{\alpha_{j+1}^\diamondsuit}) & \text{otherwise.}
\end{array} \right.
\end{eqnarray*}

Let ${\mathcal A}_{\mathit{qc}}$ be a B\"{u}chi automaton with
\begin{itemize}
\item its set of states consisting of pairs of the form $(\Theta,e)$ as above, and the $\mathit{reject}$ state,

\item $\Lambda_\dep$ its alphabet,

\item $\delta_\Theta$ its transition function, and

\item $(\emptyset,e_0)$ its initial state.
\end{itemize}
By construction, for a caterpillar word $\mathbf{w}$, ${\mathcal A}_{\mathit{qc}}$ does not reject $\bf w$ iff $\bf w$ encodes a free proto-caterpillar $\diamondsuit$ that satisfies condition (2) of Definition~\ref{caterpillar}, i.e., $\diamondsuit$ is a quasi-caterpillar.

\medskip

\noindent
\paragraph{\underline{The Automaton ${\mathcal A}_{\mathit{cc}}$}}

\smallskip

\noindent First, we need to define a function $\delta_{\mathit{pos}}$ that will let the automaton remember some terms. Given $\Pi \subseteq \{1,\ldots,\ar{\dep}\}$ and $w = (\sigma,\gamma,P) \in \Lambda_\dep$, let $\delta_{\mathit{pos}}(\Pi,w)$ be the set of integers
\[
\{i \in \{1,\ldots,\ar{\dep}\} : \text{ there is } j \in \Pi \text{ such that } \gamma[j] = \head{\sigma}[i]\}.
\]
The purpose of $\delta_{\mathit{pos}}$ will be made clear in a while.

The states of ${\mathcal A}_{\mathit{cc}}$ will be tuples $(\Pi_1,\Pi_2,q)$, where $\Pi_1,\Pi_2 \subseteq \{1,\ldots \ar{\dep}\}$ and $q \in \{\top,\bot\}$.
Roughly, $\Pi_1$ will remember the positions where the current relay term appears (we need this information since we must make sure that the current relay term survives until the next pass-on point), and $\Pi_2$ will remember the positions where all the relay terms, current and older ones, live at the given moment (we need this to make sure that they will never appear at an immortal position).

We proceed to define the function $\delta_{\mathit{cc}}$ as follows: given a state-symbol pair $(s,w)$ with $s = (\Pi_1,\Pi_2,q)$ and $w = (\sigma,\gamma,P)$:
\begin{itemize}
\item if $\delta_{\mathit{pos}}(\Pi_1, w) = \emptyset$, or there exists $i \in \delta_{\mathit{pos}}(\Pi_2,w)$ such that $\head{\sigma}[i]$ is not marked in $\dep$, then $\delta_{\mathit{cc}}(s,w) = \mathit{reject}$;

\item otherwise, $\delta_{cc}(s,w) = (\delta_{\mathit{pos}}(\Pi_1, w),\delta_{\mathit{pos}}(\Pi_2, w),\bot)$ if $P = \emptyset$, and $\delta_{\mathit{cc}}(s,w) = (P, \delta_{\mathit{pos}}(\Pi_1, w) \cup \delta_{\mathit{pos}}(\Pi_2, w),\top)$ if $P \neq \emptyset$.
\end{itemize}
Intuitively, this means that if we are not at a pass-on point ($P = \emptyset$), then keep track of the positions occupied by the current and the old relay terms. On the other hand, if we are at a pass-on point ($P \neq \emptyset$), then forget the positions occupied by the old relay terms and remember the positions at which the new one occurs. But do not forget them completely; simply add them to the set of positions where all the relay terms appear.

Let ${\mathcal A}_{\mathit{cc}}$ be a B\"{u}chi automaton with
\begin{itemize}
\item its set of states consisting of triples as described above, and the $\mathit{reject}$ state,

\item $\Lambda_\dep$ its alphabet,

\item $\delta_{\mathit{cc}}$ its transition function,

\item $(\Pi_0,\emptyset,e_0)$ its initial state, and

\item all the states of the form $(\cdot,\cdot,\top)$ being accepting.
\end{itemize}


\subsection{Proof of Lemma~\ref{lem:main-unifying-function}}
%

%
Let $\diamondsuit = (L^\diamondsuit,B^\diamondsuit,T^\diamondsuit,G^\diamondsuit)$ be a free uniformly connected caterpillar, where $B^\diamondsuit = (\alpha_i^\diamondsuit)_{i \geq 0}$, $T^\diamondsuit = (\sigma_i^\diamondsuit,h_i^\diamondsuit)_{i > 0}$, and $G^\diamondsuit = (\gamma_i^\diamondsuit)_{i > 0})$.
Let $\ccc_0,\ccc_1,\ldots$ be the relay terms of $\diamondsuit$. Moreover, assuming that $(b_i)_{i > 0}$ are the pass-on points of $\diamondsuit$, let $d \geq 0$ be such that $b_{k+1} - b_k < d$, for each $k \geq 0$.
%
%
Our goal is to define a unifying function $h$ for $\diamondsuit$ such that $h(\diamondsuit)$ is a finitary caterpillar, which means that $h(L^\diamondsuit)$ is finite.

We first observe that no matter how a unifying function $h$ for $\diamondsuit$ is defined, $h(\diamondsuit)$ is a proto-caterpillar that satisfies condition (1) of Definition~\ref{caterpillar}. This is what the next lemma tells us:

\begin{lemma}\label{lem:finitary-1}
Consider a unifying function $h$ for $\diamondsuit$.
Then:
\begin{enumerate}
\item $h(\diamondsuit)$ is a proto-caterpillar, and
\item for each $\beta \in h(L^\diamondsuit)$ and $i > 0$, $\beta \not\crel{s} h(\alpha_i^\diamondsuit)$.
\end{enumerate}
\end{lemma}

\begin{proof}
It is clear that there is no way to violate the conditions given in Definition~\ref{proto-caterpillar} by unifying terms in the legs of a proto-caterpillar. Since, by hypothesis, $\diamondsuit$ is a proto-caterpillar, we get that $h(\diamondsuit)$ is a proto-caterpillar, and (1) follows.

Concerning (2), the claim follows by the fact that none of the relay terms $\ccc_0,\ccc_1,\ldots$ of $\diamondsuit$ occurs in $L^\diamondsuit$. This implies that none of the terms $\ccc_0,\ccc_1,\ldots$ occurs in $h(L^\diamondsuit)$, while, for each $i \geq 0$, $\fr{h(\alpha^\diamondsuit_i)}$ contains a term from $\ccc_0,\ccc_1,\ldots$. Therefore, none of the atoms of $h(L^\diamondsuit)$ can stop an atom of $(h(\alpha_i^\diamondsuit))_{i > 0}$, and the claim follows.
\end{proof}

Having the above lemma in place, it is clear that to establish Lemma~\ref{lem:main-unifying-function} it remains to construct a unifying function $h$ for $\diamondsuit$ such that $h(L^\diamondsuit)$ is finite, and $h(\diamondsuit)$ satisfies condition (2) of Definition~\ref{caterpillar}. 
The rest of the section is devoted to constructing such a function.

We first define the domain of the desired function as the set of terms ${\mathfrak V} \subseteq \adom{L^\diamondsuit}$ that occur at a position $(\alpha,i)$, for some atom $\alpha \in L^\diamondsuit$, that is not related to any immortal position $(\beta,j)$, where $\beta \in B^\diamondsuit$. Notice that none of the relay terms of $\diamondsuit$ occur in ${\mathfrak V}$.
It would be useful to be able to refer to the terms of ${\mathfrak V}$ that participate in the generation of the atoms between the first body atom and the first pass-on point, as well as the atoms between two consecutive pass-on points.
Let $B_0^\diamondsuit = \{\alpha_j^\diamondsuit : 0 < j \leq b_{1}\} \subseteq B^\diamondsuit$, that is, the set of atoms between $\alpha_0^\diamondsuit$ and the birth atom of $\ccc_1$.
Moreover, for each $i > 0$, let $B_i^\diamondsuit = \{\alpha_j^\diamondsuit : b_i < j \leq b_{i+1}\} \subseteq B^\diamondsuit$, that is, the set of atoms between the birth atom of $\ccc_i$ and the birth atom of $\ccc_{i+1}$.
We also define, for each $i \geq 0$, $L_i^\diamondsuit$ as the set of atoms
\begin{multline*}
\left\{\alpha \in L^\diamondsuit : \text { there exists } j > 0 \text{ such that }\right.\\
\left.\result{\sigma_{j}^\diamondsuit,h_{j}^\diamondsuit} \in B_i^\diamondsuit, \text{ and } \alpha \in h_j^\diamondsuit(\body{\sigma_j^\diamondsuit})\right\},
\end{multline*}
which are essentially the atoms that are needed to generate $B_i^\diamondsuit$. Then, we let ${\mathfrak V}_i = {\mathfrak V} \cap \adom{L_i^\diamondsuit}$.
%

For the codomain we need a sufficiently large {\em finite} set of new terms.
Let $m_0 \geq 0$ be greater than the maximal number of variables in a TGD of $\dep$ and $m = (d+1) \cdot m_0$; recall that $d$ is the uniform distance between two consecutive pass-on points of $\diamondsuit$. We define $\mathfrak T$, which will be the codomain of the desired function, as a set of terms such that $|\mathfrak T| = 2m$ and $\mathfrak T \cap \adom{L^\diamondsuit \cup B^\diamondsuit} = \emptyset$, i.e., $\mathfrak T$ collects $m$ new terms that do not occur in $L^\diamondsuit \cup B^\diamondsuit$.
%
We can then show the following key technical lemma:

\begin{lemma}\label{lem:unifying-function}
There exists a unifying function $\hbar : {\mathfrak V}\rightarrow \mathfrak T$  such that, for each $i \geq 0$, the unifying function $\hbar_{|{\mathfrak V}_i}$ is 1-1.
\end{lemma}

\begin{proof}
We first observe that:
\begin{enumerate}
\item[(*)] for each $i \geq 0$, $|{\mathfrak V}_i| \leq m $ -- this is a consequence of the definition of $m$;

\item[(**)] for each $i > 0$, ${\mathfrak V}_i \cap \bigcup_{j<i} {\mathfrak V}_j \subseteq {\mathfrak V}_{i-1}$ -- since all the terms that occur both in $\bigcup_{j<i} {\mathfrak V}_j$ and in ${\mathfrak V}_i$ must also occur in $\alpha_{b_i}^{\diamondsuit}$.
\end{enumerate}

We are going to build an ascending sequence $(\hbar_i)_{i \geq 0}$ of functions, where $\hbar_i: \bigcup_{ j \leq i} {\mathfrak V}_j \rightarrow \mathfrak T$, such that, for each $i \geq 0$, the function ${\hbar_i}_{|{\mathfrak V}_i}$ is 1-1. Then, $\hbar$ will be defined as
$\bigcup_{i \geq 0} \hbar_i$.

Let $\hbar_0$ be an 1-1 function of the form ${\mathfrak V}_0 \rightarrow \mathfrak T$. Notice that such a function exists since, by definition, $\mathfrak T$ is sufficiently large.
Suppose now that $\hbar_{i-1}$, as specified above, has been defined. In order to define $\hbar_i$ we need to extend $\hbar_{i-1}$ to the terms in ${\mathfrak V}_i \setminus \bigcup_{j<i} {\mathfrak V}_j$ in such a way that the newly defined function is 1-1 on ${\mathfrak V}_i$.
From (**) we know that ${\mathfrak V}_i \setminus \bigcup_{j<i} {\mathfrak V}_j={\mathfrak V}_i \setminus {\mathfrak V}_{i-1} $, and, by assumption,
${\hbar_{i-1}}_{|{\mathfrak V}_{i-1}}$ is 1-1. This means that on the subset of ${\mathfrak V}_i$ where $\hbar_i$ is already defined (since $\hbar_{i-1}$ is defined) it is 1-1. Now, to be able to extend it to an 1-1 function on the entire set ${\mathfrak V}_i$ we need to have enough terms in the codomain, which is guaranteed by (*).
\end{proof}

Let $\hbar$ be the unifying function for $\diamondsuit$ provided by Lemma~\ref{lem:unifying-function}. We proceed to show that:

\begin{lemma}\label{lem:finitary-2}
\begin{enumerate}
\item The instance $\hbar(L^\diamondsuit)$ is finite.
\item For each $0 \leq i < j$, $\hbar(\alpha_i^\diamondsuit) \not\crel{s} \hbar(\alpha_j^\diamondsuit)$.
\end{enumerate}
\end{lemma}

\begin{proof}
For (1), since $\mathfrak V$ collects all the terms of $\adom{L^\diamondsuit}$ that occur at a position $(\alpha,i)$, for some $\alpha \in L^\diamondsuit$, that is not related to any immortal position $(\beta,j)$, where $\beta \in B^\diamondsuit$, we can conclude that $\adom{L^\diamondsuit} \setminus \mathfrak V$ is finite. Therefore, $\adom{\hbar(L^\diamondsuit)}$ is finite, which in turn implies that $\hbar(L^\diamondsuit)$ is finite, as needed.

For (2), we proceed by considering the following two cases:
\begin{itemize}
\item \underline{$i \leq b_k < j$ for some $k > 0$.} In this case, there exists $\ell \geq k$ such that the relay term $\ccc_\ell$ occurs in $\fr{\alpha_j^\diamondsuit}$ but not in $\fr{\alpha_i^\diamondsuit}$. Since none of the relay terms of $\diamondsuit$ occurs in the domain or the codomain of $\hbar$, we conclude that $\ccc_\ell$ occurs in $\fr{\hbar(\alpha_j^\diamondsuit)}$ but not in $\fr{\hbar(\alpha_i^\diamondsuit)}$, which implies that $\hbar(\alpha_i^\diamondsuit) \not\crel{s} \hbar(\alpha_j^\diamondsuit)$.

\item \underline{$b_k \leq i < j \leq b_{k+1}$ for some $k \geq 0$ (with $b_0 = 0$.)} Since, by hypothesis, $\diamondsuit$ is a caterpillar, we get that $\alpha_i^\diamondsuit \not\crel{s} \alpha_j^\diamondsuit$. The fact that $\hbar$ is an 1-1 function over ${\mathfrak V}_k$ allows us to conclude that $\{\alpha_i^\diamondsuit,\alpha_j^\diamondsuit\}$ is isomorphic to $\{\hbar(\alpha_i^\diamondsuit), \hbar(\alpha_j^\diamondsuit)\}$. Therefore, $\hbar(\alpha_i^\diamondsuit) \not\crel{s} \hbar(\alpha_j^\diamondsuit)$, and the claim follows.
\end{itemize}
This completes the proof of the lemma.
\end{proof}

By Lemma~\ref{lem:finitary-1} and~\ref{lem:finitary-2}, we immediately get Lemma~\ref{lem:main-unifying-function}.

\end{document}